\documentclass[longauth]{aa}

\usepackage{subfig}
\usepackage{graphicx}
\usepackage{adjustbox}
\usepackage{hyperref}
\usepackage{txfonts}
\usepackage[dvipsnames]{xcolor}
\usepackage{longtable}
\usepackage{pdflscape}
\usepackage{booktabs}
\newcommand{\nb}[1]{\textbf{\color{ForestGreen} #1}}

\def\h2o{H$_2$O}
\def\hc3n{HC$_3$N}
\def\so2{SO$_2$}
\def\sic2{SiC$_2$}
\def\kms{km\,s$^{-1}$}

\begin{document}

   \title{ATOMIUM: Molecular inventory of 17 oxygen-rich evolved stars observed with ALMA\thanks{Tables A.1--A.5 are only available in electronic form at the CDS via anonymous ftp to cdsarc.cds.unistra.fr (130.79.128.5) or via https://cdsarc.cds.unistra.fr/cgi-bin/qcat?J/A+A/}}


   \author{S. H. J. Wallstr\"om
          \inst{1}\thanks{email: sofia.wallstrom@kuleuven.be}
          \and
          T. Danilovich\inst{1,2,3}        
          \and
          H.~S.~P.~M\"uller\inst{4}
          \and
          C.~A.~Gottlieb\inst{5}
          \and
          S.~Maes\inst{1}
          \and
          M.~Van~de~Sande\inst{6}
          \and
          L.~Decin\inst{1,7}
          \and
          A.~M.~S.~Richards\inst{8}
          \and 
          A.~Baudry\inst{9}
          \and
          J.~Bolte\inst{10}
          \and
          T.~Ceulemans\inst{1}
          \and
          F.~De~Ceuster\inst{1}
          \and
          A.~de~Koter\inst{11,1}
          \and
          I.~El~Mellah\inst{12,13}
          \and
          M.~Esseldeurs\inst{1}
          \and
          S.~Etoka\inst{8}
          \and
          D.~Gobrecht\inst{14}
          \and
          E.~Gottlieb\inst{15}
          \and
          M.~Gray\inst{8,16}
          \and
          F.~Herpin\inst{9}
          \and
          M.~Jeste\inst{17}
          \and
          D.~Kee\inst{18}
          \and
          P.~Kervella\inst{19}
          \and
          T.~Khouri\inst{20}
          \and
          E.~Lagadec\inst{21}
          \and
          J.~Malfait\inst{1}
          \and
          L.~Marinho\inst{9}
          \and
          I.~McDonald\inst{8,22}
          \and
          K.~M.~Menten\inst{18}
          \and 
          T.~J.~Millar\inst{23}
          \and
          M.~Montarg\`es\inst{20}
          \and
          J.~A.~Nuth\inst{24}
          \and
          J.~M.~C.~Plane\inst{7}
          \and
          R.~Sahai\inst{25}
          \and
          L.~B.~F.~M.~Waters\inst{26,27}
          \and
          K.~T.~Wong\inst{28,29}
          \and
          J.~Yates\inst{30}
          \and
          A.~Zijlstra\inst{8}
          }

        \institute{Institute of Astronomy, KU Leuven, Celestijnenlaan 200D, 3001 Leuven, Belgium 
        \and 
        School of Physics \& Astronomy, Monash University, Wellington Road, Clayton 3800, Victoria, Australia 
        \and
        ARC Centre of Excellence for All Sky Astrophysics in 3 Dimensions (ASTRO 3D), Clayton 3800, Australia 
         \and
         Universit\"at zu K\"oln, I. Physikalisches Institut, 50937 K\"oln, Germany 
         \and
         Harvard-Smithsonian Center for Astrophysics, 60 Garden Street, Cambridge, MA 02138, USA 
         \and
         School of Physics and Astronomy, University of Leeds, Leeds LS2 9JT, UK 
         \and
         School of Chemistry, University of Leeds, Leeds LS2 9JT, UK 
         \and
         JBCA, Department Physics and Astronomy, University of Manchester, Manchester M13 9PL, UK 
         \and
         Universit\'e de Bordeaux, Laboratoire d'Astrophysique de Bordeaux, 33615 Pessac, France 
         \and 
         Department of Mathematics, Kiel University, Heinrich-Hecht-Platz 6, 24118 Kiel, Germany 
         \and
         Anton Pannekoek Institute for Astronomy, University of Amsterdam, Science Park 904, 1098 XH Amsterdam, the Netherlands 
         \and
         Departamento de F\'isica, Universidad de Santiago de Chile, Av. Victor Jara 3659, Santiago, Chile 
         \and
         Center for Interdisciplinary Research in Astrophysics and Space Exploration (CIRAS), USACH, Chile 
         \and
         Department of Chemistry and Molecular Biology, University of Gothenburg, Kemig\aa rden 4, 412 96 Gothenburg, Sweden 
         \and
         School of Engineering and Applied Sciences and Department of Earth and Planetary Sciences, Harvard University, Cambridge, USA 
         \and
         National Astronomical Research Institute of Thailand, Chiangmai 50180, Thailand 
         \and
         Max-Planck-Institut f\"ur Radioastronomie, 53121 Bonn, Germany 
         \and
         National Solar Observatory, Makawao, HI, United States 
         \and
         LESIA, Observatoire de Paris, Universit\'e PSL, CNRS, Sorbonne Universit\'e, Universit\'e Paris Cit\'e, 5 place Jules Janssen, 92195 Meudon, France 
         \and
         Department of Space, Earth and Environment, Chalmers University of Technology, Onsala Space Observatory, 43992 Onsala, Sweden 
         \and
         Universit\'e C\^ote d’Azur, Laboratoire Lagrange, Observatoire de la C\^ote d’Azur, F-06304 Nice Cedex 4, France 
         \and
         School of Physical Sciences, The Open University, Walton Hall, Milton Keynes, MK7 6AA, UK 
         \and
         Astrophysics Research Centre, School of Mathematics and Physics, Queen's University Belfast, University Road, Belfast BT7 1NN, UK 
         \and
         NASA Goddard Space Flight Center, 8801 Greenbelt Road, Greenbelt, MD 20071, USA 
         \and
         Jet Propulsion Laboratory, MS 183-900, California Institute of Technology, Pasadena, CA 91109, USA 
         \and
         Department of Astrophysics/IMAPP, Radboud University, PO Box 9010, 6500 GL Nijmegen, The Netherlands 
         \and
         SRON Netherlands Institute for Space Research Sorbonnelaan 2, 3584 CA Utrecht, The Netherlands 
         \and
         Theoretical Astrophysics, Department of Physics and Astronomy, Uppsala University, Box 516, 751 20 Uppsala, Sweden 
         \and
         Institut de Radioastronomie Millim\'{e}trique, 300 rue de la Piscine, 38406 Saint-Martin-d'H\`{e}res, France 
         \and
         University College London, Department of Physics and Astronomy, London WC1E 6BT, United Kingdom 
             }

   \date{Received ; accepted }

 
  \abstract
   {The dusty winds of cool evolved stars are a major contributor of the newly synthesised material enriching the Galaxy and future generations of stars. However, the details of the physics and chemistry behind dust formation and wind launching have yet to be pinpointed. Recent spatially resolved observations show the importance of gaining a more comprehensive view of the circumstellar chemistry, but a comparative study of the intricate interplay between chemistry and physics is still difficult because observational details such as frequencies and angular resolutions are rarely comparable.}
   {Aiming to overcome these deficiencies, ATOMIUM is an ALMA Large Programme to study the physics and chemistry of the circumstellar envelopes of a diverse set of oxygen-rich evolved stars under homogeneous observing conditions at three angular resolutions between $\sim$0.02$\arcsec - 1.4\arcsec$. Here we summarize the molecular inventory of these sources, and the correlations between stellar parameters and molecular content.}
   {Seventeen oxygen-rich or S-type asymptotic giant branch (AGB) and red supergiant (RSG) stars have been observed in several tunings with ALMA Band~6, targeting a range of molecules to probe the circumstellar envelope and especially the chemistry of dust formation close to the star. We systematically assigned the molecular carriers of the spectral lines and measured their spectroscopic parameters and the angular extent of the emission of each line from integrated intensity maps.}
   {Across the ATOMIUM sample, we detect 291 transitions of 24 different molecules and their isotopologues. This includes several first detections in oxygen-rich AGB/RSG stars: PO $\varv=1$, \so2 $\varv_1=1$ and $\varv_2=2$, and several high energy H$_2$O transitions. We also find several first detections in S-type AGB stars: vibrationally excited HCN $\varv_2=2,3$ and SiS $\varv=4,5,6$, as well as first detections of the molecules SiC, AlCl, and AlF in W~Aql. Overall, we find strong correlations between the following molecular pairs: CS and SiS, CS and AlF, NaCl and KCl, AlO and SO, \so2 and SO, and \so2 and H$_2$O; meaning both molecules tend to have more detected emission lines in the same sources. The measured isotopic ratios of Si and S are found to be consistent with previous measurements, except for an anomalously high $^{29}$Si/$^{30}$Si ratio of $4 \pm 1$ in the RSG VX~Sgr.}
   {This paper presents the overall molecular inventory and an initial analysis of the large ATOMIUM dataset, laying the groundwork for future work deriving molecular abundances and abundance profiles using radiative transfer modeling which will provide more rigorous tests for chemical models.}


   \maketitle
%

\section{Introduction}

Cool evolved stars are a major contributor of the gas and dust returned to the interstellar medium \citep{tielens_physics_2005} through their dusty winds which can reach mass-loss rates of up to 10$^{-4}$~M$_\odot$\,yr$^{-1}$ \citep{hofner_mass_2018}. Low- and intermediate-mass ($\sim$0.8 -- 8~M$_\odot$) asymptotic giant branch (AGB) stars are known to have dust-driven winds, while the mass-loss mechanism of the rarer, more massive red supergiant (RSG) stars remains uncertain. In both cases, however, the details of the physics and chemistry behind dust formation and wind launching have yet to be pinpointed, a vital step in understanding how newly synthesised material from AGB and RSG stars enriches the Galaxy and future generations of stars.

These AGB and RSG stars and their accompanying circumstellar envelopes (CSEs) provide useful chemical laboratories for studying dust and molecule formation due to their relatively simple overall spherical structure and their (mostly) radial velocity fields. Other chemically interesting cool environments, such as star-forming regions or (proto)planetary disks, tend to have more complex spatial and velocity structures, making it more difficult to disentangle the effects of chemistry. 
The overall chemical composition of an AGB CSE depends on the C/O ratio of the central star \citep[see, for example,][]{habing_asymptotic_2003}, as most of the less abundant element is locked up in CO. 
An oxygen-rich or M-type star (C/O < 1) has more free oxygen to form for example silicate dust and oxygen-bearing molecules -- such as SO, TiO, AlO -- while a carbon-rich star (C/O > 1) has more free carbon to form for example amorphous carbon dust and molecules such as HC$_3$N and SiC$_2$. Stars with C/O $\sim$ 1 are called S-type stars and have a mixed chemistry. 
The more massive RSGs are all oxygen-rich stars, due to extra nucleosynthesis processes such as hot-bottom burning \citep{herwig_evolution_2005}, resulting in oxygen-rich CSEs.

To date more than 100 molecules and 15 dust species have been detected around AGB and RSG stars \citep{decin_evolution_2021}. The difficult-to-observe H$_2$ is by far the most abundant molecule, but it is followed in abundance by CO, whose lower energy rotational transitions are easily excited over much of the CSE and readily observable with ground-based telescopes.
Observations of CO and other molecular lines provide dynamical and physical information about the CSE, and help constrain chemical models of dust and molecule formation.
A very efficient means of determining the molecular contents of diverse Galactic sources are spectral line surveys, covering wide frequency bands in the millimeter and sub-millimeter domain. A range of surveys has been carried out for evolved stars with single-dish telescopes, which have discovered a wealth of new molecules. For example, \citet{tenenbaum_arizona_2010} report the first detections of PO, AlO and AlOH in space, in the RSG VY~CMa, as well as the first detections of NaCl, PN, NS, and HCO$^+$ in an oxygen-rich source; \citet{velilla_prieto_millimeter_2017} report the first detections of NO and H$_2$CO in an oxygen-rich source, namely IK Tau (a.k.a. NML~Tau); \citet{de_beck_surprisingly_2020} report the first detections of silicon- and carbon-bearing species SiC$_2$, SiN, C$_2$H, and HC$_3$N towards an S-type star, W~Aql; and in the extensively studied, high mass-loss rate carbon-rich star CW~Leo (a.k.a. IRC+10216) refractory metal-containing species such as AlCl, AlF, MgNC, MgCN, NaCN, NaCl, and KCl are reported by \citet{cernicharo_2_2000, he_spectral_2008, pardo_ultra-deep_2022}, alongside cyanopolyynes of the family HC$_{2n+1}$N, carbon chain radials up to C$_8$H, and cyclic compounds such as c-C$_3$H$_2$.

However, these single-dish observations at low angular resolution (typically $10\arcsec - 40\arcsec$) struggle to probe the spatial distributions of molecules in the CSE, which provide further vital constraints on chemical models. Greatly increased angular resolution is achieved with interferometry, alongside increased sensitivity allowing fainter molecular lines to be detected. This results in more accurate information on the relative excitation and formation of different species, and is especially useful for detecting rarer or short-lived species which are typically only seen close to the star \textendash{} especially those suspected to play a part in dust condensation, such as AlO or carbon chains \citep{agundez_chemical_2020}. To date, only a handful of interferometric molecular line surveys of AGB stars have been carried out due to the large observing time requirements. These include studies of some sources that also were covered by single-dish surveys: the carbon-rich CW~Leo \citep{patel_interferometric_2011} and the two oxygen-rich stars R Dor and IK Tau \citep{decin_study_2017,decin_alma_2018}. The RSG VY CMa has also been studied in this way \citep{kaminski_interferometric_2013}.

\citet{patel_interferometric_2011} observed the carbon-rich AGB star CW~Leo with a 3$\arcsec$ beam using the Submillimeter Array (SMA). They detect $\sim$200 new lines of the molecules known from previous single-dish surveys, including narrow vibrationally excited lines (many from HCN) close to the star \citep[see also][]{patel_submillimeter_2009}. They note that many of their unassigned lines are also narrow, possibly produced by vibrationally excited transitions of polyatomic molecules for which laboratory measurements of rest frequencies were not yet available.
With measurements of both emission sizes and velocities for a range of lines they are also able to characterise the wind acceleration, parameterised by a beta law velocity profile \citep{lamers_introduction_1999} with a fast acceleration ($\beta=0.5$).

\citet{kaminski_interferometric_2013} observed the very high mass-loss rate ($4\times10^{-4}$~M$_\odot$\,yr$^{-1}$) and asymmetric RSG VY~CMa with a 0.9$\arcsec$ beam, again using the SMA. They detect over 200 lines from 19 molecules, including AlCl, rotational emission from TiO for the first time in an oxygen-rich evolved star, and TiO$_2$ for the first time in space. A large number of lines of oxygen-rich molecules like SO$_2$ and SO are seen, as well as many lines of the relatively refractory NaCl including vibrationally excited transitions up to $\varv=3$. Most of the observed lines show multiple velocity components, and the molecular emission shows similar spatial complexity. 

\citet{decin_study_2017,decin_alma_2018} observed two archetypal oxygen-rich AGB stars, the high mass-loss rate ($5\times10^{-6}$~M$_\odot$\,yr$^{-1}$) IK~Tau and the low mass-loss rate ($1\times10^{-7}$~M$_\odot$\,yr$^{-1}$) R~Dor, at $\sim$0.15$\arcsec$ angular resolution with the Atacama Large Millimeter and sub-millimeter Array (ALMA). They detect $\sim$200 lines from 15 different molecules, including dust precursors such as SiO, AlO, AlOH, TiO, and TiO$_2$. Highly vibrationally excited SiO (up to $\varv=5$) is detected in both sources close to the star, and AlO is detected far beyond the dust condensation radius showing that this molecule does not become entirely locked up in dust grains. 
The data is also used to characterise the wind acceleration of both sources, showing much slower accelerations than for the carbon-rich CW~Leo ($\beta \sim 5-10$).

These interferometric surveys demonstrate the importance of spatially resolved observations for obtaining a more comprehensive view of the circumstellar chemistry, as different molecules and transitions can show emission at different spatial scales and with different morphologies. However, it is difficult to draw more general conclusions about AGB or RSG stars from the study of single sources. Even in aggregate, various studies have different sensitivities, frequency coverage, and angular resolutions, and hence are difficult to compare. The next step to elucidate the circumstellar chemistry is to observe a sample of stars with a range of stellar parameters, with identical observing setups. By performing homogeneous high-angular-resolution observations, we can directly compare the chemical inventories of different sources, and compare them with stellar parameters. 

ALMA Tracing the Origins of Molecules formIng dUst in oxygen-rich M-type stars (ATOMIUM, Program ID: 2018.1.00659.L) is an ALMA Large Programme to observe 17 oxygen-rich AGB and RSG stars at high angular resolution in a range of molecular transitions  \citep{decin_substellar_2020,gottlieb_atomium_2022}.
The ATOMIUM targets were chosen to represent a range of AGB mass-loss rates, chemical types (M-type and S-type stars), and pulsation behaviors (semi-regular variables (SR), long-period variables (LPV), and regular Mira variables). The sample also includes three RSG stars, and is summarised in Table~\ref{tab:sources}.
ATOMIUM is the first ALMA Large Programme for stellar evolution, and consists of a set of homogeneous high-resolution observations that allow unambiguous comparison of the physicochemical properties of the winds of the 17 evolved stars, 
expanding the sample of evolved stars studied at such high angular resolution by a factor of four. It also provides a detailed picture of the chemical and dynamical processes throughout the stellar wind, including measurements of isotopic ratios which provide an important tracer of nucleosynthesis in the core of the star.
Alongside the ATOMIUM overview paper \citep{gottlieb_atomium_2022}, several in-depth studies of individual sources \citep[]{homan_atomium_2020, homan_atomium_2021, danilovich_chemical_2023}, specific molecules \citep{danilovich_atomium_2021, baudry_atomium_2023}, and comparisons with optical polarized light \citep{montarges_vltsphere_2023} have already resulted from this data.

This paper presents the molecular inventory of the 17 oxygen-rich AGB and RSG sources of the ATOMIUM sample.
Section~2 briefly describes the observations undertaken with ALMA as part of the ATOMIUM Large Programme, and we summarize how the carriers of the spectral lines were assigned to specific molecules and how the spectroscopic parameters of the rotational lines (flux density, line width, and integrated area) were determined. 
In Section~3 we present our results, and Section~4 contains the conclusions.
Additionally, the accompanying online tables include full lists of the molecular lines identified in the sample and in each source, including line parameters and molecular emission sizes, and all the spectra.

\begin{table*}[ht] 
\caption{ATOMIUM sources and their stellar parameters.} 
\label{tab:sources} 
\centering 
\begin{tabular}{lccccccccc}
\hline\hline
Source & Stellar & Pulsation & $P$ & $T_\mathrm{eff}$ & Diameter$^{(1)}$ & $R_\ast ^{(2)}$ & MLR$^{(3)}$ & $v_\mathrm{exp}$ & $D^{(4)}$ \\ 
 & type & type & [days] & [K] & [mas] & [cm] & [M$_\odot$/yr] & [km/s] & [pc] \\ 
\hline 
S Pav & M-type & SRa & 381 \rlap{$^{(a)}$} & 3100 \rlap{$^{(b)}$} & 12.0 \rlap{$^{(c)}$} & 1.7E+13 & 1.3E-07 \rlap{$^{(a)}$} & 13.0 \rlap{$^{(c)}$} & \phantom{0}190 \rlap{$^{(d,e,f)}$} \\ 
RW Sco & M-type & Mira & 389 \rlap{$^{(g)}$} & 3300 \rlap{$^{(b)}$} & \phantom{0}4.9 \rlap{$^{(c)}$} & 2.1E+13 & 1.3E-07 \rlap{$^{(g)}$} & 18.5 \rlap{$^{(c)}$} & \phantom{0}560 \rlap{$^{(e)}$} \\ 
T Mic & M-type & SRb & 347 \rlap{$^{(a)}$} & 3300 \rlap{$^{(b)}$} & \phantom{0}9.3 \rlap{$^{(c)}$} & 1.2E+13 & 1.4E-07 \rlap{$^{(a)}$} & 12.7 \rlap{$^{(c)}$} & \phantom{0}175 \rlap{$^{(f)}$} \\ 
R Hya & M-type & Mira & 366 \rlap{$^{(h)}$} & 2100 \rlap{$^{(i)}$} & 23.0 \rlap{$^{(c)}$} & 2.2E+13 & 1.8E-07 \rlap{$^{(i)}$} & 22.2 \rlap{$^{(c)}$} & \phantom{0}126 \rlap{$^{(f)}$} \\ 
SV Aqr & M-type & LPV & -- & 3400 \rlap{$^{(b)}$} & \phantom{0}4.4 \rlap{$^{(c)}$} & 1.5E+13 & 2.7E-07 \rlap{$^{(a)}$} & 15.9 \rlap{$^{(c)}$} & \phantom{0}445 \rlap{$^{(f)}$} \\ 
U Her & M-type & Mira & 402 \rlap{$^{(h)}$} & 3100 \rlap{$^{(c)}$} & 11.0 \rlap{$^{(c)}$} & 2.2E+13 & 3.2E-07 \rlap{$^{(j)}$} & 19.7 \rlap{$^{(c)}$} & \phantom{0}266 \rlap{$^{(k,f)}$} \\ 
U Del & M-type & SRb & 119 \rlap{$^{(d)}$} & 2800 \rlap{$^{(c)}$} & \phantom{0}7.9 \rlap{$^{(c)}$} & 2.0E+13 & 3.7E-07 \rlap{$^{(a)}$} & 14.6 \rlap{$^{(c)}$} & \phantom{0}330 \rlap{$^{(d,e,h)}$} \\ 
V PsA & M-type & SRb & 148 \rlap{$^{(a)}$} & 2400 \rlap{$^{(a)}$} & 13.0 \rlap{$^{(c)}$} & 2.7E+13 & 4.8E-07 \rlap{$^{(a)}$} & 18.8 \rlap{$^{(c)}$} & \phantom{0}278 \rlap{$^{(d,e)}$} \\ 
$\pi^1$ Gru & S-type & SRb & 150 \rlap{$^{(i)}$} & 2300 \rlap{$^{(i)}$} & 21.0 \rlap{$^{(c)}$} & 2.6E+13 & 9.2E-07 \rlap{$^{(l)}$} & 64.5 \rlap{$^{(c)}$} & \phantom{0}164 \rlap{$^{(e)}$} \\ 
R Aql & M-type & Mira & 268 \rlap{$^{(h)}$} & 2800 \rlap{$^{(i)}$} & 12.0 \rlap{$^{(c)}$} & 2.1E+13 & 1.6E-06 \rlap{$^{(j)}$} & 12.8 \rlap{$^{(c)}$} & \phantom{0}230 \rlap{$^{(d,e,h)}$} \\ 
GY Aql & M-type & Mira & 468 \rlap{$^{(h)}$} & 3100 \rlap{$^{(b)}$} & 21.0 \rlap{$^{(c)}$} & 6.4E+13 & 2.3E-06 \rlap{$^{(m)}$} & 15.0 \rlap{$^{(c)}$} & \phantom{0}410 \rlap{$^{(f)}$} \\ 
W Aql & S-type & Mira & 479 \rlap{$^{(h)}$} & 2800 \rlap{$^{(c)}$} & 11.0 \rlap{$^{(c)}$} & 3.1E+13 & 2.7E-06 \rlap{$^{(n)}$} & 24.6 \rlap{$^{(c)}$} & \phantom{0}375 \rlap{$^{(d,e,f)}$} \\ 
KW Sgr & RSG & SRc & 647 \rlap{$^{(o)}$} & 3700 \rlap{$^{(c)}$} & \phantom{0}3.9 \rlap{$^{(c)}$} & 7.0E+13 & 3.6E-06 \rlap{$^{(p)}$} & 27.7 \rlap{$^{(c)}$} & 2400 \rlap{$^{(q)}$} \\ 
IRC-10529 & M-type & Mira & 680 \rlap{$^{(i)}$} & 2700 \rlap{$^{(i)}$} & \phantom{0}6.5 \rlap{$^{(c)}$} & 4.5E+13 & 1.0E-05 \rlap{$^{(i)}$} & 21.8 \rlap{$^{(c)}$} & \phantom{0}930 \rlap{$^{(i,f)}$} \\ 
AH Sco & RSG & SRc & 738 \rlap{$^{(r)}$} & 3700 \rlap{$^{(c)}$} & \phantom{0}5.8 \rlap{$^{(c)}$} & 9.8E+13 & 1.0E-05 \rlap{$^{(s)}$} & 35.4 \rlap{$^{(c)}$} & 2260 \rlap{$^{(t)}$} \\ 
IRC+10011 & M-type & Mira & 660 \rlap{$^{(i)}$} & 2700 \rlap{$^{(i)}$} & \phantom{0}6.5 \rlap{$^{(c)}$} & 3.6E+13 & 1.9E-05 \rlap{$^{(i)}$} & 23.1 \rlap{$^{(c)}$} & \phantom{0}740 \rlap{$^{(f)}$} \\ 
VX Sgr & RSG & SRc & 732 \rlap{$^{(u)}$} & 3500 \rlap{$^{(c)}$} & \phantom{0}8.8 \rlap{$^{(c)}$} & 1.0E+14 & 6.0E-05 \rlap{$^{(i)}$} & 32.9 \rlap{$^{(c)}$} & 1560 \rlap{$^{(v)}$} \\ 
\hline 
\multicolumn{10}{l}{\footnotesize 
\textbf{Notes:} $^{(1)}$ Optical angular diameter. } \\ 
\multicolumn{10}{l}{\footnotesize 
$^{(2)}$ Calculated from the stellar angular diameter and distance. } \\ 
\multicolumn{10}{l}{\footnotesize 
$^{(3)}$ Scaled to the new distance estimate, see Section~\ref{sect:mlrs}. } \\ 
\multicolumn{10}{l}{\footnotesize 
$^{(4)}$ Several distances have been updated with respect to the values given in \citet{gottlieb_atomium_2022}. } \\ 
\multicolumn{10}{l}{\footnotesize 
\textbf{References:} $^{(a)}$ \citet{olofsson_mass_2002}; $^{(b)}$ \citet{marigo_evolution_2008}; $^{(c)}$ \citet{gottlieb_atomium_2022}; $^{(d)}$ \citet{gaia_collaboration_gaia_2018};  } \\ 
\multicolumn{10}{l}{\footnotesize 
$^{(e)}$ \citet{bailer-jones_estimating_2021}; $^{(f)}$ \citet{andriantsaralaza_distance_2022}; $^{(g)}$ \citet{groenewegen_millimeter_1999}; $^{(h)}$ \citet{perryman_hipparcos_1997};  } \\ 
\multicolumn{10}{l}{\footnotesize 
$^{(i)}$ \citet{de_beck_probing_2010}; $^{(j)}$ \citet{young_co3--2_1995}; $^{(k)}$ \citet{vlemmings_improved_2007}; $^{(l)}$ \citet{doan_extended_2017}; $^{(m)}$ \citet{loup_co_1993};  } \\ 
\multicolumn{10}{l}{\footnotesize 
$^{(n)}$ \citet{ramstedt_circumstellar_2017}; $^{(o)}$ \citet{wittkowski_vltiamber_2017}; $^{(p)}$ \citet{mauron_mass-loss_2011}; $^{(q)}$ \citet{arroyo-torres_atmospheric_2013};  } \\ 
\multicolumn{10}{l}{\footnotesize 
$^{(r)}$ \citet{kiss_variability_2006}; $^{(s)}$ \citet{jura_mass-losing_1990}; $^{(t)}$ \citet{chen_structure_2008}; $^{(u)}$ \citet{samus_general_2017}; $^{(v)}$ \citet{chen_measuring_2007}  } \\ 
\end{tabular} 
\end{table*}

\section{Data}
\subsection{ALMA observations} \label{sect:data_alma}

The ATOMIUM ALMA Large Programme covers a frequency range of $\sim$213--270 GHz in 16 spectral tunings, with a total bandwidth of 27 GHz. Observations were carried out at three different spatial resolutions -- low-resolution with beam sizes of $\sim0.8\arcsec - 1.4\arcsec$, medium-resolution with beam sizes of $\sim0.2\arcsec - 0.5\arcsec$, and high-resolution with beam sizes of $\sim0.02\arcsec - 0.05\arcsec$. The maximum scales which can be imaged (maximum recoverable scale, MRS) for each resolution are approximately $8 - 10\arcsec$, $1.5 - 4\arcsec$, and $0.4 - 0.6\arcsec$, respectively \citep[see][Table E.3]{gottlieb_atomium_2022}. The field of view (to half maximum sensitivity) is $22\arcsec - 26\arcsec$, depending on frequency. 
Only half the tunings, covering CO and some of the other lines with the most extended emission, were observed at the lowest spatial resolution. There is, however, missing flux for at least some molecules (notably CO) in some of our sources. This can occur both for lines with angular extent larger than the field of view, and for emission which is smooth on scales larger than the MRS. For example, in R~Hya about 40\% of the CO flux is resolved out \citep{homan_atomium_2021}, $\pi^1$~Gru is missing $\sim$30--70\% of the CO flux \citep{homan_atomium_2020}, and W~Aql is missing 66\% of the CO flux \citep{danilovich_chemical_2023}. There is likely also missing flux for other molecules such as HCN and some transitions of SO$_2$, in at least some sources. The full extent of missing flux has not been quantified as single-dish spectra of the same frequency ranges are not available for most of our sources.

Observations took place between 2018--2021. All the high resolution data were taken during June--July in 2019, and hence essentially samples, for each source, only one part of the stellar phase \citep[see][for details]{gottlieb_atomium_2022}. 
Calibration and imaging were performed in CASA \citep{mcmullin_casa_2007} following standard ALMA procedures. The line-free continuum (mostly stellar emission) was identified and used for self-calibration, which was applied to all data, ensuring that the data for each star had a consistent position and flux density scale (measured in Jy/beam for the image cubes, and Jy for the extracted spectra). See Table E.2 in \citet{gottlieb_atomium_2022} for details of the continuum image properties.
Image cubes were made (after continuum subtraction) for each separate spatial resolution. The channel spacing of 0.9765625 MHz gives a velocity resolution of 1.1--1.4 km\,s$^{-1}$ depending on frequency. The rms noise $\sigma_\mathrm{rms}$ per spectral channel is $\sim1-5$ mJy, depending on target elevation and requested noise level; the high-resolution data usually have the lowest noise and are the most sensitive to compact or clumpy emission, while the low-resolution data has the greatest surface brightness sensitivity. 
The observations and data reduction are described fully in \citet{gottlieb_atomium_2022}, and the characteristics such as the rms noise level, $\sigma_\mathrm{rms}$, and angular resolution of each spectral cube are given in their Table~E.3. Note that in some cases (such as very bright, compact emission or weak, extended emission) the standard reduction parameters may leave low-level artefacts in the images for individual lines; papers on specific lines or targets, using optimised imaging and combination of the different spatial resolutions, may give slightly different values from those reported here \citep[and in][]{gottlieb_atomium_2022}. To allow for this and for the propagation of errors when combining data taken at different epochs, we adopt a conservative flux scale uncertainty of 15\% (although in some cases it may be much better than this).

\subsection{Processing and analysis}\label{sect:analysis}

We used spectra of each source at high, medium, and low angular resolution to identify the various molecular transitions.
Spectra were extracted in circular apertures, centered on the continuum peak, with diameters of 0.04$\arcsec$, 0.08$\arcsec$, and 0.12$\arcsec$ for the high-resolution data; diameters of 0.4$\arcsec$, 1.2$\arcsec$, and 3.6$\arcsec$ for the medium-resolution data; and diameters of 1.2$\arcsec$, 3.6$\arcsec$, and 10.8$\arcsec$ for the low-resolution data. At each resolution, the smallest aperture was chosen to be the average beam size rounded up to one significant figure, for consistency.

In each spectrum the line identification was carried out in a systematic way. Starting from a list of expected lines based on the line survey of IK~Tau and R~Dor by \citet{decin_alma_2018}, at the position of each potential line a soft parabola function \citep[as defined in][]{de_beck_probing_2010}
was fit with a least squares fitting algorithm. A visual inspection then ruled out spurious fits, added missing lines, and adjusted the fit of weak and blended lines. The rms ($\sigma$) was measured from the parts of the spectrum more than three times the wind expansion velocity ($v_\mathrm{exp}$) away from any line, and any potential lines below 2.5$\sigma$ are considered non-detections. Any previously unknown lines were added to the line list, and identified if possible. Inspection of integrated intensity (moment 0) maps by eye further ensured any weak lines correspond to coherent emission across several channels rather than a noise fluctuation. We have also checked the intrinsic brightness (at 300~K) of the various transitions to ensure that the detected lines of each molecule are among the brightest lines in our frequency range. We do not exclude the possibility that further lines may be detected with a different data reduction method or by stacking data; however, we leave this to future papers.

\begin{table*}[ht]
\caption{Extract of Table~A.5 which lists all measured lines in each source. } \label{tab:example_AH_Sco}
\centering
\begin{tabular}{llllrrrrrrr} 
\hline \hline
Source & Line & $\nu_0$ & Res.$^{(1)}$ & Ap.$^{(2)}$ & $F_{peak}$ & $I$ & RW$^{(3)}$ & BW$^{(4)}$ & $w_{vel}$ & $r$$^{(5)}$ \\ 
& & {\tiny [GHz]} & & {\tiny [$\arcsec$]} & {\tiny [Jy]} & {\tiny [Jy km\,s$^{-1}$]} & {\tiny [km\,s$^{-1}$]} & {\tiny [km\,s$^{-1}$]} & {\tiny [km\,s$^{-1}$]} & {\tiny [$\arcsec$]} \\ 
\hline
AH Sco & SiO $\varv=2$ $J=5-4$ & 214.089 & high & 0.04 & 1.103 & 7.765 & 23.5 & -21.6 & 45.1 & 0.08 \\
AH Sco & SiO $\varv=2$ $J=5-4$ & 214.089 & high & 0.12 & 1.679 & 11.892 & 23.5 & -18.9 & 42.4 & 0.08 \\
AH Sco & SiO $\varv=2$ $J=5-4$ & 214.089 & high & 0.08 & 1.644 & 11.594 & 23.5 & -18.9 & 42.4 & 0.08 \\
AH Sco & SiO $\varv=2$ $J=5-4$ & 214.089 & med & 3.6 & 2.604 & 15.148 & 15.3 & -14.7 & 30.1 & 0.37 \\ 
AH Sco & SiO $\varv=2$ $J=5-4$ & 214.089 & med & 1.2 & 2.616 & 15.449 & 18.1 & -14.7 & 32.8 & 0.37 \\ 
AH Sco & SiO $\varv=2$ $J=5-4$ & 214.089 & med & 0.4 & 2.552 & 14.964 & 20.8 & -20.2 & 41.0 & 0.37 \\
\multicolumn{11}{c}{$\cdots$} \\
\hline
\multicolumn{11}{l}{\footnotesize 
  \textbf{Notes:} $^{(1)}$Spatial resolution of the data; $^{(2)}$Aperture diameter used to extract the spectrum; $^{(3)}$ Velocity of the red line wing; } \\ 
  \multicolumn{11}{l}{\footnotesize   
  $^{(4)}$ Velocity of the blue line wing; $^{(5)}$ Angular emission radius. } \\ 
  \multicolumn{11}{l}{\footnotesize
  The full version of the table (which also includes columns for uncertainties) is available at the CDS. } \\
\end{tabular}
\end{table*}

Example medium and high angular resolution spectra of R~Hya are shown in Appendix~C, with molecular identifications, while Appendix~A contains more detailed information about the identified lines. 
Additional information about the lines and molecular data can be found in Appendix~\ref{sect:app:lab-spec-data},
Table~A.1 gives a full list of identified molecules by source, 
Table~A.2 gives the maximum radial extent of each molecule in each source, 
Table A.3 lists all identified molecular 
lines with references for the line parameters, and 
Table~A.4 lists all the unidentified lines. 
Table~A.5 list all detected lines in each source, and gives measurements taken at each spatial resolution (high, medium, low) where the line was detected. Each line was measured separately in the spectra extracted with the three different apertures given above.

Line parameters were extracted from each spectrum, and
for each detected line its peak flux, integrated intensity, velocity width (full width at zero power), and angular extent (from its integrated intensity map, see below) was measured. 
The peak flux was taken to be the flux in the brightest channel included within the line extent. 
The spectral extent of a line was measured between the channels where each line wing reaches 2.5$\sigma$, with a minimum width of four channels ($\approx$5 \kms), thus defining the velocity width and integrated intensity. As this was done automatically, broad weak lines may have their velocity widths and hence integrated intensities underestimated. 
The uncertainty on the peak flux was taken to be the rms plus the 15\% absolute flux uncertainty (see Section~\ref{sect:data_alma}), and the uncertainty on the line width was taken to be the channel width. 

Potentially blended lines and lines with absorption or double-peaked profiles were checked manually to ensure that the correct line region was chosen. For lines which are blended or truncated by the edge of an observing band, 
only a peak flux is reported, as measured at the line center, and for lines with absorption the "peak" flux value is actually the positive or negative extremum value. Note that some lines show (partial) absorption only at certain apertures, most commonly the smaller apertures. 
The measured line width was used to make integrated intensity (moment 0) maps by summing the intensities of all the spectral channels covered by a line. From these maps the maximum radial extent of the emission from each line was measured, defined as the maximum radial extent of the 3$\sigma$ contour enclosing the star (centered on the peak of the continuum emission). The contours were azimuthally smoothed (in bins with at least ten samples, corresponding to angular ranges of $\geq$60$^\circ$) to limit deviations due to noise, following the method in \citet{danilovich_atomium_2021}. Note that for transitions with clumpy emission, this method may underestimate the extent of the emission. For transitions with no measured extent from this method an upper limit is given, equal to half the major axis of the restoring beam ($b_{maj}$/2). 
We also note that there may be diffuse extended emission of various molecules which is not properly imaged, or which is filtered out if its size is similar to (or greater than) the maximum recoverable scale. Such emission may be brought out with different data reduction techniques or spectral averaging.

A majority of detected lines are seen in the data at multiple angular resolutions, as expected; however some transitions are only seen in the high-resolution data. We note that this data is more sensitive so faint lines are more likely to be detected. Furthermore, very compact emission (less than a few hundred milliarcsec in diameter) is best detected in the highest angular resolution data, as it may otherwise be diluted in large extraction apertures or with larger beams, or have its apparent flux reduced by imaging artefacts.

\subsection{Mass-loss rate uncertainties} \label{sect:mlrs}

Mass-loss rates ($\dot{M}$) for the ATOMIUM sample were gathered from literature studies (see references in Table~\ref{tab:sources}). However, these mass-loss rate determinations often assumed different distances to the stars than is used in this paper, and hence the mass-loss rate must be scaled to the new distance. This has been done using a simple formula to account for the dilution factor stemming from the distance to the star:
\begin{equation}
    \dot{M}_\mathrm{new} = \dot{M}_\mathrm{old} \left( \frac{D_\mathrm{new}}{D_\mathrm{old}} \right)^2.
\end{equation}
Table~\ref{tab:mlr_update} lists the factor by which this has changed the literature mass-loss rate, alongside more details about how the mass-loss rates were derived.

\begin{table}[ht]
\caption{Mass-loss rates from literature. }
\label{tab:mlr_update} 
\centering 
\begin{tabular}{lrcccc}
\hline\hline
Source & $D_\mathrm{old}$ & $\dot{M}_\mathrm{old}^{(1)}$& Deriv.$^{(2)}$ & N$_\mathrm{obs}$  & $\frac{\dot{M}_\mathrm{new}}{\dot{M}_\mathrm{old}}$ \\ 
\hline 
AH Sco & 2000 & 8.0E-06 & a & 1 & 1.3 \\ 
GY Aql & 540 & 4.0E-06 & b & 1 & 0.6 \\ 
IRC-10529 & 620 & 4.5E-06 & c & 2 & 2.2 \\ 
\nb{IRC+10011} & 740 & 1.9E-05 & c & \nb{16} & 1.0 \\ 
KW Sgr & 3000 & 5.6E-06 & a & 1 & 0.6 \\ 
\nb{$\pi^1$ Gru} & 150 & 7.7E-07 & d & \nb{4} & 1.2 \\ 
R Aql & 190 & 1.1E-06 & e & 2 & 1.5 \\ 
\nb{R Hya} & 118 & 1.6E-07 & c & \nb{7} & 1.1 \\ 
RW Sco & 700 & 2.1E-07 & f & 2 & 0.6 \\ 
S Pav & 150 & 8.0E-08 & g & 1 & 1.6 \\ 
SV Aqr & 470 & 3.0E-07 & g & 2 & 0.9 \\ 
T Mic & 130 & 8.0E-08 & g & 2 & 1.8 \\ 
U Del & 210 & 1.5E-07 & g & 2 & 2.5 \\ 
U Her & 360 & 5.9E-07 & e & 1 & 0.5 \\ 
V PsA & 220 & 3.0E-07 & g & 1 & 1.6 \\ 
\nb{VX Sgr} & 1570 & 6.1E-05 & c & \nb{3} & 1.0 \\ 
\nb{W Aql} & 395 & 3.0E-06 & g & \nb{21} & 0.9 \\ 
\hline
\multicolumn{6}{l}{\footnotesize 
 $^{(1)}$ These values typically have large uncertainties of a factor $\sim$3--10. } \\ 
\multicolumn{6}{l}{\footnotesize 
$^{(2)}$ Description of the various mass-loss rate derivations: } \\ 
\multicolumn{6}{l}{\footnotesize 
a: Empirical formula for 60~$\mu$m flux from \citet{jura_mass-losing_1990}. } \\ 
\multicolumn{6}{l}{\footnotesize 
b: Empirical formula for CO flux based on \citet{knapp_mass_1985}, } \\ 
\multicolumn{6}{l}{\footnotesize 
\quad updated using \citet{mamon_photodissociation_1988}. } \\ 
\multicolumn{6}{l}{\footnotesize 
c: Empirical formula for CO flux derived from a grid of 1D } \\ 
\multicolumn{6}{l}{\footnotesize 
\quad radiative transfer models from \citet{de_beck_probing_2010}. } \\ 
\multicolumn{6}{l}{\footnotesize 
d: 3D non-LTE radiative transfer modeling using } \\ 
\multicolumn{6}{l}{\footnotesize 
\quad SHAPEMOL \citep{santander-garcia_shapemol_2015}. } \\ 
\multicolumn{6}{l}{\footnotesize 
e: 1D radiative transfer modeling, LVG model from \citet{morris_molecular_1980}. } \\ 
\multicolumn{6}{l}{\footnotesize 
f: Empirical formula for CO flux from \citet{olofsson_study_1993}. } \\ 
\multicolumn{6}{l}{\footnotesize 
g: 1D non-LTE radiative transfer modeling using } \\ 
\multicolumn{6}{l}{\footnotesize 
\quad MCP \citep{schoier_models_2001}. } \\ 
\end{tabular} 
\end{table}

Updating the distances to these sources has changed the derived mass-loss rate by at most a factor of 2.5. However, we expect the literature mass-loss rate values to have uncertainties larger than this (though they are seldom quantified), with determinations frequently based on only one or two independent observations which makes it difficult to disentangle the degeneracies between for example the mass-loss rate, temperature profile, and the outer radius of the wind. Sources with derived mass-loss rates based on more than two independent observational data points are highlighted in green, which is the case for only five sources: IRC+10011, $\pi^1$~Gru, R~Hya, VX~Sgr, and W~Aql. W~Aql has undoubtedly the best constrained mass-loss rate as it is based on direct modeling of 21 individual CO lines, and even here the model has an estimated uncertainty of a factor $\sim$3 \citep{danilovich_detailed_2014, ramstedt_circumstellar_2017}. 
Furthermore, many of the mass-loss rates are derived using empirical formulae, based on CO line fluxes and some stellar parameters, rather than direct radiative transfer modelling of CO observations. These formulae involve a large number of assumptions resulting in uncertainties of at least a factor $\sim$3--5 \citep{de_beck_probing_2010, olofsson_study_1993}, when they are quantified at all. Two of the mass-loss rates are derived from empirical formulae using the observed 60~$\mu$m flux (labeled "a" in Table~\ref{tab:mlr_update}), which requires further assumptions about the wind velocity, dust properties, and gas-to-dust ratio.  

The derived mass-loss rates also generally assume a constant and spherically symmetric mass loss, but the ATOMIUM observations have shown this is not true for any of our sources \citep{decin_substellar_2020}: they all show significant structure in their winds, and some \textendash{} like $\pi^1$~Gru and R~Hya \textendash{} are highly asymmetric. This is interpreted as the likely presence of a companion. If the companion shapes the wind into an equatorial density enhancement, this could change the derived mass-loss rate by a factor of up to three if the system is seen face-on, and potentially much more if the system is seen edge-on \citep{el_mellah_wind_2020}. Furthermore, simply the assumption of a constant mass-loss rate has been shown to change the derived mass-loss rate by a factor of a few \citep{kemper_mass_2003, decin_variable_2007}. 

Taken together, these factors result in typical uncertainties on the mass-loss rates of up to an order of magnitude. However, the mass-loss rate still provides a useful way to order the sources to look for general trends, and in the absence of more precise estimates we will continue to use the values in Table~\ref{tab:sources} in our analysis.
Improving the mass-loss rate estimates for the ATOMIUM sources will be undertaken in a future publication, using radiative transfer modelling of homogeneous single-dish observations of at least four CO transitions per source, combined with the spatial information from the ALMA observations.

\section{Results and discussion} \label{sect:results}

This section presents an overview of the molecular inventory in Section~\ref{sect:atomium_stats}, molecular emission sizes in Section~\ref{sect:extents}, correlations between various stellar parameters and their molecular content in Section~\ref{sect:corr}, an analysis of the spatial distributions of SO and \so2 in Section~\ref{sect:so_so2}, isotopic ratios for a subset of the molecules in Section~\ref{sect:isotopic_ratios}, and a discussion of the unidentified lines in Section~\ref{sect:ulines}

\subsection{Overview of molecular inventory}\label{sect:atomium_stats}

Across the variety of 17 AGB and RSG sources in the ATOMIUM sample, we detect 287 molecular lines of which 29 remain unidentified.
Emission from a total of 24 molecules has been identified, namely:
AlCl, AlF, AlO, AlOH, CO, CN, CS, H$_2$O, H$_2$S, HC$_3$N, HCN, KCl, NaCl, OH, PO, SO, SO$_2$, SiC, SiC$_2$, SiN, SiO, SiS, TiO, and TiO$_2$; and 19 isotopologues thereof containing one or more of atoms of $^{13}$C, $^{17}$O, $^{29}$Si, $^{30}$Si, $^{33}$S, $^{34}$S, and $^{37}$Cl.

\noindent This includes some first detections in oxygen-rich AGB and RSG stars: 
\begin{itemize}
    \item PO in the vibrationally excited $\varv=1$ transition, detected in the RSG AH~Sco and the AGB stars IRC+10011, IRC-10529, R~Hya, and S~Pav.
    \item SO$_2$ in the high-energy vibrationally excited $\varv_2=2$ and $\varv_1=1$ transitions, detected in the RSGs AH~Sco and VX~Sgr, respectively. 
    \item Of the ten detected rotational transitions from various vibrational states of H$_2$O, all but one are the first identifications in space, as is one high energy transition of OH \citep[see][who also detect further OH transitions by stacking data]{baudry_atomium_2023}.
\end{itemize}

\noindent and first detections in S-type AGB stars:
\begin{itemize}
    \item High vibrational level transitions of HCN $\varv_2 = 2,3$ in $\pi^1$~Gru and W~Aql. Vibrationally excited HCN has been studied for C-rich AGB stars by \citet{jeste_vibrationally_2022}.
    \item High vibrational level transitions of SiS $\varv=4,5,6$ in $\pi^1$~Gru and W~Aql. Such high vibrational levels have previously been observed in CW~Leo by \citet{patel_interferometric_2011}.
    \item Two transitions of the SiS double isotopologue $^{29}$Si$^{33}$S in W~Aql, previously observed in CW~Leo by \citet{patel_interferometric_2011}.
    \item Four transitions of HC$_3$N in W~Aql, confirming the tentative stacking detection by \citet{de_beck_surprisingly_2020}.
    \item One transition of SiC in W~Aql, to be more thoroughly discussed in \citet{danilovich_chemical_2023}. SiC has previously been detected in 12 C-rich sources by \citet{massalkhi_abundance_2018}.
    \item Several transitions of AlCl and one transition of AlF in W~Aql, as discussed in \citet{danilovich_atomium_2021}.
\end{itemize}

\begin{figure*}[ht]
    \includegraphics[width=\textwidth]{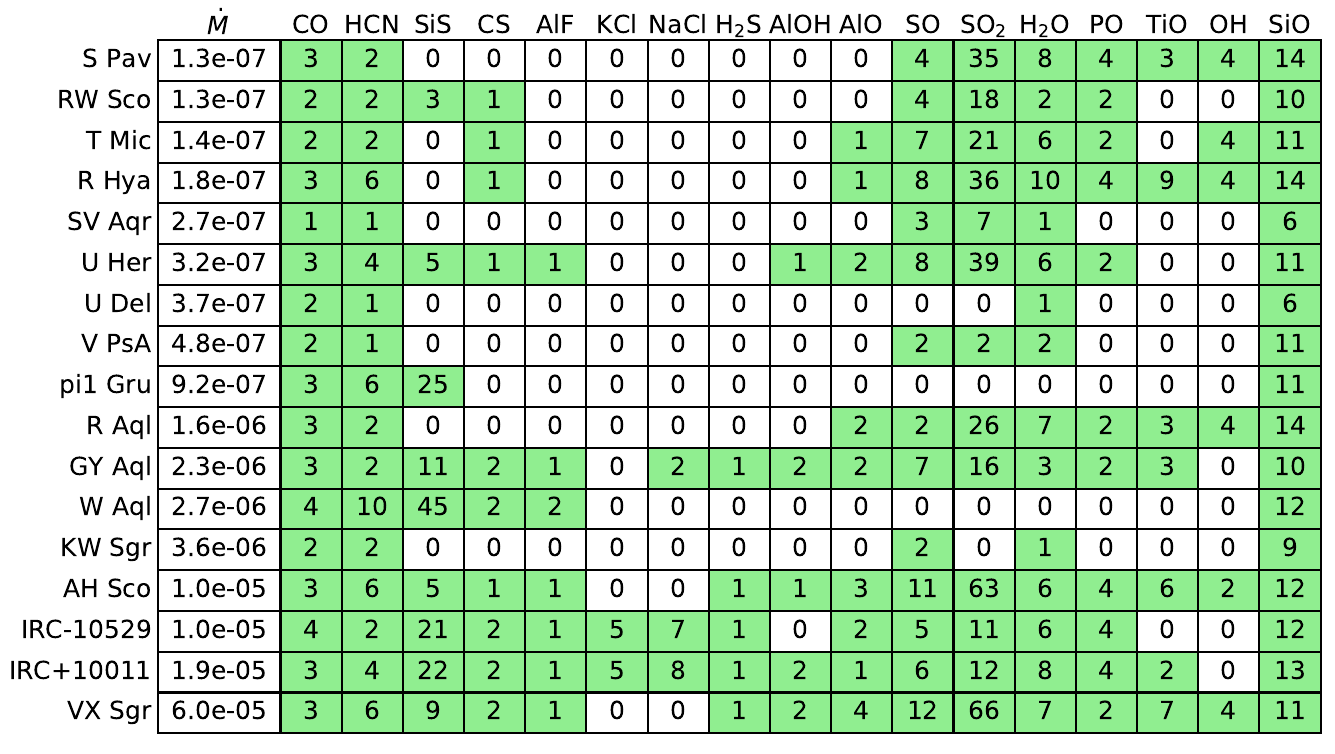}
    \caption{Overview of how many lines of each molecule are detected in each source. The sources are listed in order of increasing mass-loss rate ($\dot{M}$), and molecules found in only a single source are not included.}
    \label{fig:mol_MLR}
\end{figure*}

A visual overview of the molecules detected in each source can be seen in Figure~\ref{fig:mol_MLR}, where the sources are arranged in order of increasing mass-loss rate (note that molecules detected in only one source \textendash{} TiO$_2$ in VX~Sgr; and AlCl, $^{13}$CN, SiN, SiC, SiC$_2$, and HC$_3$N in W~Aql \textendash{} are not included).

Looking at the overall molecular content, we can group together some similar sources:
\begin{itemize}
\item The two S-type sources, $\pi^1$~Gru and W~Aql, are unsurprisingly different from the other, oxygen-rich sources. They are notably lacking in oxygen-bearing molecules, such as SO and SO$_2$, and show a large number of SiS lines, however W~Aql is significantly more molecule-rich than $\pi^1$~Gru.
\item IRC+10011 and IRC-10529 are very similar in molecular content, showing, for example, a range of Al- and Cl-bearing molecules, relatively few SO$_2$ lines, and a relatively large number of SiS lines. 
GY~Aql also has a similar molecular content to IRC+10011 and IRC-10529, though it shows comparatively more SO$_2$ and fewer SiS lines. 
\item Two sources are very line-poor: U~Del and the RSG KW~Sgr. Beyond the molecules seen in all sources (CO, SiO, HCN) these two sources only contain a single H$_2$O line each and KW~Sgr exhibits two SO lines. An explanation may lie in the lack of detection of high-$\varv$ SiO lines (see below) in both sources, which may indicate weak shocks in the inner wind and hence post-shock temperatures too low to drive a rich chemistry and excite many of the high-$\varv$ molecular lines. 
Furthermore, KW~Sgr is a younger RSG than AH~Sco or VX~Sgr, with a spectral classification of M2, similar to Betelgeuse which is also known to be relatively poor in molecules \citep{huggins_neutral_1994}. 
\end{itemize}

From Figure~\ref{fig:mol_MLR} we can see that CO, HCN, and SiO are ubiquitous, as expected from previous studies, with similar numbers of lines detected in almost all sources (excepting the large number of vibrationally excited HCN lines in W~Aql). 
All sources show vibrationally excited (up to $\varv=4$) lines of SiO except the line-poor KW~Sgr (up to $\varv=2$) and U~Del (up to $\varv=1$), and also SV~Aqr (up to $\varv=1$).  
Three sources also show $^{30}$SiO lines up to $\varv=5$: R~Aql, R~Hya, and S~Pav. Note that the $\varv=5$ transitions of the main SiO isotopologue were not covered by our observations. 
Many of the SiO lines have maser components, especially the high-$\varv$ lines but also some $\varv=0$ lines (Pimpanuwat et al., in prep.).

SO and SO$_2$ are detected in a majority of sources, excepting the line-poor U~Del and KW~Sgr (in which SO but not SO$_2$ is seen), and the S-type sources $\pi^1$~Gru and W~Aql. 
A large number of SO$_2$ $\varv=0$ and $\varv_2=1$ lines are detected in many sources, and the two RSGs AH~Sco and VX~Sgr also show a few high-energy $\varv_1=1$ or $\varv_2=2$ transitions. SO$_2$ is one of the most widespread species in our sample, partially due to its nature as an asymmetric top which exhibits many different energy levels throughout its energy ladder, which are fairly easily populated by radiation and collisions. 
The other sulfur-bearing molecules \textendash{} H$_2$S, SiS, and CS \textendash{} are expected to have abundances that scale with mass-loss rate, and hence are more likely to be seen in high mass-loss-rate sources \citep{decin_alma_2018, danilovich_sulphur-bearing_2017, danilovich_sulphur-bearing_2018}. Indeed, we only see H$_2$S in the highest mass-loss-rate sources, but for SiS and CS there is no clear trend with mass-loss rate.
\citet{danilovich_sulphur_2016, danilovich_rotational_2020} also found a difference in the spatial distributions of SO and SO$_2$ in low- and high-mass-loss-rate sources, which is discussed further in Section~\ref{sect:so_so2}.

SiS shows transitions from several isotopologues, and high-$\varv$ lines (up to $\varv=5-6$) are seen in both S-type sources, $\pi^1$~Gru and W~Aql, which are expected to have higher SiS abundances \citep{schoier_abundance_2007, danilovich_sulphur-bearing_2018}. All other sources only show lines of SiS up to $\varv=1$, except IRC+10011 and IRC-10529 where $\varv=3$ is reached.

H$_2$O is detected in all the sources in our sample except the two S-type stars, in several high-lying rotational transitions with lower state energy levels above $\sim$3900~K, including high vibrational energy states. Four OH hyperfine split transitions with similar high energies ($E_\mathrm{low} \gtrsim 4800$~K) are also detected. 
This includes many first detections of these transitions in space; they are discussed in detail in \citet{baudry_atomium_2023} and hence will not be extensively discussed here.

The halide-bearing molecules \textendash{} AlF, NaCl, and KCl \textendash{} are mostly found in high mass-loss-rate sources, although AlF (and AlOH) are also seen in the intermediate mass-loss-rate source U~Her. 
Specifically the chlorine-bearing molecules KCl and NaCl are found only in IRC+10011, IRC-10529, and GY Aql. It is perhaps unexpected that the chlorides are missing in the RSGs AH Sco and VX Sgr, as NaCl is very strong in the RSG VY CMa \citep{tenenbaum_arizona_2010, quintana-lacaci_history_2023}, however VY CMa is a more evolved RSG with a higher mass-loss rate than the RSGs in our sample.  
Furthermore, another chlorine-bearing molecule, AlCl, is only found in the S-type star W Aql, in which we do not detect NaCl or KCl, so AlCl seems to require different formation and/or excitation conditions.

W~Aql is also the only source to show transitions of $^{13}$CN, SiN, SiC, SiC$_2$, and HC$_3$N (note $^{12}$CN was not covered by our observations). This is unsurprising as it is a bright, nearby S-type star with a fairly carbon-rich outflow \citep{de_beck_surprisingly_2020}.
The other molecule only seen in a single source is TiO$_2$ in VX~Sgr, which may be because VX~Sgr is the highest-mass-loss-rate source in our sample and so its denser wind may allow less abundant molecules to be detected.

\subsection{Angular extents of molecules}\label{sect:extents}


\begin{table}[ht]
\caption{Maximum measured radial extent of each molecule across all sources.}
\label{tab:molext_permol} 
\centering 
\begin{tabular}{lcccr}
\hline\hline
Mol. & Range in & Median & Max & N$_\mathrm{sou}$ \\ 
 & $r_\mathrm{max}$ ($R_\star$) & $r_\mathrm{max}$ ($R_\star$) & $r_\mathrm{max}$ ($\arcsec$) &  \\ 
\hline 
SiC & \llap{<}2.1 & -- & 0.01 & 1 \\ 
TiO$_2$ & 14 & -- & 0.06 & 1 \\ 
OH & \llap{<}3.0 -- 17 & 6 & 0.20 & 6 \\ 
AlOH & \llap{<}1.3 -- 45 & 4 & 0.20 & 5 \\ 
AlF & \llap{<}2.8 -- 81 & 12 & 0.44 & 7 \\ 
PO & \llap{<}2.0 -- 120 & 26 & 0.33 & 11 \\ 
AlCl & 130 & -- & 0.69 & 1 \\ 
AlO & \llap{<}1.2 -- 150 & 3 & 0.67 & 9 \\ 
KCl & 140 -- 150 & 145 & 0.50 & 2 \\ 
TiO & \llap{<}1.5 -- 160 & 6 & 0.47 & 7 \\ 
HC$_3$N & 210 & -- & 1.14 & 1 \\ 
H$_2$O & 5.3 -- 230 & 36 & 0.66 & 15 \\ 
SiC$_2$ & 270 & -- & 1.47 & 1 \\ 
NaCl & \llap{<}1.5 -- 280 & 250 & 0.90 & 3 \\ 
SiN & 390 & -- & 2.14 & 1 \\ 
H$_2$S & 7.2 -- 490 & 140 & 1.60 & 5 \\ 
HCN & 22 -- 590 & 120 & 1.92 & 17 \\ 
SiS & 5.5 -- 610 & 210 & 1.97 & 10 \\ 
CS & \llap{<}13 -- 640 & 150 & 3.51 & 10 \\ 
SO$_2$ & 4.3 -- 710 & 290 & 2.31 & 13 \\ 
SO & 5.1 -- 780 & 340 & 2.53 & 14 \\ 
SiO & 140 -- 1100 & 440 & 3.42 & 17 \\ 
CO & 29 -- 2400\rlap{$^\dagger$} & 1300 & 7.87\rlap{$^\dagger$} & 17 \\ 
\hline
\multicolumn{5}{l}{\footnotesize 
\textbf{Notes:} Maximum extents in CO are likely underestimated } \\ 
\multicolumn{5}{l}{\footnotesize 
and are marked with $^\dagger$. Transitions with no measurable} \\ 
\multicolumn{5}{l}{\footnotesize 
extent are noted as upper limits equal to the radius of the } \\ 
\multicolumn{5}{l}{\footnotesize
beam ($b_{maj}$/2). The molecules are arranged according to } \\ 
\multicolumn{5}{l}{\footnotesize
the maximum measured $r_\mathrm{max}$ in $R_\star$.} \\ 
\end{tabular} 
\end{table}

While a full analysis of the spatial distributions of various molecules is beyond the scope of this paper, we found it useful as a first step to simply compare their observationally measured extents, as defined in Section~\ref{sect:analysis}.
Table~\ref{tab:molext_permol} shows the measured angular extent of each molecule across the 17 ATOMIUM sources. For each molecule, we have taken the maximum measured angular extent ($r_\mathrm{max}$) in each source, and then give the range and median of these measurements in units of the stellar radius, $R_\star$ (see Table~\ref{tab:sources} for $R_\star$ values of each source). The maximum $r_\mathrm{max}$ for each molecule is also given in arcseconds. 
The maximum extents of CO are almost certainly underestimates because the emission of this molecule typically extends beyond the field of view of the observations. 

Some of the smallest emission extents are found for the molecules TiO$_2$, OH, AlOH, and AlF. It is not surprising that these molecules are mainly seen close to the star, in the densest part of the wind, as this is predicted by chemical models \citep{agundez_chemical_2020, mangan_kinetic_2021}, especially for the Al-bearing molecules which are expected to participate in dust formation. However, OH is a special case as the observed transitions all have very high lower state energies ($E_\mathrm{low} \geq 4800$~K) which will only be excited close to the star. In fact, 1.6~GHz OH maser emission has been measured up to $\sim$$100-1000$~R$_\star$ in for example IRC+10011 and R~Aql \citep{bowers_circumstellar_1983}, U~Her \citep{chapman_oh_1994}, and the RSG VX~Sgr \citep{szymczak_polarization_1997}. 

The largest angular extents, $>$1.5$\arcsec$, are mainly found for molecules which are seen in a majority of sources: CO, SiO, SO, SO$_2$, CS, SiS, and HCN. However, for almost all molecules there are some sources where the maximum measured extent is very small. Which source it is varies by molecule, and its value is often an outlier among the sample. For example, in CO the smallest $r_\mathrm{max}$ of 29~$R_\star$ in the line-poor RSG KW~Sgr is much smaller than the next smallest value of 530~$R_\star$ in AH~Sco. V~PsA has the smallest measured $r_\mathrm{max}$ in HCN, SO, and SO$_2$; R~Hya shows the smallest extents in SiS and CS; and the smallest measured $r_\mathrm{max}$ in SiO is in $\pi^1$~Gru.

There are some interesting observations to be made from this summary from a chemical standpoint. First, AlO extends further out than AlOH, though this is only true for the RSGs VX~Sgr and AH~Sco, which have much larger AlO extents than the other sources: $\geq$100~$R_\star$. They also have the largest $r_\mathrm{max}$ in AlOH, at 45 and 8~$R_\star$, respectively \textendash{} smaller than their AlO extents. This is unexpected as AlOH is formed from AlO (by reaction with H$_2$O and H$_2$), although AlOH is also easily photolysed back to AlO \citep{mangan_kinetic_2021}. The other three AGB sources with detections of both AlOH and AlO show slightly larger extents in AlOH. 
So here we have a dichotomy between the RSG and AGB sources, where perhaps the photolysis of AlOH to AlO is more efficient in the RSGs explaining their large AlO extents. 

Another surprising observation is that TiO extends further out than TiO$_2$ in the only star where TiO$_2$ is detected: the RSG VX~Sgr. VX~Sgr shows a maximum extent in TiO of 65~$R_\star$, larger than the 14~$R_\star$ extent in TiO$_2$. This might indicate that TiO$_2$ is depleted because it is taking part in the formation of dust particles in the inner wind.

\subsection{Correlations}\label{sect:corr}

\begin{figure*}
    \includegraphics[width=\textwidth]{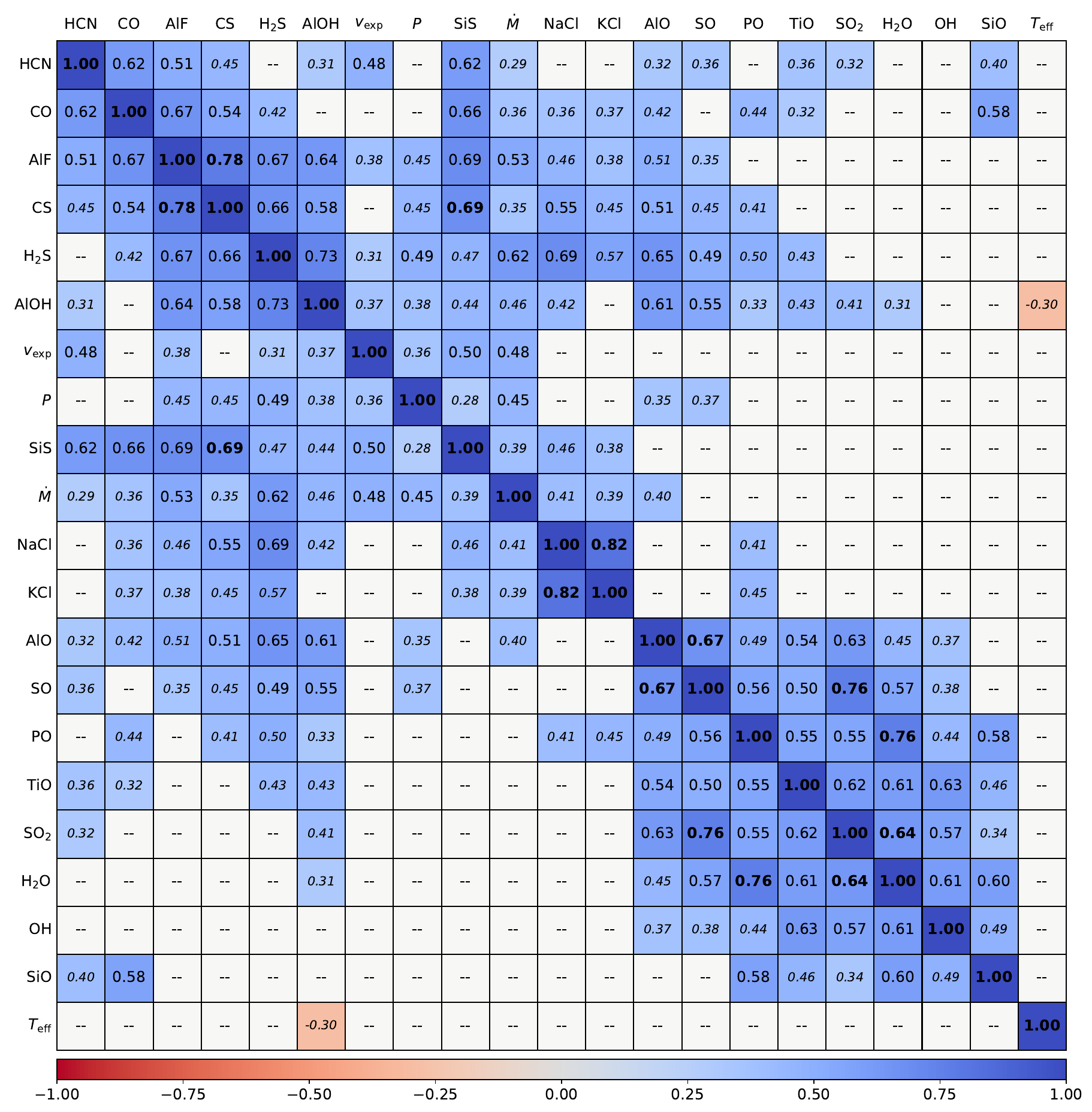}
    \caption{Kendall's $\tau_b$ rank correlation coefficients between the expansion velocity (v$_\mathrm{exp}$), pulsation period (P), mass-loss rate ($\dot{M}$), effective temperature (T$_\mathrm{eff}$), and number of detected lines of various molecules in each source. Correlation coefficients with 3$\sigma$ significance are in boldface, 2$\sigma$ in regular font, and 1$\sigma$ in small italics. Coefficients with lower than 1$\sigma$ significance are not included, and instead given as \-\-. The cells are colorized from perfect correlation (1) in blue, to perfect anti-correlation (-1) in red.}
    \label{fig:corr_table}
\end{figure*}

To get an overview of how the molecular content varies by source, we have calculated Kendall's $\tau_b$ rank correlation coefficients \citep{kendall_treatment_1945} for various stellar parameters and the molecular content of each source. These are shown in Figure~\ref{fig:corr_table}. Kendall's rank correlations are chosen as a non-parametric measure of correlation, that does not assume linear relationships between variables and is applicable to ordinal data. 
This allows us to use the differing number of detected lines of a given molecule as a (very rough) proxy for its relative abundance in different sources. The number of detected lines is a function of molecular parameters and excitation conditions as well as abundance, which we cannot properly take into account without radiative transfer modeling. However, by ranking the sources by the number of detected lines we can nevertheless determine which groups of molecules tend to coincide by having relatively large numbers of lines in the same sources. We also negate the need to normalize different molecules by the absolute number of potentially detectable lines in our frequency range. Many alternative techniques, such as Pearson's correlations or principal component analysis, assume linear relationships between variables and hence are less applicable to our data at this initial analysis stage. 

To calculate the Kendall's $\tau_b$ correlation coefficients the sources were ranked (by measured value for stellar parameters and by number of detected lines for molecules) for each variable, and the ranks between every pair of variables were compared (adjusting for ties). This assesses how well the relationship between the two variables can be described by a monotonic function, that is, whether the first variable tends to increase as the second does and vice versa. 
The variables being correlated are the effective temperatures, mass-loss rates, terminal expansion velocities, and pulsation periods (as given in Table~\ref{tab:sources}), and the number of lines detected for each molecule and its isotopologues. Molecules which are found in only one source (TiO$_2$ in VX~Sgr, and AlCl, HC$_3$N, SiC$_2$, SiN, and $^{13}$CN in W~Aql) have not been included. 
Generally, we assume that a larger number of detected lines of a given molecule implies a higher molecular abundance; however there may be additional effects making the detection of, for example, highly vibrationally excited lines more likely. These will be discussed for each molecule in the following sections. 

As we have more than ten sources to compare, the sampling distribution of Kendall's $\tau_b$ is approximately a normal Gaussian distribution \citep{kendall_new_1938}. We therefore use p-values\footnote{A p-value measures the probability of obtaining the observed results "by chance", assuming that the null hypothesis of no correlation is true. Therefore, a smaller p-value indicates a greater statistical significance.} corresponding to 1$\sigma$ (p < 0.15865), 2$\sigma$ (p < 0.02275), and 3$\sigma$ (p < 0.00135) to determine which correlations are significant. 
We consider a correlation coefficient $\gtrsim$0.5 to be a strong correlation, which is generally only seen for correlations with at least 2$\sigma$ significance. Correlations with 3$\sigma$ significance have coefficients $\geq$0.64 and hence will be termed very strong correlations. For simplicity, we use a single term to refer to both significance and strength of correlation, as they are strongly related, and will refer to correlations with 2$\sigma$ significance as strong correlations, and those with 1$\sigma$ significance as weak correlations. 
Figure~\ref{fig:corr_table} is colorized by the value of the correlation coefficient, with stronger positive correlations in a darker blue. For further visual differentiation, 3$\sigma$ correlation coefficients are written in boldface, 2$\sigma$ in regular font, and 1$\sigma$ in small italics. Correlation coefficients with p-values above 0.15865 (that is lower than 1$\sigma$ significance) are not included in our analysis or in the figure.

\begin{figure}
    \includegraphics[width=\hsize]{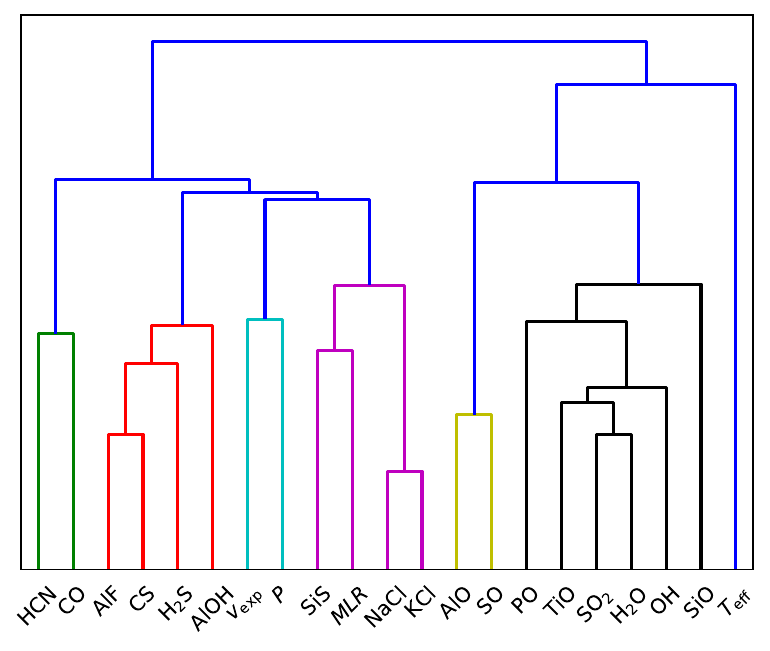}
    \caption{Dendrogram of the correlations in Figure~\ref{fig:corr_table}, showing the hierarchical relationships between the different parameters. A closer connection point (lower down in the diagram) between two parameters signals a stronger relationship. The dark blue lines indicate the top-level division into seven clusters, and the other colors are simply to guide the eye in distinguishing between the different clusters. }
    \label{fig:dendrogram}
\end{figure}

The molecular correlations with at least a 1$\sigma$ significance were then used to calculate a dendrogram (see Figure~\ref{fig:dendrogram}) of the hierarchical relationships between the different parameters,
using the farthest point (a.k.a. complete linkage) algorithm for hierarchical clustering. 
This dendrogram shows which parameters are most similar to each other in terms of their correlations to all the other parameters, and groups them into seven clusters shown with different colors. Note that this grouping may split up pairs of parameters that are strongly correlated with each other but have dissimilar correlations with other parameters. These clusters are reflected in Figure~\ref{fig:corr_table}, and the calculated correlations will now be discussed cluster by cluster for simplicity.

Cluster 1: HCN and CO is shown with green lines in Figure~\ref{fig:dendrogram} and discussed in Sect~\ref{sect:cluster1}; Cluster 2: AlF, CS, H$_2$S, and AlOH is shown with red lines and discussed in Sect~\ref{sect:cluster2}; Cluster 3: $v_\mathrm{exp}$ and $P$ is shown with cyan lines and discussed in Sect~\ref{sect:cluster3}; Cluster 4: SiS, $\dot{M}$, NaCl, and KCl is shown with purple lines discussed in Sect~\ref{sect:cluster4}; Cluster 5: AlO and SO is shown with yellow lines and discussed in Sect~\ref{sect:cluster5}; Cluster 6: PO, TiO, SO$_2$, H$_2$O, OH, and SiO is shown with black lines and discussed in Sect~\ref{sect:cluster6}; and Cluster 7: $T_\mathrm{eff}$ is discussed in Sect~\ref{sect:cluster7}.

\subsubsection{Cluster 1: HCN, CO}\label{sect:cluster1}
The first cluster contains the correlated molecules HCN and CO. For these molecules the $\varv=0$ transitions of the main isotopologues are seen in all sources, so more detected lines means the detection of isotopologues ($^{13}$CO, C$^{17}$O, H$^{13}$CN) or vibrationally excited transitions (CO $\varv=1$; HCN $\varv_2=1,2,3$; H$^{13}$CN $\varv_2=1$).

The isotopologues of HCN and CO tend to be detected in the same sources, as one might expect for both isotopologues with $^{13}$C. 
$^{13}$CO is detected in all sources except SV~Aqr, and H$^{13}$CN is detected in all sources except SV~Aqr, U~Del, and V~PsA. C$^{17}$O is only detected in two sources: IRC-10529 and W Aql. 
From this we can assume that SV~Aqr, U~Del, and V~PsA either have relatively low $^{13}$C abundances or faint HCN emission, while IRC-10529 may have relatively high $^{17}$O abundances and W~Aql is the source with by far the brightest emission in the main CO isotopologue.
Similarly, the vibrationally excited transitions in HCN and CO tend to coincide. Vibrationally excited HCN is detected in six sources: IRC+10011, $\pi^1$~Gru, R~Hya, W~Aql, and the RSGs AH~Sco and VX~Sgr, four of which also show vibrationally excited H$^{13}$CN. These six sources are also among the 11 which show emission in the CO $\varv=1$ line. 

Detecting more lines of HCN and CO is correlated with more detected lines of AlF, CS, SiS, AlO, TiO, and SiO, as well as a higher mass-loss rate ($\dot{M}$).
Additionally, CO is correlated with detecting more lines of H$_2$S, NaCl, KCl, and PO; while HCN is correlated with detecting more lines of AlOH, SO, and SO$_2$, as well as a larger expansion velocity ($v_\mathrm{exp}$).  The detection of more isotopologues and vibrationally excited lines of CO and HCN at higher mass-loss rates is due to a denser wind, and most of the correlated molecules are also preferentially seen in denser winds.
Some of these molecules, such as SiS and H$_2$S, are known tracers of higher mass-loss rate winds \citep{danilovich_sulphur-bearing_2017, danilovich_sulphur-bearing_2018}. Others, such as AlF, AlOH, AlO, and KCl, are expected to have fairly low abundances \citep{agundez_chemical_2020} and hence would only be detectable in a sufficiently dense wind. 
To further explain the observed correlations, a higher $\dot{M}$ is also correlated with a higher expansion velocity, and some of the molecules like SiO, PO, and SO$_2$ show vibrationally excited transitions in similar sources as vibrationally excited CO and HCN, which may imply these molecules are excited in similar regions of the wind.

\subsubsection{Cluster 2: AlF, CS, H$_2$S, AlOH}\label{sect:cluster2}
The second cluster contains the molecules AlF, CS, H$_2$S, and AlOH.
AlF is detected in 7/17 sources \textendash{} GY~Aql, IRC+10011, IRC-10529, U~Her, W~Aql, and the RSGs AH~Sco and VX~Sgr \textendash{} of which W~Aql also shows the corresponding $\varv=1$ transition.
CS is detected in one transition in 10/17 sources, of which five sources also show a $^{13}$CS line: GY~Aql, IRC+10011, IRC-10529, VX~Sgr, and W~Aql.
H$_2$S has one detected transition, in 5/17 sources: AH~Sco, GY~Aql, IRC+10011, IRC-10529, and VX~Sgr.
AlOH has two $\varv=0$ transitions, and is detected in 5/17 sources: AH~Sco, GY~Aql, IRC+10011, U~Her, and VX~Sgr.

AlF and CS are very strongly correlated, with all five sources showing $^{13}$CS also detected in AlF. 
Further, all four molecules in this cluster are strongly correlated with each other and tend to be detected in the same sources. 4/5 sources showing H$_2$S also show AlOH, and all sources with either molecule also show AlF. 
All four molecules are also correlated with a longer pulsation period ($P$), higher mass-loss rate ($\dot{M}$), and detecting more lines of SiS, NaCl, AlO, and SO. Three-out-of-four molecules are further correlated with detecting more lines of HCN, CO, KCl, and PO, as well as a larger expansion velocity ($v_\mathrm{exp}$). 

These correlations can largely be explained by the similarities between the four sources AH~Sco, GY~Aql, IRC+10011, and VX~Sgr. These sources have the second to eighth longest periods, and highest to seventh highest mass-loss rates. They all show a fair number of SiS lines, and include two of the three sources with detections of NaCl (see Sect~\ref{sect:cluster4}). The RSGs AH~Sco and VX~Sgr are the two sources which show the most lines of both AlO and SO (see further discussion in Sect~\ref{sect:cluster5}). 
Regarding HCN and CO: all four sources show their $^{13}$C isotopologues and the CO $\varv=1$ transition, three show vibrationally excited HCN and two also show vibrationally excited H$^{13}$CN. IRC+10011 is one of two sources showing KCl emission. All four sources are among the 11 showing PO emission, and the group contains 2/5 sources showing vibrationally excited transitions of PO. Finally, these four sources have the third to ninth largest expansion velocities.

The strong correlation between AlOH and AlO is expected due to their close chemical coupling \citep{mangan_kinetic_2021, gobrecht_bottom-up_2022}.
The strong correlation between AlOH and AlF probably comes from AlF's formation reaction AlOH + HF $\rightarrow$ AlF + H$_2$O; and
the correlation between AlO and AlF also follows from the lesser formation reaction AlO + HF $\rightarrow$ AlF + OH \citep{danilovich_atomium_2021}.

\subsubsection{Cluster 3: v$_\mathrm{exp}$, $P$}\label{sect:cluster3}
The third cluster contains expansion velocity (v$_\mathrm{exp}$) and pulsation period ($P$), which are weakly correlated with each other. The clustering of these parameters is not unexpected, as for AGB stars both expansion velocity and pulsation period tend to increase as they evolve \citep{habing_asymptotic_2003}, along with the mass-loss rate. In fact, for a constant and spherical mass loss we expect a perfect correlation between v$_\mathrm{exp}$ and the mass-loss rate, so the correlation of 0.48 seen here is further evidence of wind asymmetries. 

A larger expansion velocity and longer pulsation period is correlated with detecting more lines of AlF, H$_2$S, AlOH, and SiS, as well as a larger mass-loss rate. 
The expansion velocity is also correlated with detecting more lines of HCN, while the pulsation period is correlated with detecting more lines of CS, AlO, and SO. As discussed in Section~\ref{sect:cluster1}, most of these molecules are preferentially detected in a dense (higher $\dot{M}$) wind.

\subsubsection{Cluster 4: SiS, $\dot{M}$, NaCl, KCl}\label{sect:cluster4}
The fourth cluster contains the molecules SiS, NaCl, and KCl, along with the mass-loss rate ($\dot{M}$). 
NaCl is detected in three sources: GY~Aql, IRC+10011, and IRC-10529, of which IRC+10011 and IRC-10529 also  show vibrationally excited transitions and transitions of Na$^{37}$Cl. KCl is detected only in these same two sources: IRC+10011 and IRC-10529. 
SiS is detected in 10/17 sources, while its various isotopologues of $^{29}$Si, $^{30}$Si, $^{33}$S, and $^{34}$S are detected in between five and seven sources, including GY~Aql, IRC+10011, and IRC-10529. GY~Aql also shows vibrationally excited transitions of SiS up to $\varv=1$, while IRC+10011 and IRC-10529 show vibrationally excited transitions up to $\varv=3$. The only sources showing higher vibrationally excited transitions of SiS (up to $\varv=6$) are the two S-type sources $\pi^1$~Gru and W~Aql, which are expected to have higher SiS abundances \citep{danilovich_sulphur-bearing_2018}, and also show transitions in all the isotopologues. 
Furthermore, IRC+10011, IRC-10529, GY~Aql, $\pi^1$~Gru, and W~Aql have between the second and ninth highest mass-loss rates in our sample. 

Detecting more lines of SiS and a higher mass-loss rate are correlated with detecting more lines of HCN, CO, AlF, CS, H$_2$S, AlOH, and of course NaCl and KCl, along with larger expansion velocities (v$_\mathrm{exp}$) and longer pulsation periods ($P$). 
These correlations can largely be explained by the molecular content of GY~Aql, IRC+10011, IRC-10529, $\pi^1$~Gru, and W~Aql. As noted above, these sources show many lines of SiS and high mass-loss rates. Most of them are detected in vibrationally excited transitions of HCN, and IRC-10529 and W~Aql are the only sources to show a transition of the $^{17}$O isotopologue of CO. All but $\pi^1$~Gru show transitions of AlF and CS, and they make up 3/5 and 2/5 of the sources showing transitions of H$_2$S and AlOH, respectively. When it comes to expansion velocities $\pi^1$~Gru has by far the largest, and the other sources are up to the tenth largest. They also have the fourth to eighth longest periods, except $\pi^1$~Gru which instead has one of the shortest periods in our sample. Furthermore, as noted in Section~\ref{sect:cluster3}, AGB stars tend to increase in mass-loss rate, expansion velocity, and pulsation period as they evolve, explaining their correlation.

Within this cluster, the chemically similar KCl and NaCl are very strongly correlated with each other. They are also correlated with detecting more lines of CO, AlF, CS, H$_2$S, and PO. NaCl is additionally correlated with detecting more lines of AlOH. 
The strong correlation between KCl and NaCl is because they are both formed from reactions of the metal atom with HCl \citep{plane_experimental_1989,helmer_experimental_1993}.
The correlation with another halogen-bearing molecule, AlF, implies that these chemically related molecules may coexist in similar environments in the stellar wind, though note that the only source where AlCl is detected, W~Aql, does not show any transitions of KCl or NaCl. 
Through AlF the correlations with AlOH and the sulfur-bearing CS and H$_2$S can be explained (see Section~\ref{sect:cluster2}), and the correlation with PO may be due to the main sources containing NaCl and KCl \textendash{} IRC+10011 and IRC-10529 \textendash{} also showing all the detected PO transitions, although they are only 2/5 sources to do so.

\subsubsection{Cluster 5: AlO, SO}\label{sect:cluster5}
The fifth cluster contains the molecules AlO and SO, which are strongly correlated with each other.
AlO is detected in 9/17 sources 
of which four also show vibrationally excited transitions: R~Aql, U~Her, and the RSGs AH~Sco and VX~Sgr. 
SO is detected in 14/17 sources, ten of which also show $^{34}$SO. The less abundant isotopologue $^{33}$SO is detected in five sources, and the single detected SO $\varv=1$ transition is observed in seven sources. Of the sources showing transitions of AlO, the vast majority are detected in either both isotopologues or one isotopologue and the $\varv=1$ transition of SO. Three of these sources \textendash{} AH~Sco, U~Her, and VX~Sgr \textendash{} are seen in both SO isotopologues and its $\varv=1$ transition.

The detection of more lines of AlO and SO is correlated with detecting more lines of HCN, AlF, CS, H$_2$S, AlOH, PO, TiO, SO$_2$, H$_2$O, and OH, as well as longer pulsation periods ($P$). Additionally, detecting more lines of AlO is correlated with detecting more lines of CO and a higher mass-loss rate ($\dot{M}$). 
Overall, AlO and SO are correlated with the other oxygen-bearing molecules, as expected, but also with the molecules in Cluster 1 (HCN, CO) and Cluster 2 (AlF, CS, H$_2$S, AlOH). 
As explained in Sect~\ref{sect:cluster2}, the strong correlation between AlO and AlOH stems from their close chemical coupling, but the chemical links to SO and H$_2$S are less clear. One might suspect a connection through the sulfur chemistry, as both AlO and SO are also correlated with CS, but neither is correlated with SiS. However, it may be that the conditions required to populate the higher energy levels of SiS (and hence have more detected lines of SiS) differ from the conditions that excite more transitions of (vibrationally excited) AlO and (vibrationally excited and/or isotopologues of) SO.
Overall, both Cluster 1 and Cluster 2 molecules tend to have many detected lines in the three sources showing the most lines of AlO and SO \textendash{} AH~Sco, U~Her, and VX~Sgr \textendash{} which may explain the correlations. The grouping of the low mass-loss rate AGB star U~Her with the high mass-loss rate RSGs AH~Sco and VX~Sgr here is unexpected, and will be explored further in a future paper.

\subsubsection{Cluster 6: PO, TiO, SO$_2$, H$_2$O, OH, SiO}\label{sect:cluster6}
The sixth cluster contains the remaining oxygen-bearing molecules: PO, TiO, SO$_2$, H$_2$O, OH, and SiO.
A PO doublet is detected in 11/17 sources, of which five sources \textendash{} IRC+10011, IRC-10529, R~Hya, S~Pav, and the RSG AH~Sco \textendash{} also show the corresponding $\varv=1$ doublet.
TiO is detected in 7/17 sources, with vibrationally excited transitions detected in three sources: R~Hya, and the RSGs AH~Sco and VX~Sgr.
SO$_2$ is detected in 13/17 sources, with $^{34}$SO$_2$ in nine sources and vibrationally excited transitions detected in seven sources. Six of these sources have detections of both $^{34}$SO$_2$ and vibrationally excited SO$_2$: AH~Sco, R~Aql, R~Hya, RW~Sco, U~Her, and VX~Sgr.
H$_2$O is detected in 15/17 sources, with intrinsically fainter and/or higher energy transitions in progressively fewer sources.  
OH is detected in 6/17 sources, of which five show both detected $\varv=0$ doublets: R~Aql, R~Hya, S~Pav, T~Mic, and VX~Sgr.
SiO is detected in all sources: the main isotopologue in $\varv=0$ and $\varv=1$ and both $^{29}$SiO and $^{30}$SiO in a $\varv=0$ transition. So more detected transitions implies the detection of more highly vibrationally excited transitions: up to $\varv=4$ in the main isotopologue, which is seen in 13/17 sources; up to $^{29}$SiO $\varv=3$ which is seen in six sources \textendash{} IRC+10011, IRC-10529, R~Aql, R~Hya, S~Pav, and W~Aql; and up to $^{30}$SiO $\varv=5$ which is seen in four sources \textendash{} IRC+10011, R~Aql, R~Hya, and S~Pav. 

The molecules in this cluster are all correlated with each other, as expected, and most also show correlations with the other oxygen-bearing molecules. PO, TiO, and SiO are correlated with CO; PO, TiO, SO$_2$, and H$_2$O are correlated with AlOH; and all except SiO are correlated with AlO and SO. 

PO is very strongly correlated with H$_2$O, which supports experimental findings that the reaction of excited P atoms with H$_2$O is a major formation route for PO \citep{douglas_experimental_2022}. 
The five sources with the most detected PO lines \textendash{} that is the ones showing the $\varv=1$ transition: AH~Sco, IRC+10011, IRC-10529, R~Hya, and S~Pav \textendash{} are also among the sources with the most detected H$_2$O lines, including R~Hya which is the only source to show all ten H$_2$O transitions. Conversely, the six sources with no detected PO transitions \textendash{} KW~Sgr, $\pi^1$~Gru, SV~Aqr, U~Del, V~PsA, and W~Aql \textendash{} also show the fewest detected H$_2$O lines. The two S-type sources, $\pi^1$~Gru and W~Aql, show neither PO nor H$_2$O lines, and the others only have one to two H$_2$O lines and are overall quite line-poor. 
However, from the dendrogram in Figure~\ref{fig:dendrogram} we can see that PO is less closely grouped with H$_2$O than SO$_2$, TiO, and OH are. This is likely due to PO showing correlations that H$_2$O lacks with a range of molecules \textendash{} CO, CS, H$_2$S, NaCl, and KCl \textendash{} in addition to their shared correlations with the other oxygen-bearing molecules. This may be mostly due to IRC+10011 and IRC-10529 which are, for example, two of the three sources with detected NaCl or KCl, and two of the five sources showing H$_2$S. Conversely, in R~Hya and S~Pav, which show the most H$_2$O lines, we do not detect any transitions of NaCl, KCl, or H$_2$S.  

Detecting more lines of TiO is correlated with detecting more lines of HCN, CO, and H$_2$S, alongside its correlations with the other oxygen-bearing molecules. 
There are up to nine TiO lines, with vibrationally excited transitions in $\varv=1$, with $E_\mathrm{low}$ values around 1500~K, and $\varv=2$, with $E_\mathrm{low}$ values around 3000~K.
Hence detecting more lines of TiO implies more vibrationally excited lines, as is also the case for HCN and CO. The three sources with the most TiO lines \textendash{} AH~Sco, R~Hya, and VX~Sgr \textendash{} are also the only sources to show vibrationally excited TiO transitions. The $^{13}$C isotopologues of HCN and CO are detected in all three of these sources, as are the CO $\varv=1$ and HCN $\varv_2=1$ transitions. 
The vibrationally excited CO $\varv=1$ has an $E_\mathrm{low}$ value around 3000~K, just as the $\varv=2$ transitions of TiO do,  
and the HCN $\varv_2=1$ transitions have $E_\mathrm{low}$ values around 1000~K, roughly similar to the $\varv=1$ transitions of TiO. This implies that these vibrationally excited transitions may originate in similar regions of the circumstellar envelope, potentially explaining their correlation.

SO$_2$ and H$_2$O are very strongly correlated with each other, and SO$_2$ is also very strongly correlated with SO. 
The four sources with the most transitions of SO$_2$ \textendash{} AH~Sco, R~Hya, U~Her, and VX~Sgr \textendash{} are also the four with the most SO transitions: this includes transitions of its $^{33}$S and $^{34}$S isotopologues, as well as the one vibrationally excited SO transition. These four sources are also in the half of the sample with the most detected H$_2$O transitions, as indeed are all seven of the sources with the most SO$_2$ transitions. 

SO$_2$ and SO are chemically connected by the fast reaction SO~+~OH~$\rightarrow$~SO$_2$~+~H \citep{demore_chemical_1997}. As SO$_2$ is created from SO, one would expect these molecules to be anti-correlated, but this is probably ameliorated by photolysis of SO$_2$ back to SO. This reaction also helps explain the strong correlation between SO$_2$ and OH, and its correlation with H$_2$O as observationally the photodissociation of H$_2$O into OH is linked with an abundance peak in SO \citep{danilovich_sulphur_2016}.
We detect up to 66 lines of SO$_2$ in our sources, with vibrational levels up to $\varv_2=2$. 
A majority ($\sim$65\%) of all the detected SO$_2$ lines are in the $\varv=0$ state, and most of the rest are in the $\varv_2=1$ state. The energy of the lower transition level ($E_\mathrm{low}$) values are up to 2300~K, with the $\varv=0$ transitions spanning the whole range, while the $\varv_2=1$ transitions have $E_\mathrm{low}$ values above 750~K, and the four $\varv_1=1$ and three $\varv_2=2$ transitions all have energies above 1500~K. The single vibrationally excited transition of SO has an $E_\mathrm{low}$ value around 1600~K. 
The ten detected H$_2$O lines have much higher $E_\mathrm{low}$ values ($\geq$3900~K) and many are in vibrationally excited states. 
There is some indication that in sources with more detected transitions of SO$_2$, SO, and H$_2$O, these transitions are probing a hotter gas as they generally have high $E_\mathrm{low}$ values. However, the H$_2$O transitions we can detect have much higher energies than any SO$_2$ or SO transitions, implying they are probing a different region of the wind, and furthermore these sources also show more transitions of $^{33}$S and $^{34}$S isotopologues. Therefore, the strong correlations between SO$_2$ and SO, and SO$_2$ and H$_2$O seem to also reflect increased abundances of all three molecules in certain sources.

OH only shows significant correlations with the other oxygen-bearing molecules: strong correlations with SO$_2$, H$_2$O, and TiO, and weak correlations with AlO, SO, PO, and SiO.
The strong correlation between OH and SO$_2$ can be explained by the fast SO$_2$ formation reaction described above. 
OH and H$_2$O are expected to be chemically related, we only detect transitions at high energies \textendash{} $\geq$3900~K for H$_2$O, and $\geq$4700~K for OH \textendash{} which are likely to be excited in similar regions around the star. 
The strong correlation between OH and TiO is surprising from a chemical standpoint as they would be expected to be anti-correlated due to the fast reaction TiO~+~OH~$\rightarrow$~TiO$_2$~+~H. However, the two molecules are not necessarily co-located even if they are found in the same sources, so this reaction may not be very prominent. The detection of TiO$_2$ does not help resolve this issue as its transitions are inherently fairly weak, and hence not unexpectedly TiO$_2$ is detected in only the highest mass-loss rate source, VX~Sgr,

SiO shows strong correlations with CO, H$_2$O, and PO; and weak correlations with larger pulsation amplitudes, and detecting more lines of HCN, SO$_2$, TiO, and OH. SiO is seen in every source due to its strong $\varv=0$ and $\varv=1$ lines, but some sources show more highly vibrationally excited transitions up to $\varv=5$, and lines from the $^{29}$SiO and $^{30}$SiO isotopologues. 
The five sources with the most PO transitions \textendash{} AH~Sco, IRC+10011, IRC-10529, R~Hya, and S~Pav \textendash{} are among the seven sources with the most SiO transitions. These sources also have among the most CO and H$_2$O transitions. 
Most of the SiO transitions with $\varv>0$ are dominated by maser emission and weak masers are seen even in $\varv=0$ for some stars (Pimpanuwat et al., in prep). The strong correlations with CO, H$_2$O and PO may indicate these high-$\varv$ SiO masers, with $E_\mathrm{low}$ values up to $\sim$8600~K, form in similar regions around the star as the high-energy lines of H$_2$O and PO, and the $\varv=1$ line of CO. Similarly, sources with more detected transitions of HCN, SO$_2$, TiO, and OH also tend to show vibrationally excited lines of these molecules, which may also form in similar regions.

\subsubsection{Cluster 7: $T_\mathrm{eff}$}\label{sect:cluster7}
Finally, we have a cluster containing only the effective temperature of the star, $T_\mathrm{eff}$, which shows the only significant negative correlation: a weak negative correlation with detecting more lines of AlOH. 
There are two detected lines of AlOH, and both are low energy $\varv=0$ transitions. Two lines of AlOH are detected in GY Aql, IRC+10011, and the RSG VX Sgr, while a single line is detected in U Her and the RSG AH Sco. GY Aql, IRC+10011, and U Her have effective temperatures at or below the median value for our sources, while the RSGs VX Sgr and AH Sco have the highest effective temperatures. So this negative correlation between effective temperature and detections of AlOH seems to mainly hold for AGB sources.
The correlation might potentially be explained by a relative lack of alumina dust (and hence more free Al to form AlOH) in these sources. There is some evidence for alumina dust being the major dust component in warmer semi-regular variable AGB stars, as it can form and survive at higher temperatures than silicate dust \citep[][and references therein]{gobrecht_dust_2016}, and GY Aql, IRC+10011, and U Her are all relatively cool Mira variables.

\subsection{SO and SO$_2$ spatial distributions}\label{sect:so_so2}

While a full analysis of the spatially resolved molecular data, including a radiative transfer analysis to obtain molecular abundance distributions, is beyond the scope of this paper, we have performed an initial analysis of the spatial distributions of SO and SO$_2$ and how they relate to the mass-loss rate of the source.
As was first put forward by \cite{danilovich_sulphur_2016}, there are two classes of SO distributions around AGB stars. For the lower mass-loss rate stars, the SO abundance peak is centred on the star, whereas for the higher mass-loss rate AGB stars, the relative SO abundance has a shell-like distribution. \cite{danilovich_sulphur_2016} found that the $e$-folding radius (the radius at which the abundance has dropped by a factor of $1/e$) of SO for the low mass-loss rate stars and the radius of the abundance peak for the higher mass-loss rate stars closely corresponded to the photodissociation radius of H$_2$O \citep[that is, the radius at which the abundance of OH, the photodissociation product of H$_2$O, peaks -- see the empirical relation from][]{netzer_mass_1987,maercker_hifi_2016}. This was interpreted to be the result of different SO formation pathways with varying wind density. A similar analysis could not be performed in that work for SO$_2$ since there were not enough detected lines for the higher mass-loss rate stars.

Here, we carefully examine the data to ascertain whether our observations are consistent with the distributions first proposed by \cite{danilovich_sulphur_2016}. Hence we must define the two classes of SO and SO$_2$ distributions observationally. 
We consider a centralized SO or SO$_2$ distribution to be one where, for the majority of detected lines, the peak of the molecular emission in centred on the continuum peak and decreases monotonically with distance from the star. Conversely, for a shell-like distribution, the molecular emission for most lines does not decrease uniformly with angular distance from the star: the radial distribution is either flat or shows a second peak offset from the continuum peak and often associated with a (broken) ring-like structure. The shell-like distributions may be apparent in integrated intensity maps, but are generally more clearly seen in channel maps (see Appendix~\ref{app:so_so2_distr}). They are also more clearly detectable in lower-energy lines.


SO and SO$_2$ were detected for most of the stars in the ATOMIUM sample. The only stars without SO and SO$_2$ detections are the two S-type stars, W~Aql and $\pi^1$~Gru, and the line-poor U~Del. The line-poor RSG KW~Sgr was detected in SO but not in SO$_2$. Henceforth, when referring to "all" stars in discussing SO and SO$_2$ detections, these sources are not included.

In Appendix~\ref{app:so_so2_distr} we show channel maps of the bright SO$_2$ $\varv=0$ $J_{K_a,K_c}=14_{0,14}-13_{1,13}$ line at 244.254~GHz in all the sources. The channel maps of R~Aql, GY~Aql, IRC-10529, and IRC+10011 (Figures~\ref{fig:raql_so2}--\ref{fig:irc+10011_so2}), as well as VX~Sgr (Figure~\ref{fig:vxsgr_so2}), clearly show an overall shell-like structure.
A similar shell-like structure is also seen for SO in these sources (shown around R~Aql in the $\varv=0$ $N_J=5_{5}-4_{4}$ line at 215.221~GHz in Fig. \ref{fig:raql_so}), but the SO emission is brighter closer to the continuum peak \textendash{} that is, the star itelf \textendash{} with fainter extended structures distributed similarly to the \so2.
These minor differences suggest that \so2 is predominantly found further from the star than SO is, which could be the result of \so2 being a daughter species (formed from SO) for this group of stars.
A similar trend is not seen for the stars with centralized \so2 emission, where SO and \so2 are both distributed similarly close to the continuum peaks. If anything, there is a slight tendency among the centralized \so2 sources for the SO emission to be more extended than the \so2 emission, but this could be a result of different abundances, some SO lines being inherently brighter than many \so2 lines, or differing excitation conditions between the examined lines.

\subsubsection{AGB stars}

The two brightest SO lines covered by our frequency setup are $N_J=5_5-4_4$ at 215.221~GHz and $N_J=6_5-5_4$ at 251.826~GHz, the two lines with $\Delta N = \Delta J = -1$ and $\varv=0$. These two lines are detected for all 14 stars with any SO detections. Overall six SO lines in the $\varv=0$ vibrational ground state are detected in our sample, with $E_\mathrm{low} \sim 10 - 90$~K. 
The most energetic line detected in our sample is the $N_J=6_7-5_6$ line at 259.857~GHz ($E_\mathrm{low}=1635$~K) in the first vibrationally excited state ($\varv=1$), which was detected for ten stars. The only other line in $\varv=1$ that was covered by our observations has a predicted intensity \citep[following the intensity calculations of][at 300~K]{pickett_submillimeter_1998} almost three orders of magnitude lower than the detected line, so we do not expect to detect it. 
With such a small number of SO lines covered over a relatively narrow range of energies, aside from the single line in $\varv=1$, it is difficult to draw any firm conclusions about trends across the sample. Without a more detailed analysis involving radiative transfer modelling, which is beyond the scope of the present work, we cannot easily determine which lines were not detected for a particular source because of excitation conditions in the CSE or because of the sensitivity of our observations.

SO$_2$ gives us more opportunity for such an analysis since many more lines, coming from a wide range of energy levels, fall in the covered frequency range and many of these were detected by our observations. 
One source, V~PsA shows only two detected \so2 lines: $J_{K_a,K_c} = 14_{0,14}-13_{1,13}$ at 244.254 GHz and $J_{K_a,K_c} = 30_{4,26}-30_{3,27}$ at 259.599 GHz. These are the lines with the highest predicted intensities (at 300~K) suggesting that sensitivity is the main limitation to detecting further \so2 lines. With so few lines, we cannot draw any further conclusions for V~PsA and exclude it from further discussion of \so2.
The other ten AGB stars for which we detect \so2 (GY~Aql, IRC+10011, IRC$-$10529, R~Aql, R~Hya, RW~Sco, S~Pav, SV~Aqr, T~Mic, and U~Her), all have detections of at least six lines in the ground vibrational state (with $30~\mathrm{K} \leq E_\mathrm{low} \leq 280~\mathrm{K}$). 
Four stars (R~Aql, R~Hya, U~Her and S~Pav) also have detections of at least four lines in the $\varv_2=1$ vibrational state.

The observational categories defined above, with shell-like or centralized \so2 distributions, approximately correspond to the categories put forward by \cite{danilovich_sulphur_2016} for SO, and are applicable to \so2 as the emission of these two molecules for the same star tends to be broadly similar.
Ordering the sources with \so2 detections by mass-loss rate, as in Table~\ref{tab:so2E}, shows a general tendency for the low mass-loss-rate sources to have centralized distributions and higher mass-loss-rate sources to have shell-like distributions. 
We note, however, that the mass-loss rates for many of the stars in the ATOMIUM sample are uncertain (see Section~\ref{sect:mlrs}).

\begin{table}[t]
\caption{\so2 spatial distributions, radial extents, and median lower level energies in all sources.} \label{tab:so2E}
\centering 
\begin{tabular}{lcccl} 
\hline\hline
Source & Max $r$ & Med. $E_\mathrm{low}$ & N & SO$_2$ \\ 
& R$_\ast$ & [K] & & distribution \\ 
\hline
S Pav & 144 & 443 & 35 & centralized \\ 
RW Sco & 451 & \phantom{0}83 & 17 & centralized$^\dagger$ \\ 
T Mic & 214 & 119 & 21 & centralized \\ 
R Hya & \phantom{0}51 & 281 & 33 & centralized \\ 
SV Aqr & 282 & \phantom{0}61 & \phantom{0}7 & centralized$^\dagger$ \\ 
U Her & 289 & 489 & 38 & centralized \\ 
R Aql & 510 & 138 & 23 & shell-like \\ 
GY Aql & 188 & \phantom{0}82 & 15 & shell-like  \\ 
IRC-10529 & 711 & \phantom{0}66 & 10 & shell-like \\ 
IRC+10011 & 692 & \phantom{0}82 & 11 & shell-like  \\ 
\hline
AH Sco & 342 & 816 & 55 & centralized  \\ 
VX Sgr & 400 & 816 & 57 & shell-like  \\ 
\hline 
\multicolumn{5}{l}{\footnotesize 
\textbf{Notes:} Spatial distributions with uncertain classifications due } \\ 
\multicolumn{5}{l}{\footnotesize 
to faint \so2 emission are marked with $^\dagger$. AGB stars are listed } \\ 
\multicolumn{5}{l}{\footnotesize 
above the horizontal line and RSGs below. } \\ 
\end{tabular} 
\end{table}

From an examination of the detected \so2 lines for each AGB source, we also found a tendency for the sources with shell-like emission to be detected in lower-energy \so2 lines, while the sources with centralized emission tended to be detected in more higher-energy lines (in addition to the lower-energy lines). Three out of the six stars with centralized distributions (R~Hya, U~Her and S~Pav) also show vibrationally excited emission. In contrast, only one star with shell-like emission (R~Aql) has vibrationally excited \so2 detections. Furthermore, some of the highest energy \so2 lines in the ground vibrational state (such as $J_{K_a,K_c} = 45_{6,40} - 44_{7,37}$ at 229.750~GHz with $E_\mathrm{low} = 1034$~K) are only detected in the centralized AGB sources.
To quantify this trend, Table \ref{tab:so2E} includes the average of the lower energy levels of each detected line for each star. 
Overall, there is a trend for the sources with a shell-like distribution to have lower median $E_\mathrm{low}$ values than the sources with a centralized distribution. This trend is confounded by RW~Sco and SV~Aqr, two centralized sources with median $E_\mathrm{low}$ values below 100~K. However, these sources both show faint \so2 emission, making the classification of their emission distribution more difficult. They also have the fewest \so2 detections among the centralized sources, tending to show only the brightest \so2 lines which generally have low $E_\mathrm{low}$ values.

The observed trend between \so2 line energy and spatial distribution is not a consequence of naturally having greater spatial resolution for the nearest sources, since our two most distant sources (IRC$-$10529 and IRC+10011 see Table~\ref{tab:sources}) have shell-like distributions of \so2 lines, while two of the nearest sources (T~Mic and R~Hya) have centralized \so2 lines. It is also not caused by lines with different energies being excited at different distances from the star, since this dichotomy holds if we use only the $J_{K_a,K_c}=14_{0,14}-13_{1,13}$ line at 244.254~GHz, which is one of the brightest and most frequently detected lines, to make the determination. We also note that there is a general tendency for centralized sources to have more \so2 lines detected than the shell-like sources: the maximum radial extent of \so2 is found to be negatively correlated ($r \sim -0.3$) with the median $E_\mathrm{low}$ values and the number of \so2 lines detected in each source.
This is most likely a consequence of the excitation conditions. The \so2 in centralized sources is found closer to the star and hence in a warmer environment, while a significant portion of the shell-like emission originates further out in a cooler region of the wind. At higher temperatures where there are more potential \so2 transitions to be excited, and hence unsurprising that more (and more energetic) \so2 lines are detected.

\begin{figure}
    \includegraphics[width=\hsize]{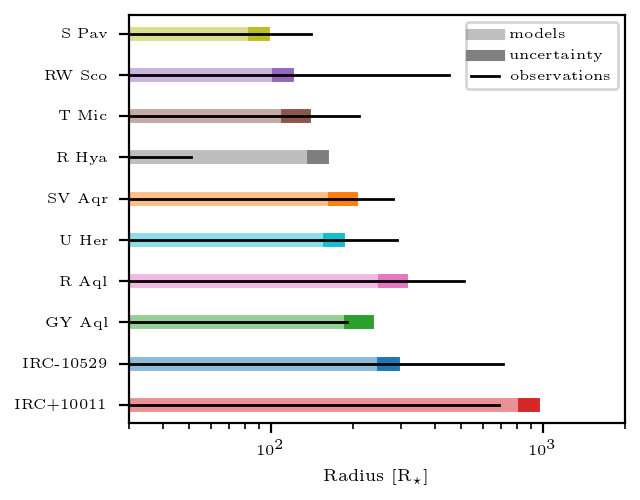}
    \caption{Emission sizes of SO$_2$ as expected from chemical models, and our maximum measured line extents, for each source.}
    \label{fig:so2_chem_models}
\end{figure}

We can also compare the measured extents of SO$_2$ in our AGB sources with what is expected from the chemical model of \citet{van_de_sande_determining_2018}, for which calculations for each ATOMIUM source (using the stellar parameters in Table~\ref{tab:sources}) is shown in Figure~\ref{fig:so2_chem_models}. The sources are arranged as in Table~\ref{tab:so2E}, in order of increasing mass-loss rate, and for each the colored bar shows the modeled $e$-folding radius of the SO$_2$ abundance, and the black line the maximum measured 3$\sigma$ extent of SO$_2$ lines from our observations. 
The chemical model assumes a power-law temperature profile with exponent $\varepsilon$, as implemented in \citet{van_de_sande_determining_2018} and \citet{maes_sensitivity_2023}:
\begin{equation}
T = T_\star \left(\frac{r}{R_\star}\right)^{-\varepsilon},
\end{equation}
where $T_\star$ is the surface temperature of the AGB star and $R_\star$ the stellar radius. As the temperature profile in the outflows is generally not known, a range of models with different temperature profiles, varying $\varepsilon$ from 0.3 to 1.0 (in steps of 0.05), were calculated. This range of models is included in Figure~\ref{fig:so2_chem_models} as an uncertainty on the model result. 

An increase in the \so2 radius with increasing mass-loss rate is clearly seen from the model results, and the observationally measured extents broadly follow the same trend. In general, the measured \so2 extents are larger than the modeled $e$-folding radii, and some discrepancy is expected when comparing modeled abundances with measured emission intensities, which does not take, for example, excitation effects into account.   
There are, however, two significant outliers: RW~Sco which has a much larger measured \so2 extent than expected from the models, and R~Hya whose measured extent is much smaller than expected. 
There are several reasons to suspect the mass-loss rate we use for RW~Sco is underestimated, which would explain its larger measured extent. First, its \so2 distribution is shell-like rather than centralized, suggesting it should have a relatively high mass-loss rate to follow the \so2 distribution trend. 
Second, the mass-loss rate value is taken from \citet{groenewegen_millimeter_1999}, who use an empirical formula from \citet{olofsson_study_1993} linking the integrated CO intensity with the mass-loss rate. \citet{olofsson_study_1993} see evidence that their formula underestimates the mass-loss rates for low mass-loss rate objects, like RW~Sco. This is in addition to the inherent uncertainties in the formula, which are a factor of $\sim$5. The discrepancy between the measured extent of \so2 and the prediction from the chemical model is further evidence that the mass-loss rate of RW~Sco may be underestimated.
As for the unexpectedly small observed extent of \so2 in R~Hya, its geometry may provide an explanation: \citet{homan_atomium_2021} find evidence for an equatorial density enhancement, and possible rotating disk, in the inner 0.4$\arcsec$ of the wind, which corresponds to $\sim$35~$R_\star$. This is similar to the maximum measured extent of \so2 (51~$R_\star$), so it may be that \so2 is largely confined to this disk, or not abundant enough to be detected outside it. Furthermore, the equatorial density enhancement increases the uncertainty of the calculated mass-loss rate by a factor of a few \citep{el_mellah_wind_2020}.
Some discrepancy may also be caused by the limitations of the chemical model: it assumes a smooth, spherically symmetric outflow, with a constant expansion velocity and mass-loss rate. R~Hya, with its equatorial density enhancement, deviates significantly from these assumptions. Additionally, most of the ATOMIUM sources show additional structure or deviations from spherical symmetry \citep{decin_substellar_2020, gottlieb_atomium_2022} which may affect the chemistry in the wind.

\subsubsection{RSG stars}

The two red supergiants, VX Sgr and AH Sco, are detected in seven and six SO transitions, respectively, out of the seven total detected in ATOMIUM. They are also extensively detected in \so2, with more than 45 lines seen for each star, and approximately a third of this count lying in the $\varv_2=1$ excited state. There are additionally three tentative detections of \so2 towards AH~Sco in the $\varv_2=2$ state and four tentative detections towards VX~Sgr in the $\varv_1=1$ state. Previous studies of the more massive and highly evolved red supergiants VY~CMa and NML~Cyg \citep{adande_sulfur_2013,Andrews2022} report asymmetric and localised emission for both stars, including multiple components of SO and \so2. In contrast, when considering only the lower-energy ground vibrational state SO and \so2 emission observed by ATOMIUM, we do not see directed outflows for VX~Sgr or AH~Sco. Based on observations of other molecular lines, it is likely that both of these stars have complex circumstellar structures. However, when considering only the SO and \so2 lines, we can place both RSGs into the categories we have defined here for AGB stars: VX~Sgr has shell-like \so2 and SO emission, while AH~Sco has more uniform emission resembling the centralized AGB stars. Since both stars have high mass-loss rates and many high-energy \so2 line detections, making them the two stars with the highest mean energy levels across the \so2 lines, they do not fit the trends observed for AGB stars. In light of this and the previous studies of SO and \so2 for RSGs, it is likely that different factors contribute to RSG SO and \so2 distributions, possibly including different formation mechanisms, compared with the AGB stars. However, a more detailed analysis of the ATOMIUM RSGs is beyond the scope of this paper.


\subsection{Isotopic ratios}\label{sect:isotopic_ratios}

In order to calculate molecular abundances from our observational data, extensive radiative transfer modeling would be required. However, with a few assumptions we can calculate isotopic ratios from observations of different isotopologues in the same transitions. These ratios provide constrains on the nucleosynthesis within AGB and RSG stars, their stellar mass and age, and the Galactic environment in which they were born. 
We have limited the calculation of isotopic ratios to pairs of transitions in $\varv=0$ with at least one minor isotope in both molecules, to avoid the problem of potential missing flux or high optical depth in the main isotopologue lines. Hence we assume all the lines used to calculate isotopic ratios are optically thin and have no missing flux; assumptions which are supported by their narrow line shapes and limited angular extents. We have also checked that the chosen lines don't appear to be masing, as none are excessively bright or narrow. In practise, this limits us to calculations with isotopologues of SiS, as some lines of for example SiO isotopologues fall outside our observed frequency ranges. 

We also need to account for the differences in line strength between different isotopologues, so to calculate the isotopic ratio between example isotopologues $^{a}X$ and $^{b}X$ we have used the following formula \citep{danilovich_alma_2020}:
\begin{equation}
\frac{^{a}X}{^{b}X} = \frac{I(^{a}X)}{I(^{b}X)} \left( \frac{\nu_{^{a}X}}{\nu_{^{b}X}} \right) ^{-2},
\end{equation}
where $I$ is the integrated intensity and $\nu$ the transition frequency for each isotopologue. 
For each pair of transitions we measured the integrated intensity in the spectrum where the sum of the intensities of both transitions is maximized. By taking the integrated intensities from spectra extracted at the same angular resolution and aperture size, the beam filling factors should be the same for both transitions. The calculated ratios are given in Table~\ref{tab:isorat}.
In cases where an isotopic ratio can be calculated from multiple pairs of transitions for a single source, Table~\ref{tab:isorat} contains the weighted average of these ratios. 

\begin{table}[t]
\caption{Isotopic ratios calculated from single and double isotopologues of SiS.}
\label{tab:isorat} 
\centering 
\begin{tabular}{llr}
\hline\hline
Source & Isotopologues & Ratio \\ 
\hline 
GY Aql & $^{29}$Si/$^{30}$Si & $1.3 \pm 0.3$ \\ 
IRC+10011 & $^{29}$Si/$^{30}$Si & $1.7 \pm 0.1$ \\ 
IRC-10529 & $^{29}$Si/$^{30}$Si & $1.2 \pm 0.1$ \\ 
$\pi^1$ Gru & $^{29}$Si/$^{30}$Si & $0.9 \pm 0.5$ \\ 
VX Sgr & $^{29}$Si/$^{30}$Si & $4 \pm 1$ \\   
W Aql & $^{29}$Si/$^{30}$Si & $1.5 \pm 0.5$ \\ 
\hline
W Aql & $^{33}$S/$^{34}$S & $0.4 \pm 0.3$ \\ 
\hline
IRC+10011 & $^{34}$S/$^{32}$S & $0.11 \pm 0.06$ \\ 
IRC-10529 & $^{34}$S/$^{32}$S & $0.04 \pm 0.02$ \\ 
W Aql & $^{34}$S/$^{32}$S & $0.06 \pm 0.02$ \\ 
\hline
\end{tabular} 
\end{table}

The silicon $^{29}$Si/$^{30}$Si isotopic ratios are found to be generally in the range 1--2 for our oxygen-rich and S-type AGB stars, with the RSG VX~Sgr showing a larger ratio of $\sim$4. Previously measured values in oxygen-rich sources are also in the range 1--2: 1.7 in IK Tau \citep{danilovich_alma_2019}; 1.58 in R Dor \citep{de_beck_circumstellar_2018}; between 0.99 and 1.35 in a sample of ten M-type AGB stars, and around 1.5 in two RSG stars \citep{peng_silicon_2013}. The solar value of 1.52 \citep{asplund_chemical_2021} is similarly in the same range as the AGB stars.

The high $^{29}$Si/$^{30}$Si ratio of $4\pm1$ measured in VX~Sgr is outside the range of almost all measured Si ratios in evolved stars and in the local galaxy in general. $^{29}$Si/$^{30}$Si ratios measured from presolar SiC grains are all close to the solar ratio of 1.52 \citep{zinner_silicon_2006}, as are measurements of various sources at different Galactic radii \citep{monson_uniform_2017}. The only literature ratio we were able to find that matches VX~Sgr are ratios $\sim$1--10 from infrared observations of the red giant EU Del \citep{pavlenko_analysis_2020}. EU Del is a very low-mass and metal-poor star, located below the tip of the red giant branch \citep{mcdonald_eu_2016}, so it is not similar to the high mass-loss rate RSG VX~Sgr. 
$^{29}$Si/$^{30}$Si ratios above 2 have also been measured in the AGB stars $\chi$ Cyg (2.4) and V1111 Oph (2.9) by \citet{ukita_sio_1988}, as well as a measured value of $3 \pm 1.5$ in IK~Tau by \citet{decin_circumstellar_2010}, consistent with the value of 1.7 cited above. 
We note that VX~Sgr has been shown to have some puzzling characteristics \citep{tabernero_nature_2021}, so this anomalous $^{29}$Si/$^{30}$Si ratio may be another signature of its odd nature.

For most oxygen-rich AGB stars we do not expect the $^{29}$Si/$^{30}$Si ratio to change during the AGB phase \citep{zinner_silicon_2006}, so the measured ratios reflect those from the stars' natal clouds. The primary isotope of silicon, $^{28}$Si, is an $\alpha$-process element, while the two isotopes $^{29}$Si and $^{30}$Si form largely from $^{25}$Mg and $^{26}$Mg during Ne burning, which creates similar amounts of both isotopes \citep[within a factor $\sim$1.5,][]{woosley_evolution_1995},
as well as during core-collapse Type II supernovae. According to the models of \citet{kobayashi_evolution_2011}, $^{29}$Si/$^{30}$Si ratios of $\sim$2--4 can be formed in the core-collapse supernovae of 25--30~M$_\odot$ progenitors, though the supernovae of both lower and higher mass progenitors produce much lower ratios. Some $^{29}$Si and $^{30}$Si also forms from $^{28}$Si in the He-burning shells of AGB stars, but in small and similar amounts 
\citep{monson_uniform_2017}.

The sulfur $^{33}$S/$^{34}$S isotopic ratio is only measured in the S-type star W~Aql, where it is found to be $0.4\pm0.3$. This ratio has been measured to be 0.19 $\pm$ 0.03 in IK Tau \citep{danilovich_alma_2019}, and 0.17 $\pm$ 0.02 in R Dor \citep{danilovich_alma_2020}. The solar value is 0.17 \citep{asplund_chemical_2021}, consistent with the two oxygen-rich AGB stars and also with W~Aql within the uncertainties. 

The sulfur $^{34}$S/$^{32}$S isotopic ratio is found to be $0.06\pm0.02$ in the S-type star W~Aql, and $0.04\pm0.02$ and $0.11\pm0.06$ in IRC+10011 and IRC-10529, two high mass-loss rate oxygen-rich sources. A slightly lower ratio of 0.03 is found in IK Tau \citep{danilovich_alma_2019}, while a value of 0.04 is found in the oxygen-rich RN Cnc \citep{winters_molecules_2022}, and both R Dor and the Sun have similar values of 0.05 \citep{danilovich_alma_2020, asplund_chemical_2021}. Hence all our measured $^{34}$S/$^{32}$S ratios are consistent with the solar value, within uncertainties. 

Both the main sulfur isotope $^{32}$S and the second most abundant isotope $^{34}$S are primarily produced through explosive nucleosynthesis during Type II supernovae, so these values measured in AGB stars reflect the abundances in the stars' natal clouds. However, the abundance of the $^{33}$S isotope may increase during the AGB phase via the slow neutron capture process \citep{anders_abundances_1989}. This is consistent with its detection in the S-type star W~Aql, which is more evolved than the oxygen-rich stars in our sample.

\subsection{Unidentified lines}\label{sect:ulines}
Table~A.4 lists the 28 unidentified lines in the ATOMIUM sample, and in which sources they are detected. 
Four of these lines \textendash{} U221.507, U254.791, U255.023, and U259.329 \textendash{} are coincident with calculated lines of SO$_2$ in its $\varv_3$=1 state. The lines were observed toward stars with SO$_2$ lines in $\varv=0$ and $\varv_2=1$ such that the assignments appear to be reasonable. On the other hand, except for S~Pav, toward which two such lines were detected, only one line was detected for the other stars and none of these stars displayed emissions in $\varv_2=2$ or $\varv_1=1$. We would therefore have to invoke very selective excitations, which appears to be too speculative at present even if this cannot be ruled out entirely. Therefore, we refrain from viewing these assignments even as tentative ones.

U253.973, seen in AH~Sco, IRC-10529, and VX~Sgr, is possibly due to NS, a blend of its $\varv=0~J=11/2-9/2, \Omega=1/2, F=13/2-11/2^f$ transition at 253.968 GHz, and $J=11/2-9/2, \Omega=1/2, F=11/2-9/2^f$ at 253.971 GHz. Facts in favor of this assignment are that these are the brightest NS lines covered by our observations, and the NS molecule has been detected in IK~Tau \citep{velilla_prieto_millimeter_2017}. Unfortunately, the corresponding $F=13/2-11/2^e$ and $F=11/2-9/2^e$ transitions around 253.570 GHz are not covered by our observations, and the U-line is blended with an SO$_2$ line in AH~Sco and VX~Sgr, making the assignment to NS very tentative.

Comparing all the U-line frequencies with the rotational lines of the following species \textendash{} MgO, CaO, NaO, FeO, ZrO, MgOH, CaOH, NaOH, ZrS, MgS, and CaS \textendash{} we tentatively conclude that there is no evidence for new metal oxides, hydroxides, and sulfides in the ATOMIUM survey. Furthermore, we would not expect most of these species to exist at observable levels because of the high concentrations of H and H$_2$ in the winds of these stars \citep{decin_alma_2018}.

\section{Conclusions}

We have observed 17 oxygen-rich AGB and RSG sources at high angular resolution ($0.02\arcsec - 0.05\arcsec$), and complementary lower resolutions (up to 1.4$\arcsec$), with the ALMA interferometer as part of the ATOMIUM Large Programme, detecting 291 transitions of 24 different molecules and their isotopologues. 

We find a range of conclusions, both major and minor:
\begin{itemize}
    \item We have first detections in oxygen-rich AGB and RSG stars of several vibrationally excited transitions: PO $\varv=1$, \so2 $\varv_1=1$ and $\varv_2=2$, and high energy H$_2$O transitions \citep[as examined in more detail in][]{baudry_atomium_2023}. 
    \item We also have several first detections in S-type AGB stars: vibrationally excited HCN $\varv_2=2,3$ and SiS $\varv=4,5,6$; as well as first detections of the molecules SiC, AlCl, and AlF in W~Aql \citep[as examined in more detail in]{danilovich_atomium_2021,danilovich_chemical_2023}.
    \item We have calculated correlations between the molecular content of different sources, finding strong correlations (with 3$\sigma$ significance and correlation coefficients above 0.64) between sources with more detected lines of: CS and SiS, CS and AlF, NaCl and KCl, AlO and SO, \so2 and SO, and \so2 and H$_2$O. Some of these correlations are expected from previous results (for example, CS and SiS both trace more dense winds) or chemical reactions (for example, NaCl and KCl both form from reactions of the metal atom with HCl), while the correlations of, for example, CS and AlF, or AlO and SO, have less clear origins. 
    \item Two of our sources are found to be extremely line-poor: the AGB star U~Del and the RSG KW~Sgr. We speculate this may be indicative of weak shocks in the inner wind.
    \item The spatial distributions of SO and \so2 are found to be generally consistent with previous results, with a centralized distribution for low mass-loss rate sources and a shell-like distribution for high mass-loss rate sources.
    \item The isotopic ratios of Si and S are generally in line with previously measured ratios (including solar ratios) except for an anomalously high $^{29}$Si/$^{30}$Si ratio of $4 \pm 1$ in the RSG VX~Sgr.
\end{itemize}

This paper has presented the overall molecular inventory and an initial analysis of the large ATOMIUM dataset, laying the groundwork for future work deriving molecular abundances and abundance profiles using radiative transfer modeling which will provide more rigorous tests for chemical models.

\begin{acknowledgements}
We thank the referee for their helpful comments improving the manuscript, and
Kelvin Lee for his contributions to the line identification strategy. 
SHJW acknowledges support from the Research Foundation Flanders (FWO) through grant 1285221N, and the ERC consolidator grant 646758 AEROSOL. 
TD acknowledges support from the Research Foundation Flanders (FWO) through grant 12N9920N, and the Australian Research Council through a Discovery Early Career Researcher Award (DE230100183). 
HSPM acknowledges support by the Deutsche Forschungsgemeinschaft through
the collaborative research grant SFB 956 (project ID 184018867).
SM and LD acknowledge support from the ERC consolidator grant 646758 AEROSOL, from the KU Leuven C1 excellence grant C16/17/007 MAESTRO, and from the FWO research grant 6099720N.
MVdS acknowledges support from the European Union’s Horizon 2020 research and innovation programme under the Marie Sk\l{}odowska-Curie grant agreement No 882991.
AB and FH acknowledge funding from the French National Research Agency (ANR) project PEPPER (ANR-20- CE31- 0002).
DG was funded by the project grant ‘The Origin and Fate of Dust in Our Universe’ from the Knut and Alice Wallenberg Foundation.
MM acknowledges funding from the Programme Paris Region fellowship supported by the R\'egion Ile-de-France. 
JMCP was supported by STFC grant number ST/T000287/1.
RS's contribution to the research described in this publication was carried out at the Jet Propulsion Laboratory, California Institute of Technology, under a contract with NASA.
KTW acknowledges support from the ERC under the European Union's Horizon 2020 research and innovation programme (Grant agreement no. 883867, project EXWINGS).

This paper makes use of the following ALMA data: ADS/JAO.ALMA\#2018.1.00659.L. ALMA is a partnership of ESO (representing its member states), NSF (USA), and NINS (Japan), together with NRC (Canada), NSC and ASIAA (Taiwan), and KASI (Republic of Korea), in cooperation with the Republic of Chile. The Joint ALMA Observatory is operated by ESO, AUI/NRAO, and NAOJ. We also acknowledge excellent support from the UK ALMA Regional Centre (UK ARC), which is hosted by the Jodrell Bank Centre for Astrophysics (JBCA) at the University of Manchester. The UK ARC Node is supported by STFC Grant ST/P000827/1.

This project has received funding from the European Union’s Horizon 2020 research and innovation program under the Marie Sk\l{}odowska-Curie Grant agreement No. 945298, 
and was supported in part by the Australian Research Council Centre of Excellence for All Sky Astrophysics in 3 Dimensions (ASTRO 3D) through project number CE170100013.

This work has made use of Python packages Astropy \citep{robitaille_astropy_2013, price-whelan_astropy_2018, price-whelan_astropy_2022}, SciPy \citep{virtanen_scipy_2020}, pandas \citep{mckinney_data_2010}, NumPy \citep{harris_array_2020}, and Matplotlib \citep{hunter_matplotlib_2007}.
\end{acknowledgements}

\bibliographystyle{aa}
\bibliography{holger_linerefs,papers}

\begin{appendix}
\setlength\LTcapwidth{\textwidth}

\section{Molecular data, line identifications and measured parameters}\label{app:lineids}

\subsection{Laboratory spectroscopic background on line identifications}
\label{sect:app:lab-spec-data}

Spectroscopic data were retrieved from the Cologne Database for Molecular Spectroscopy, 
CDMS,\footnote{https://cdms.astro.uni-koeln.de/} \citep{muller_cologne_2001,CDMS_2005,CDMS_2016} 
unless stated otherwise. Several of the accessed entries were created, updated, or extended 
in vibrational quanta $\varv$ to support the ATOMIUM project. While all of the relevant laboratory or observational 
data were considered for these entries, other data are not always up-to-date. We mention only 
cases in which our observations may be affected. We usually list the most important references 
for species with more than one line in ATOMIUM, possibly supplemented by references for data 
that encompass the ATOMIUM frequency range or are close to it. 
We list in Table~A.3 references of lines identified in ATOMIUM; 
we may refrain from listing additional background information for species with only 
one or two lines in ATOMIUM.

We point out that Hund's case (b) quantum numbers are given in the CDMS catalog throughout because of the \texttt{spfit} and \texttt{spcat} programs \citep{spfit_1991} employed for most of the CDMS entries. 
Hund's case (a) quantum numbers may be more appropriate for some of the radical species, and these are listed in most cases in Table~A.3. 
The CDMS documentation file usually contains information on how to convert Hund's case (b) quantum numbers to Hund's case (a). 
The quantum numbers $N$, $J$, and $F$ indicate the total rotational angular momentum, the total rotational angular momentum including the electronic spin, and the total rotational angular momentum including the electronic and nuclear spin, respectively. For non-radicals, $J = N$, and $J$ is usually given instead of $N$. 
The quantum numbers $K_a$ and $K_c$ designate projection of the total rotational angular momentum onto the $a$- and $c$-axis, respectively, in the case of asymmetric top molecules such as SO$_2$, H$_2$O, TiO$_2$, and SiC$_2$. 
Rotational levels are usually designated as $J_{K_a,K_c}$. 
Vibrational quanta are usually designated with $\varv$; a subscript indicated the number of the vibrational mode. 
The triply excited bending mode of a triatomic molecule is indicated as $\varv_2 = 3$. Multiple excitations of triatomics, for example, $\varv_2 = 3$, $\varv_3 = 1$, are often given as triple $(\varv_1\varv_2\varv_3)$, where $\varv_i$ indicates the excitation in the vibrational mode $i$; in the above case (031). 
The Greek letter $\nu$ is used to designate vibrational bands; $\nu_1$ could also be witten as $(100)-(000)$ and $\nu_1 - \nu_3$ as $(100)-(001)$.

The references for CO, CS, H$_2$S, and $^{13}$CN are given in Table~A.3 only.

The HCN rest frequencies are from a combined fit to laboratory data for $\varv_2 \le 3$, 
$\varv_3 = 1$, and $\varv_1 = 1$. The rotational data for $\varv_2 = 0$ and 1 are largely 
from \citet{HCN_v0_rot_2002} and \citet{HCN_v2_2003}, respectively; data for higher 
excited rotational data are largely from \citet{HCN_high-v_rot_2003}. 
The H$^{13}$CN rest frequencies are from an equvalent fit. The $\varv_2 = 0$ and 1 data 
are from \citet{HCN-isos_rot_2004}; additional extensive $\varv = 0$ data were taken 
from \citet{HC-13-N_v0_rot_2005}. Important for both fits were IR data from 
\citet{HCN_isos_IR_2000}.

The SiO rest frequencies were derived from \citet{SiO_isos_rot_2013} who performed Fourier 
transform microwave (FTMW) spectroscopy on several isotopic species up to very high 
vibrational states along with millimeter and submillimeter measurements of several 
isotopic species in their ground vibrational states. Additional FTMW data of SiO, 
$^{29}$SiO (and Si$^{18}$O) were taken from \citet{SiO_SiS_SO_FTMW_2003}. Excited 
vibrational data of the main isotopologue were published by \citet{SiO_rot_1991}. 
Noteworthy are furthermore IR data from \citet{SiO_isos_IR_1995}.

The current rest frequencies of SO rely mostly on \citet{SO_rot_1997} and 
\citet{SO_isos_rot_1996}, those of the isotopic species mostly on the latter work. 
Additional older data were also used at lower frequencies; some references are given 
in Table~A.3. 
We point out that Hund's case (a) quantum numbers may be found in the older literature, usually designated as $J_N$, whereas Hund's case (b) quantum numbers are more common in the more recent literature, designated as $N_J$.

The SO$_2$ $\varv_2 = 0$ and 1 data were derived from \citet{SO2_v0_v2_rot_2005} with 
important additional rest frequencies for $\varv = 0$ from \citet{SO2_isos_rot_1998} 
and for both vibrational states from \citet{SO2_v0_v2_34_rot_1996} and 
\citet{SO2_v0_v2_34_rot_1985}. The $\varv_2 = 2$ and $\varv_1 = 1$ data are 
based on an unpublished fit by one of us (HSPM) and employ besides own new data 
published rotational transition frequencies from \citet{SO2_vibs_rot_1968}, 
unpublished data communicated by the late Walter Lafferty, presumably associated 
with \citet{SO2_AlPIne_1996}, and IR data from \citet{SO2_nu1_1981} and 
\citet{SO2_IR_5states_1993}. The $^{34}$SO$_2$ $\varv = 0$ data are largely based 
on \citet{SO2_isos_rot_1998} with important additional rest frequencies from 
\citet{SO2_v0_v2_34_rot_1996} and \citet{SO2_v0_v2_34_rot_1985}.

The SiS rest frequencies were derived from \citet{SiO_isos_rot_2013} who performed Fourier 
transform microwave (FTMW) spectroscopy on several isotopic species up to very high 
vibrational states along with millimeter and submillimeter measurements of several 
isotopic species mostly in their ground vibrational states. Additional FTMW data of 
SiS, $^{29}$SiS and Si$^{34}$S) were taken from \citet{SiO_SiS_SO_FTMW_2003}. 
Important were also IR data, in particular from \citet{SiS_isos_IR_1990}.

The H$_2$O data pertaining to the lowest five vibrational states were based on the 
JPL catalog \citep{pickett_submillimeter_1998,intro_JPL-catalog} entry which, in turn, is based on 
\citet{H2O_5states_rot_2012}. Besides extensive new data, this work employs additional 
rotational and rovibrational data from a plethora of sources. 
\citet{H2O_2nd-Triad_2014} presented a similar study including the next three vibrational 
states, but, unfortunately, calculated transition frequencies for these start only at 300~GHz. 
Therefore, we inspected the HITRAN2020 \citep{HITRAN2020_2022} and W2020 
\citep{H2O_W2020} compilations for highly vibrationally excited H$_2$O transitions. 
As these were deemed to be quite uncertain, we resorted to the transition frequencies 
determined in our study on H$_2$O and OH \citep{baudry_atomium_2023}.

The initial OH data were taken from the JPL catalog and bear on \citet{OH_analysis_2013}. 
This analysis is based on a plethora of laboratory spectroscopic investigations. 
The $\Lambda$-doubling transitions with high rotational quanta, however, are rather uncertain 
and display systematic deviations, see for example \citet{OH_highly-ex_2019} and \citet{baudry_atomium_2023}. In our study 
on H$_2$O and OH \citep{baudry_atomium_2023}, we combined the data gathered in 
\citet{OH_analysis_2013} with $\Lambda$-doubling transition frequencies from our own 
astronomical observations and those from \citet{OH_highly-ex_2019} to improve the 
calculations of the OH $\Lambda$-doubling transition frequencies in the upper millimeter 
and submillimeter regions \citep{baudry_atomium_2023}.

Calculations of the PO rest frequencies rely mainly on \citet{PO_vibs_rot_2002} 
with additional $\varv = 0$ data from \citet{PO_rot_1983}.

The TiO rest frequencies are based on an unpublished isotopic invariant fit by one of us 
(HSPM). The main source of experimental data are those from the minor Ti isotopologues 
from \citet{TiO-isos_rot_2016}. Additional important data relevant for our study are 
the ground state rotational data of $^{46,48,50}$TiO from \citet{TiO_isos_rot_2008}, 
those of $^{48}$TiO from \citet{TiO-rot_1998}, and the extensive IR data for all 
Ti isotopologues from \citet{TiO_isos_IR_2021}.

Calculations on the rotational spectrum of TiO$_2$ are based on \citet{TiO2_rot_2011}. 
Besides our own upper millimeter and lower submillimeter data of $^{46,48,50}$TiO$_2$, 
the analysis also employs microwave data of the same isotopologues from \citet{TiO2_rot_2008}. 
The rotational temperatures were of order of 30~K and 3~K, respectively, which limited 
the quantum number range accessed. Transition frequencies involving higher rotational 
excitation may be quite uncertain as a consequence. The $K_a$ = 6$-$5 $Q$-branch 
transitions with $J$ = 25, 27, and 29 display larger uncertainties, such that their 
true transition freaquencies may differ from the calculated ones by more than 1~MHz.

The ground state rotational data of AlO were taken from \citet{AlO_rot_1990} and from 
\citet{AlO_rot_1989}; transition frequencies of $\varv = 1$ and 2 were published by 
\citet{AlO_vibs_rot_1994}.

The AlOH data were taken from the JPL catalog; they are derived from \citet{AlOH_rot_1993}.

The AlF rest frequencies are mainly based on the rotational data of \citet{AlF_rot_1970a}. 
Additional rotational data are from \citet{AlF_rot_1970Zb}, and rovibrational data are 
mainly from \citet{AlF_IR_1992}.

Calculations of the rotational spectrum of AlCl and Al$^{37}$Cl depends mainly on 
rotational data of \citet{AlCl_rot_1972}. Very accurate frequencies of the $J = 1 - 0$ 
transitions by \citet{div-MX_rot_1993} were also employed as were rovibrational transitions 
taken from \citet{AlCl_IR_1993}.

The NaCl and KCl rest frequencies rely mainly on measurements by 
\citet{NaCl_rot_2002} and \citet{KCl_rot_2004}, respectively.

The SiN rest frequencies are based on \citet{SiN_rot_2006}, with the important lower 
frequency data taken from \citet{SiN_rot_1983}.

The SiC transition frequencies are based on \citet{SiC_isos_rot_1990} and on 
\citet{SiC_rot_det_1989}.

The current calculation of SiC$_2$ $\varv=0$ transition frequencies were derived from 
\citet{SiC2_HIFI_analysis_2012}. Important additional data besides HIFI-Herschel
data from that work come from laboratory measurements by \citet{SiC2_rot_1989} and 
from astronomical observations by \citet{cernicharo_2_2000}.

The ground state rotational data of HC$_3$N depend on \citet{HC3N_rot_2000} with 
additional contributions mainly from \citet{HC3N_rot_1995}.


\onecolumn

\clearpage
\section{SO$_2$ and SO channel maps}\label{app:so_so2_distr}

\begin{figure*}[ht]
    \includegraphics[width=\textwidth]{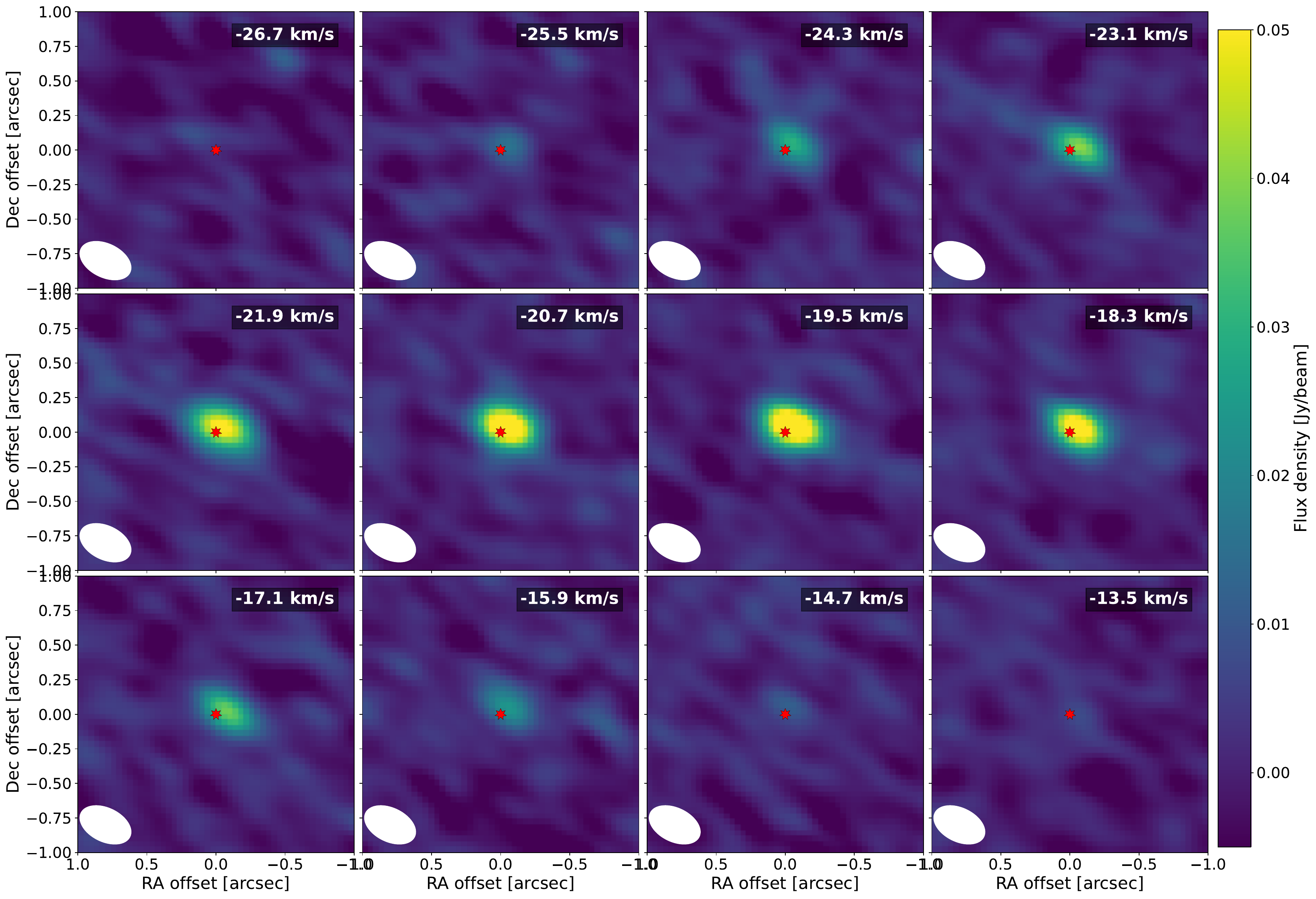}
    \caption{Channel map of the SO$_2$ $\varv=0$ $J_{K_a,K_c}=14_{0,14}-13_{1,13}$ line at 244.254~GHz observed towards S~Pav at medium angular resolution. The synthesized beam is indicated by the white ellipse in the lower left corner of each panel, the LSR velocity in \kms is indicated in the upper right corner, and the continuum peak is indicated by the position of the red star at (0,0).}
    \label{fig:spav_so2}
\end{figure*}

\begin{figure*}[ht]
    \includegraphics[width=\textwidth]{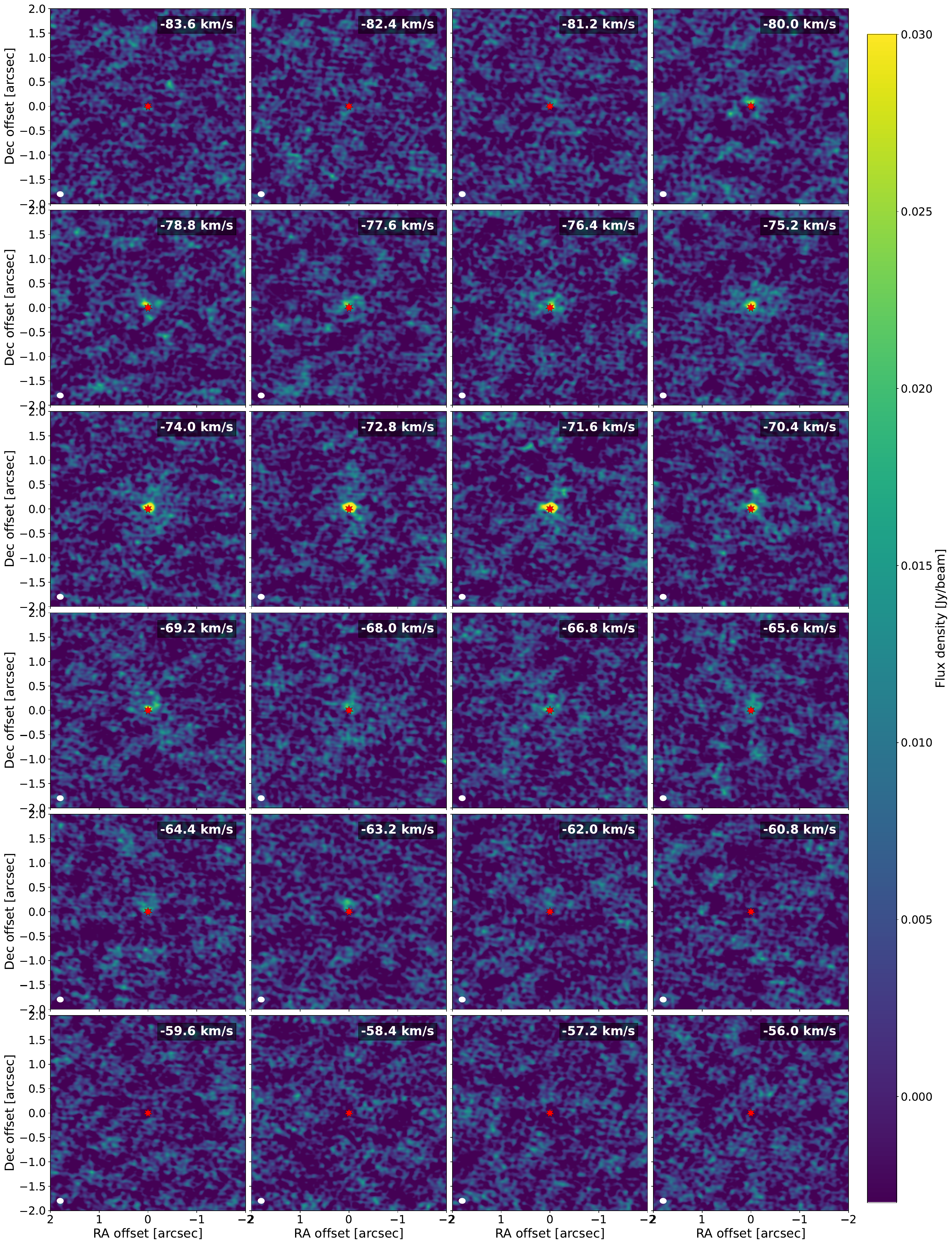}
    \caption{Channel map of the SO$_2$ $\varv=0$ $J_{K_a,K_c}=14_{0,14}-13_{1,13}$ line at 244.254~GHz observed towards RW~Sco at medium angular resolution. See caption of Figure~\ref{fig:spav_so2}.}
    \label{fig:rwsco_so2}
\end{figure*}

\begin{figure*}[ht]
    \includegraphics[width=\textwidth]{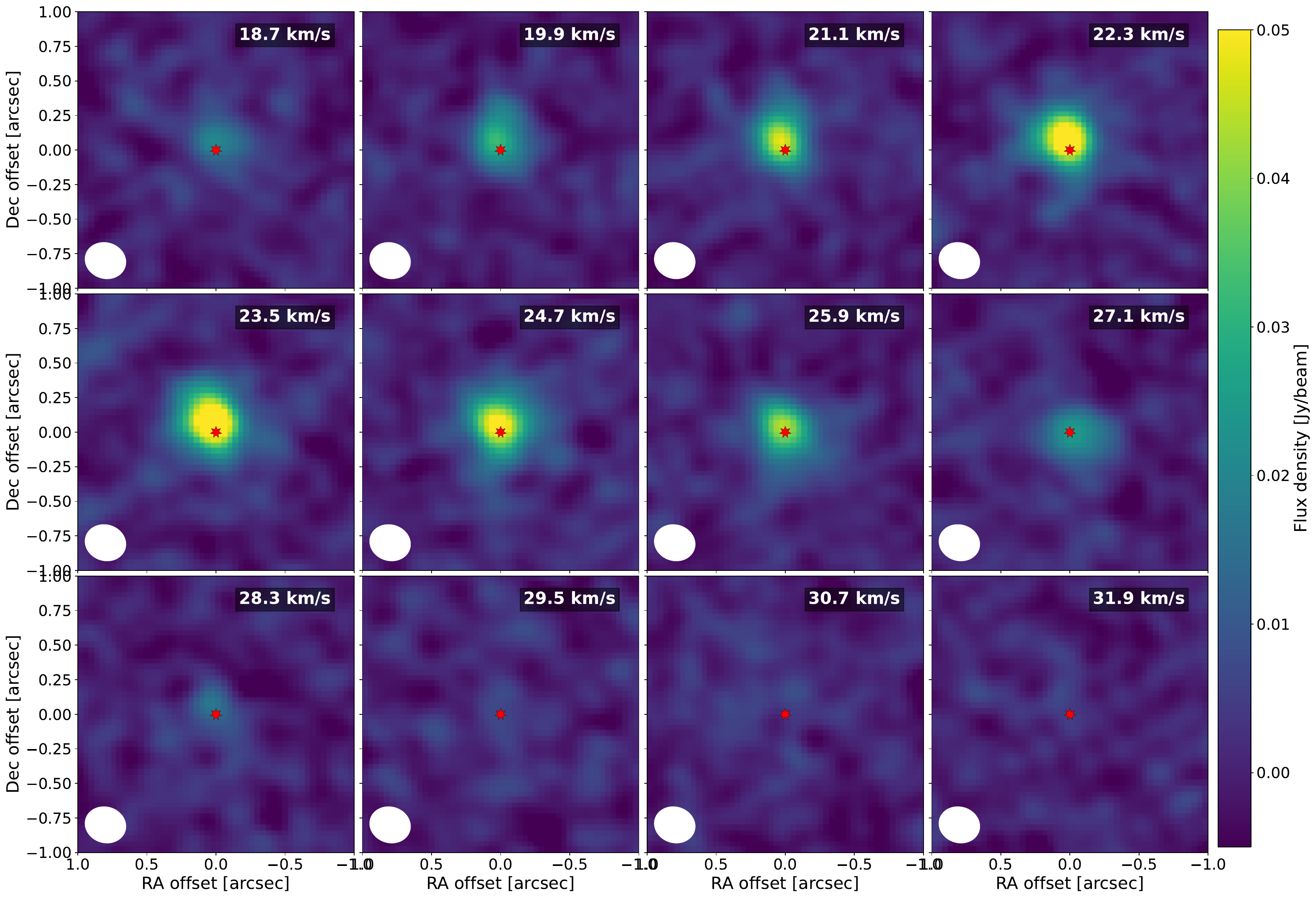}
    \caption{Channel map of the SO$_2$ $\varv=0$ $J_{K_a,K_c}=14_{0,14}-13_{1,13}$ line at 244.254~GHz observed towards T~Mic at medium angular resolution. See caption of Figure~\ref{fig:spav_so2}.}
    \label{fig:tmic_so2}
\end{figure*}

\begin{figure*}[ht]
    \includegraphics[width=\textwidth]{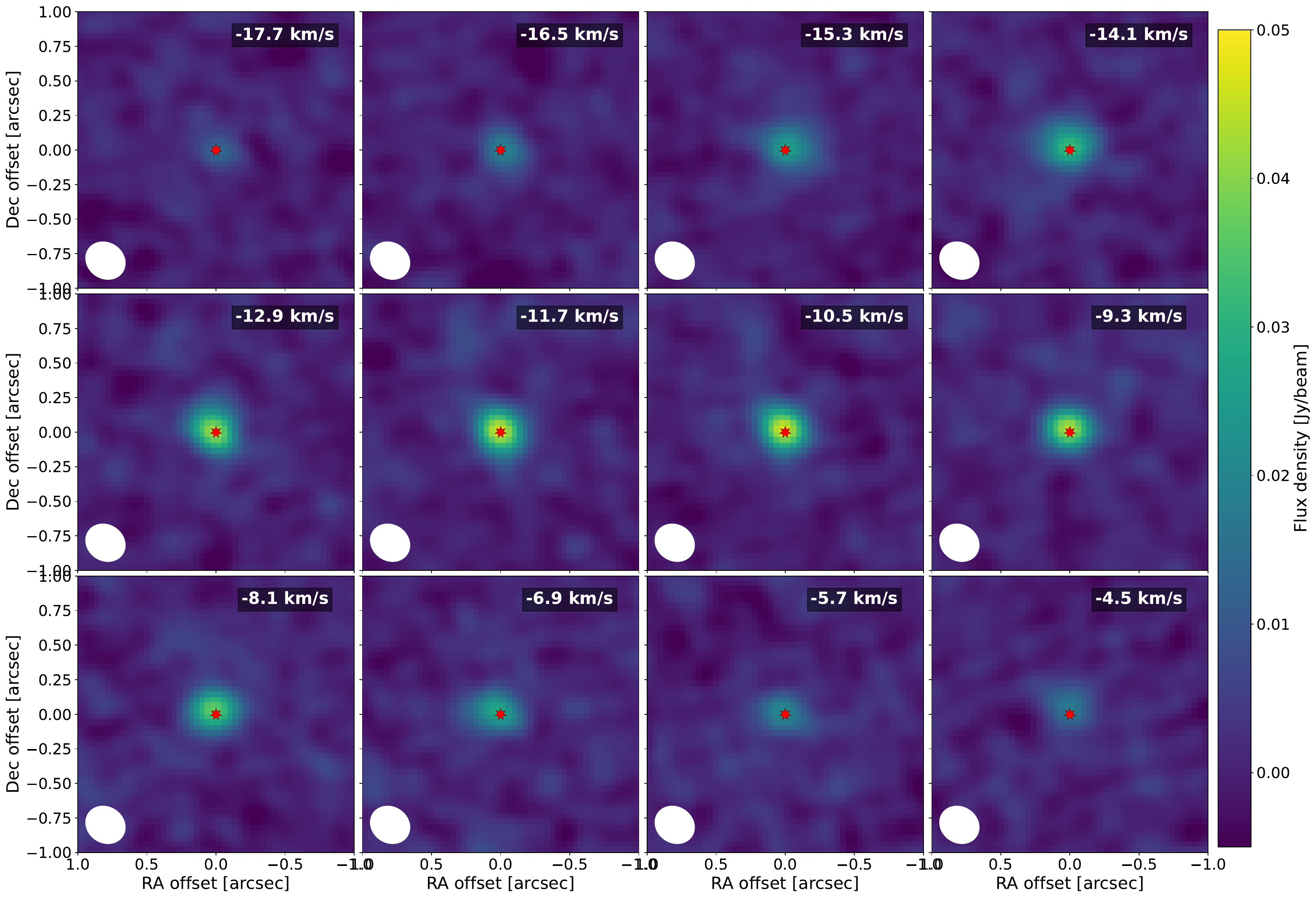}
    \caption{Channel map of the SO$_2$ $\varv=0$ $J_{K_a,K_c}=14_{0,14}-13_{1,13}$ line at 244.254~GHz observed towards R~Hya at medium angular resolution. See caption of Figure~\ref{fig:spav_so2}.}
    \label{fig:rhya_so2}
\end{figure*}

\begin{figure*}[ht]
    \includegraphics[width=0.98\textwidth]{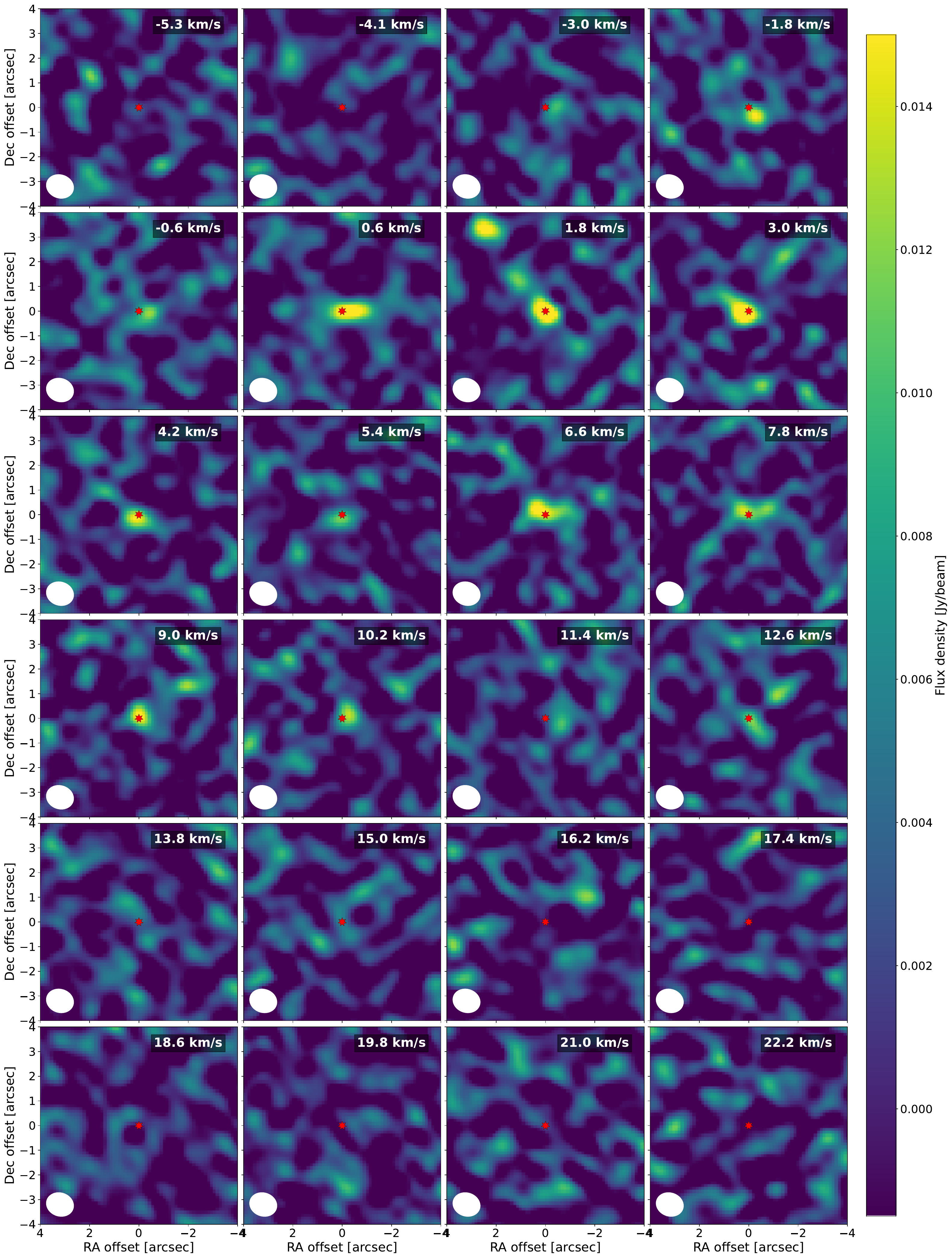}
    \caption{Channel map of the SO$_2$ $\varv=0$ $J_{K_a,K_c}=14_{0,14}-13_{1,13}$ line at 244.254~GHz observed towards SV~Aqr at low angular resolution. The medium angular resolution map is too faint to show the emission distribution. See caption of Figure~\ref{fig:spav_so2}.}
    \label{fig:svaqr_so2}
\end{figure*}

\begin{figure*}[ht]
    \includegraphics[width=\textwidth]{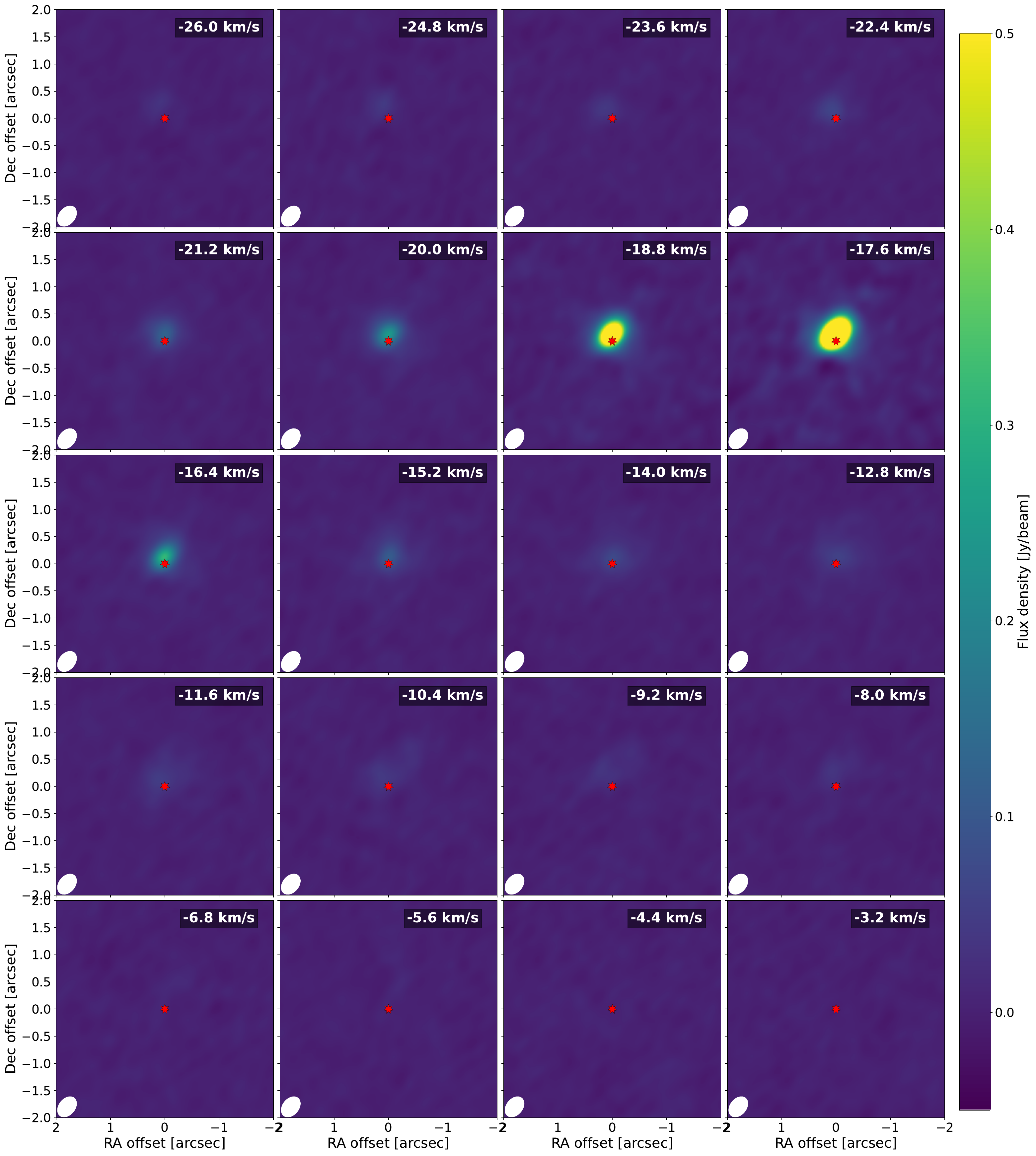}
    \caption{Channel map of the SO$_2$ $\varv=0$ $J_{K_a,K_c}=14_{0,14}-13_{1,13}$ line at 244.254~GHz observed towards U~Her at medium angular resolution. See caption of Figure~\ref{fig:spav_so2}.}
    \label{fig:uher_so2}
\end{figure*}

\begin{figure*}[ht]
    \includegraphics[width=\textwidth]{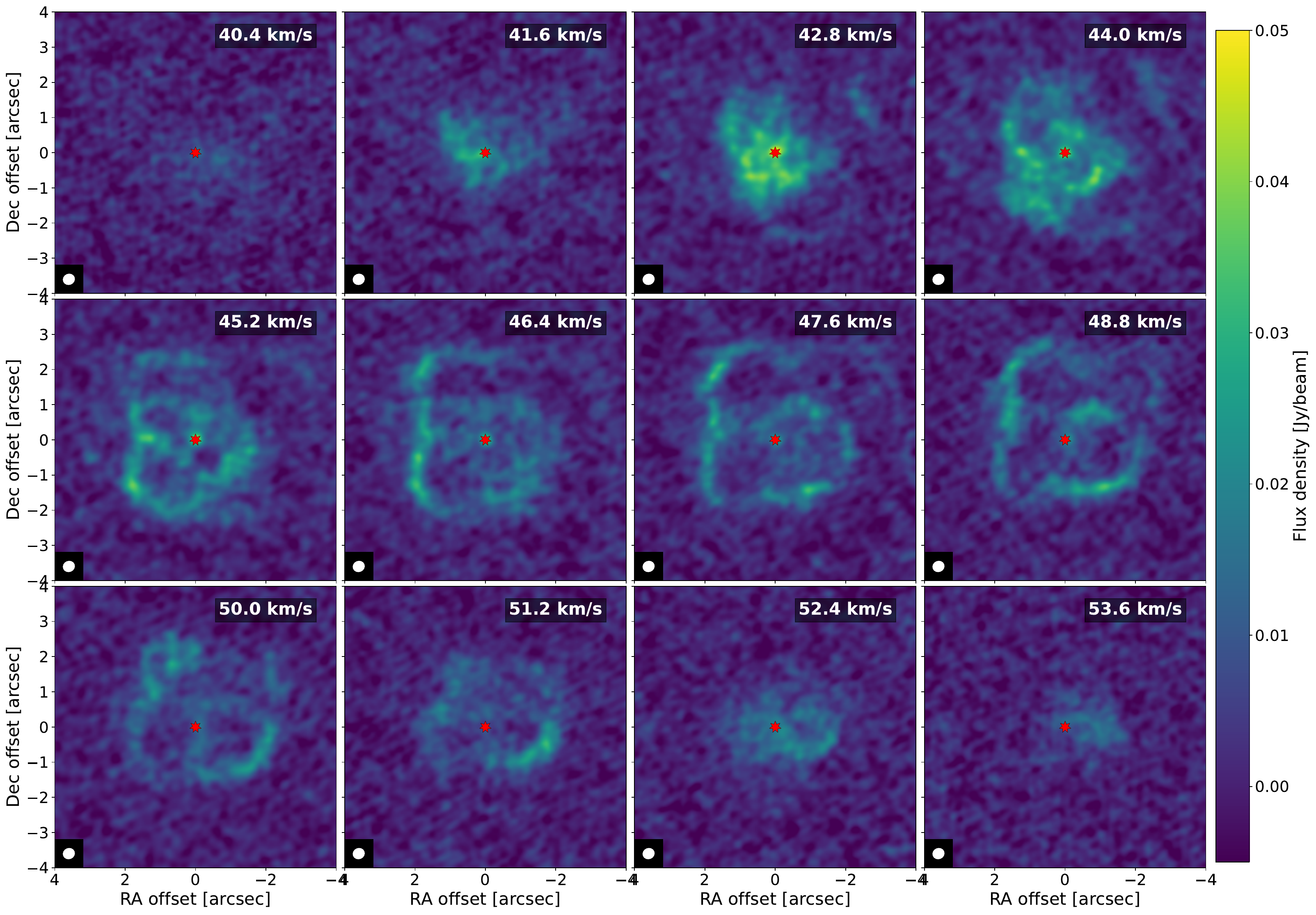}
    \caption{Channel map of the SO$_2$ $\varv=0$ $J_{K_a,K_c}=14_{0,14}-13_{1,13}$ line at 244.254~GHz observed towards R~Aql at medium angular resolution. See caption of Figure~\ref{fig:spav_so2}.}
    \label{fig:raql_so2}
\end{figure*}

\begin{figure*}[ht]
    \includegraphics[width=\textwidth]{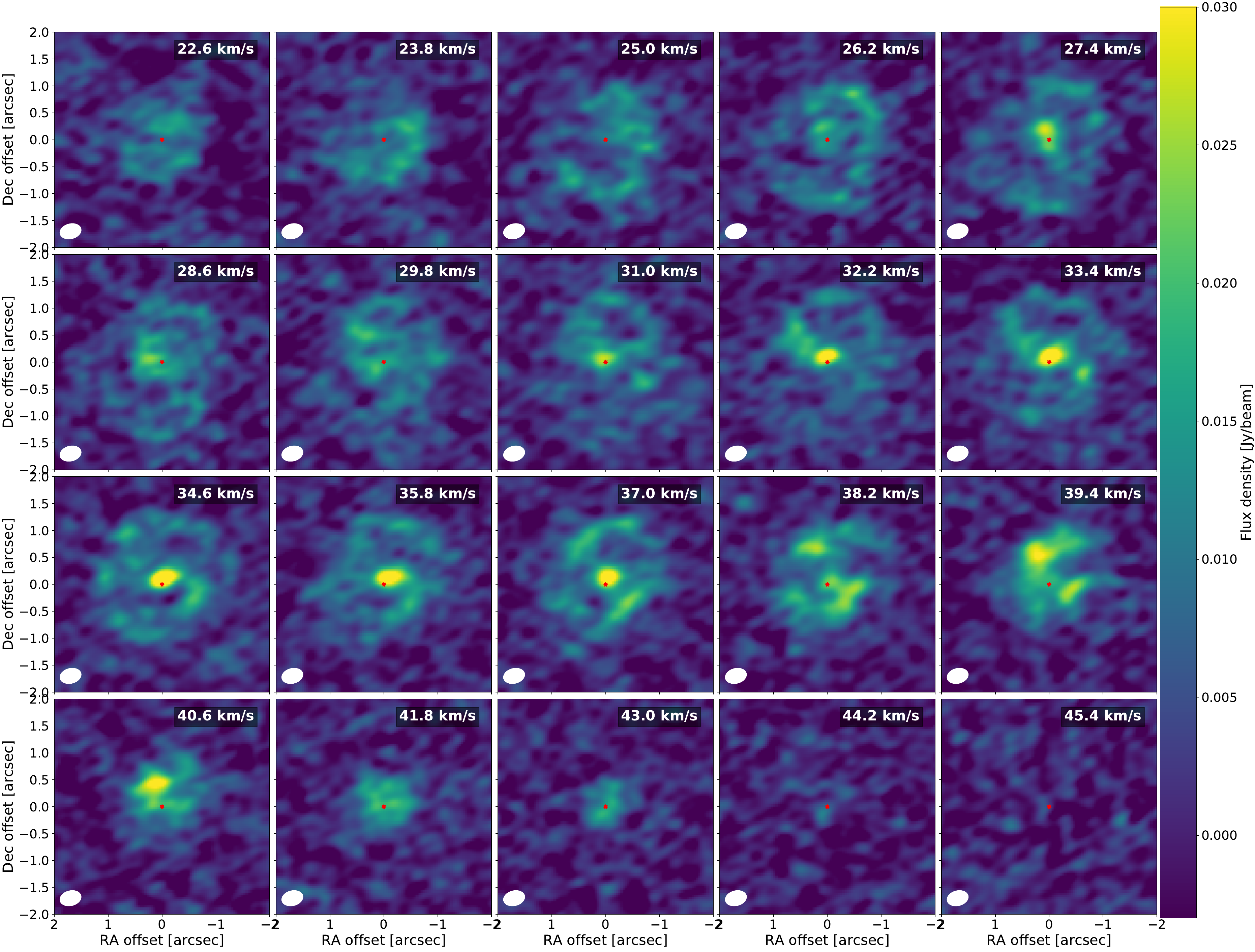}
    \caption{Channel map of the SO$_2$ $\varv=0$ $J_{K_a,K_c}=14_{0,14}-13_{1,13}$ line at 244.254~GHz observed towards GY~Aql at medium angular resolution. See caption of Figure~\ref{fig:spav_so2}.}
    \label{fig:gyaql_so2}
\end{figure*}

\begin{figure*}[ht]
    \includegraphics[width=\textwidth]{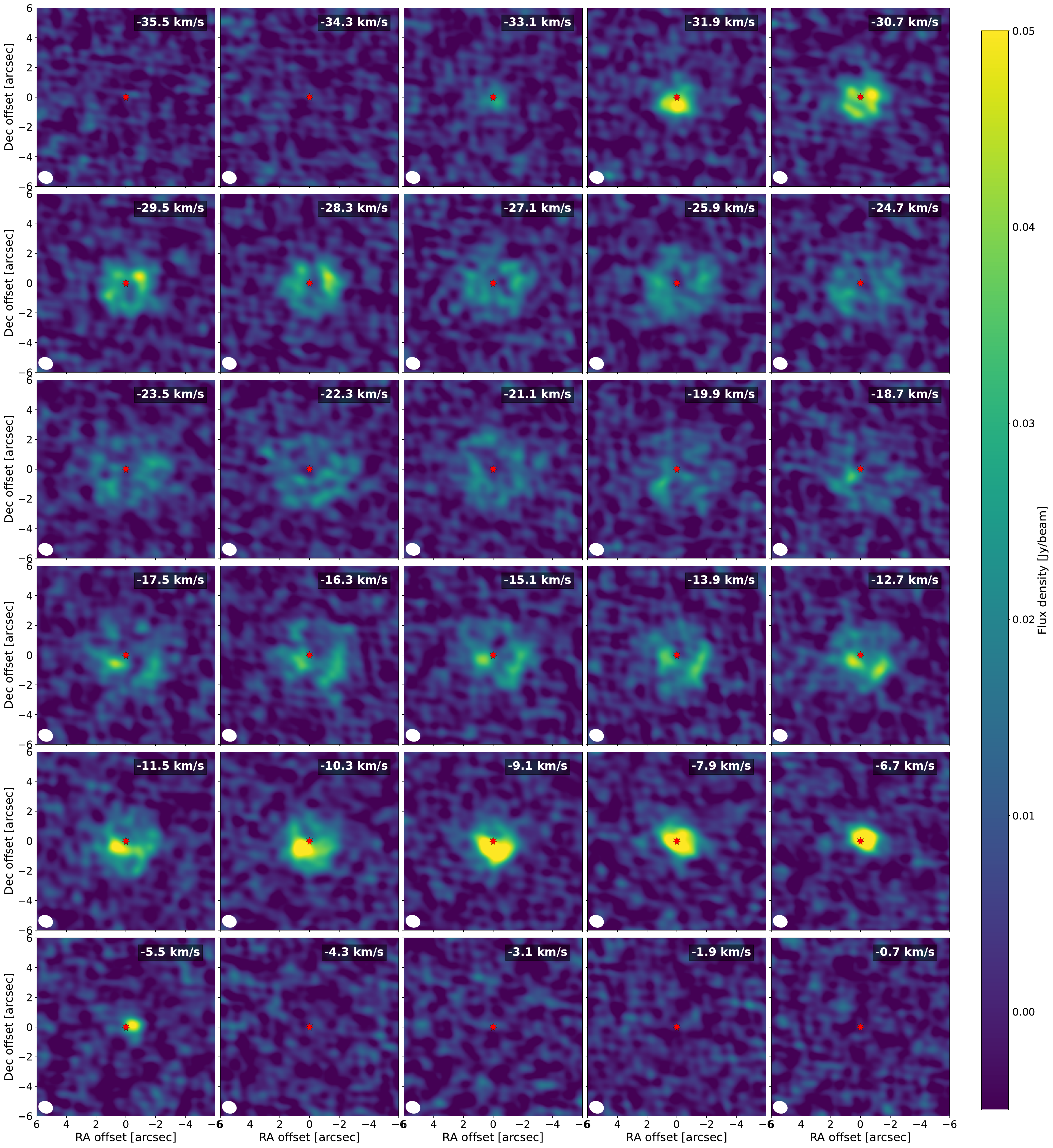}
    \caption{Channel map of the SO$_2$ $\varv=0$ $J_{K_a,K_c}=14_{0,14}-13_{1,13}$ line at 244.254~GHz observed towards IRC-10529 at low angular resolution. The medium angular resolution map has too much resolved out flux to show the emission distribution. See caption of Figure~\ref{fig:spav_so2}.}
    \label{fig:irc-10529_so2}
\end{figure*}

\begin{figure*}[ht]
    \includegraphics[width=\textwidth]{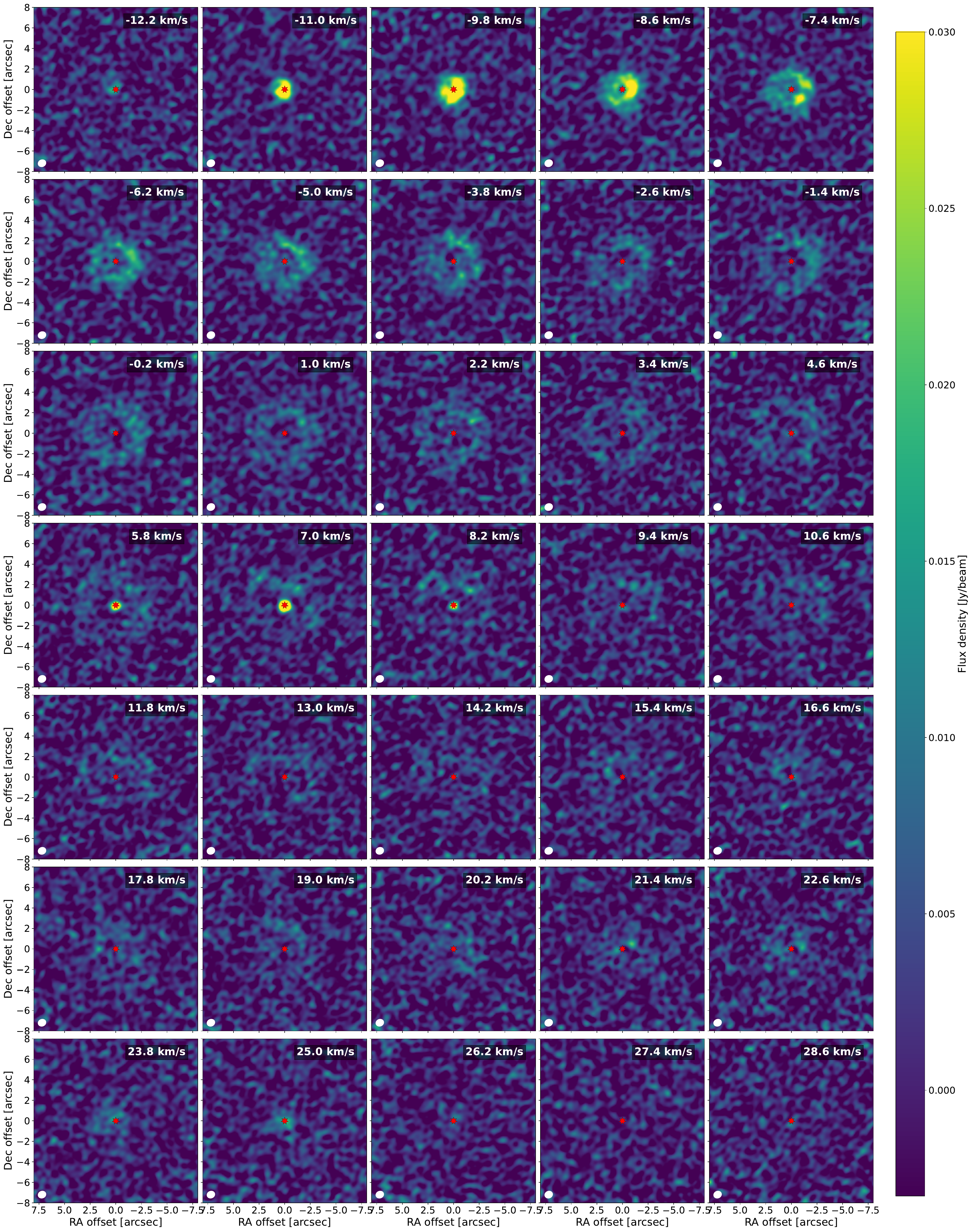}
    \caption{Channel map of the SO$_2$ $\varv=0$ $J_{K_a,K_c}=14_{0,14}-13_{1,13}$ line at 244.254~GHz observed towards IRC+10011 at low angular resolution. The medium angular resolution map has too much resolved out flux to show the emission distribution. See caption of Figure~\ref{fig:spav_so2}.}
    \label{fig:irc+10011_so2}
\end{figure*}

\begin{figure*}[ht]
    \includegraphics[width=\textwidth]{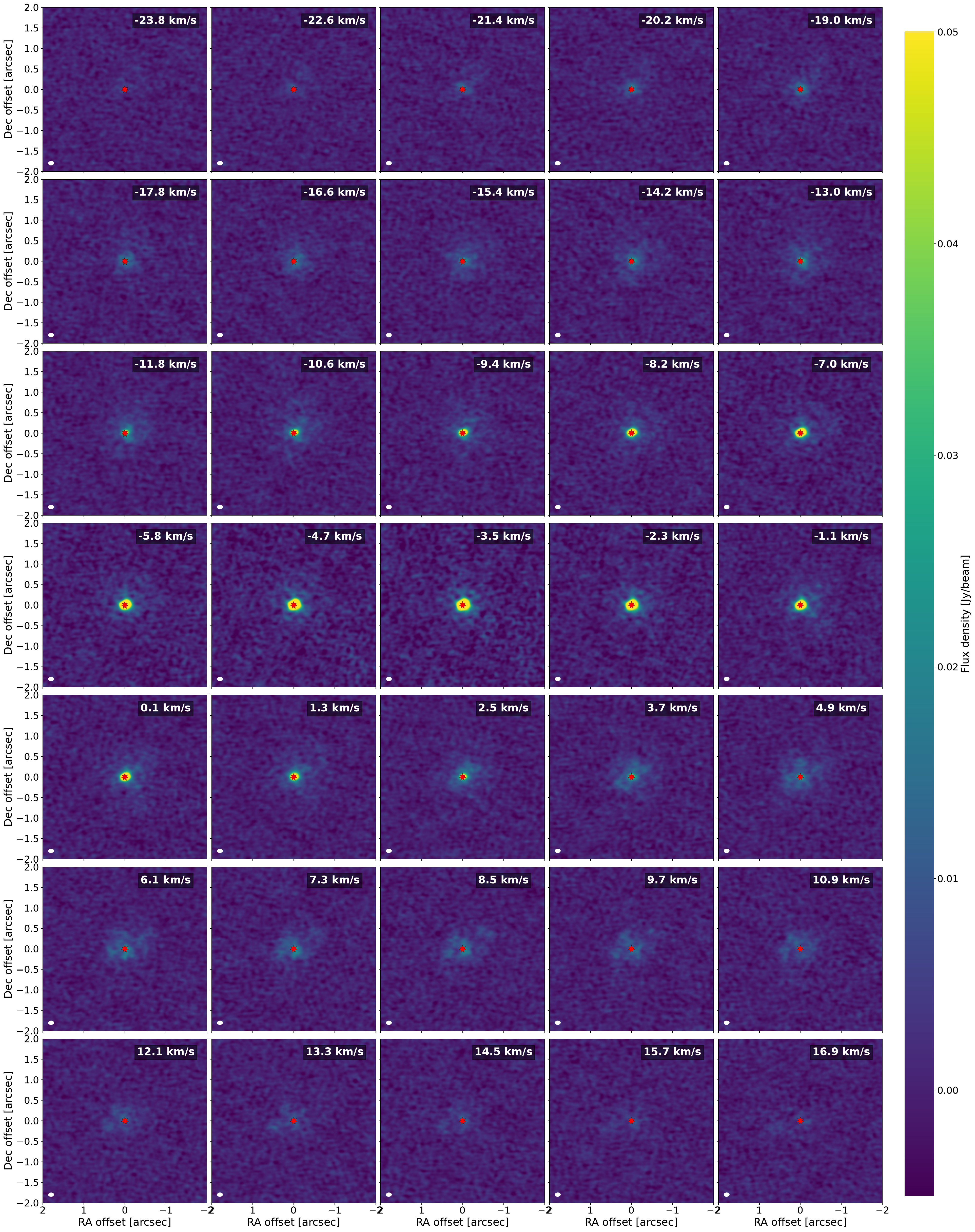}
    \caption{Channel map of the SO$_2$ $\varv=0$ $J_{K_a,K_c}=14_{0,14}-13_{1,13}$ line at 244.254~GHz observed towards AH~Sco at medium angular resolution. See caption of Figure~\ref{fig:spav_so2}.}
    \label{fig:ahsco_so2}
\end{figure*}

\begin{figure*}[ht]
    \includegraphics[width=\textwidth]{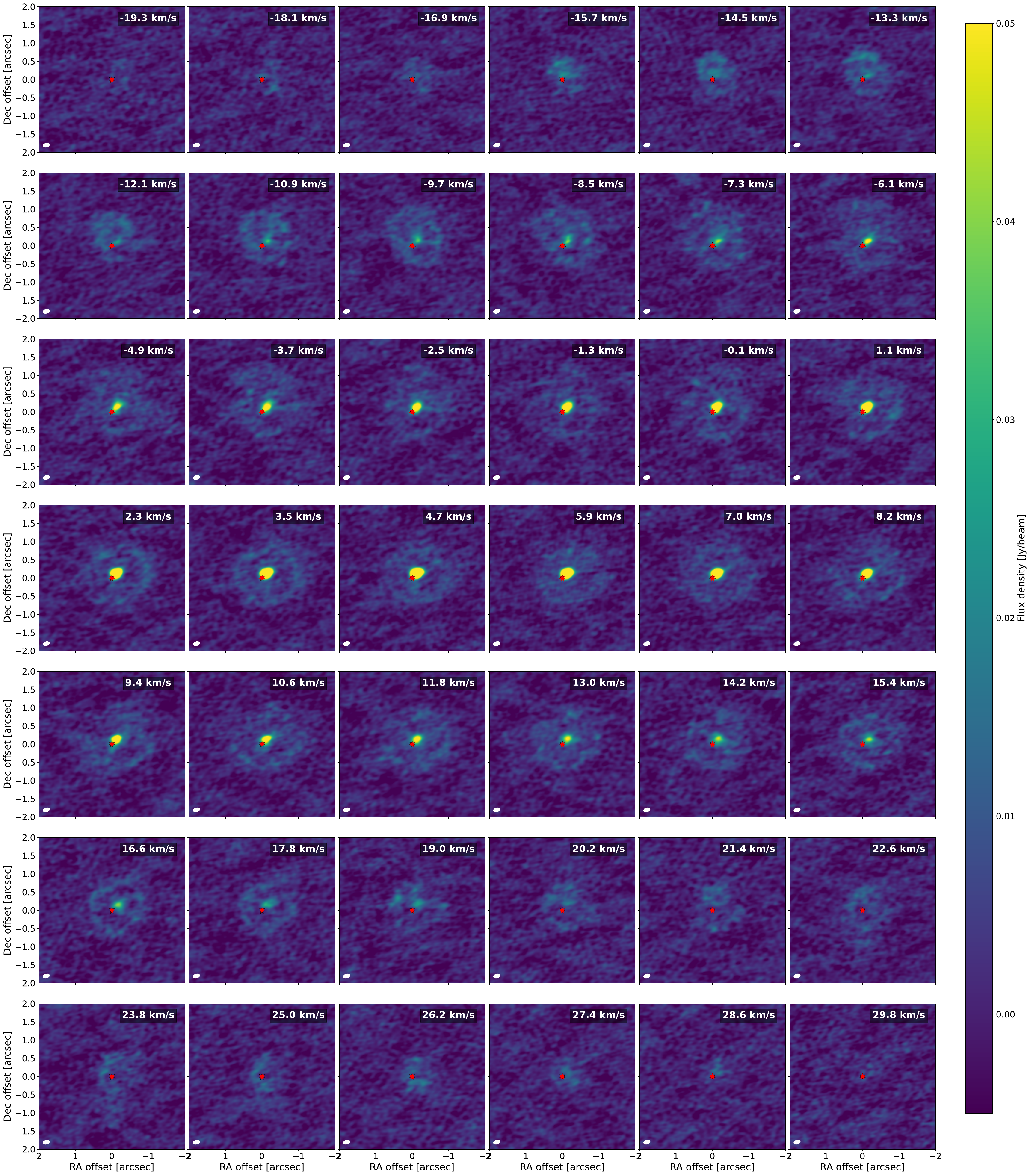}
    \caption{Channel map of the SO$_2$ $\varv=0$ $J_{K_a,K_c}=14_{0,14}-13_{1,13}$ line at 244.254~GHz observed towards VX~Sgr at medium angular resolution. See caption of Figure~\ref{fig:spav_so2}.}
    \label{fig:vxsgr_so2}
\end{figure*}

\begin{figure*}[ht]
    \includegraphics[width=\textwidth]{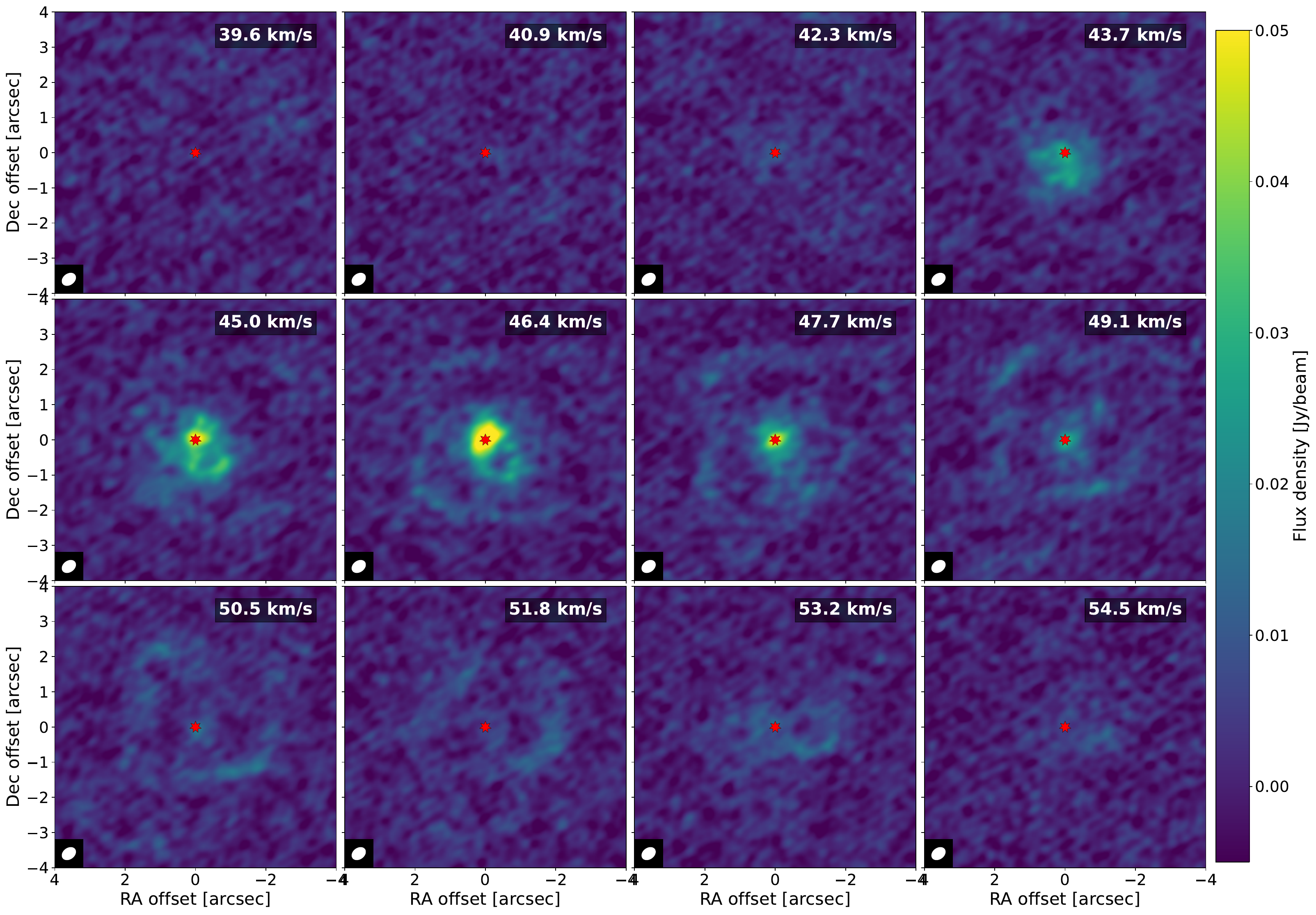}
    \caption{Channel map of the SO $\varv=0$ $N_J=5_{5}-4_{4}$ line at 215.221~GHz observed towards R~Aql at medium angular resolution. See caption of Figure~\ref{fig:spav_so2}.}
    \label{fig:raql_so}
\end{figure*}

\clearpage
\section{Spectra}

\begin{figure}[thbp]
\centering
\caption{Spectrum of R~Hya, in the 16 observed frequency bands. 
The top spectrum is extracted in a 1.2$\arcsec$-diameter circular aperture from the medium resolution spectrum, and the bottom shows the high resolution spectrum extracted in a 0.08$\arcsec$-diameter aperture. All identified lines are marked, though note that some are more easily seen at other resolutions or extraction apertures.}
\label{app:RHya_spec}
\subfloat{\includegraphics[width=\textwidth]{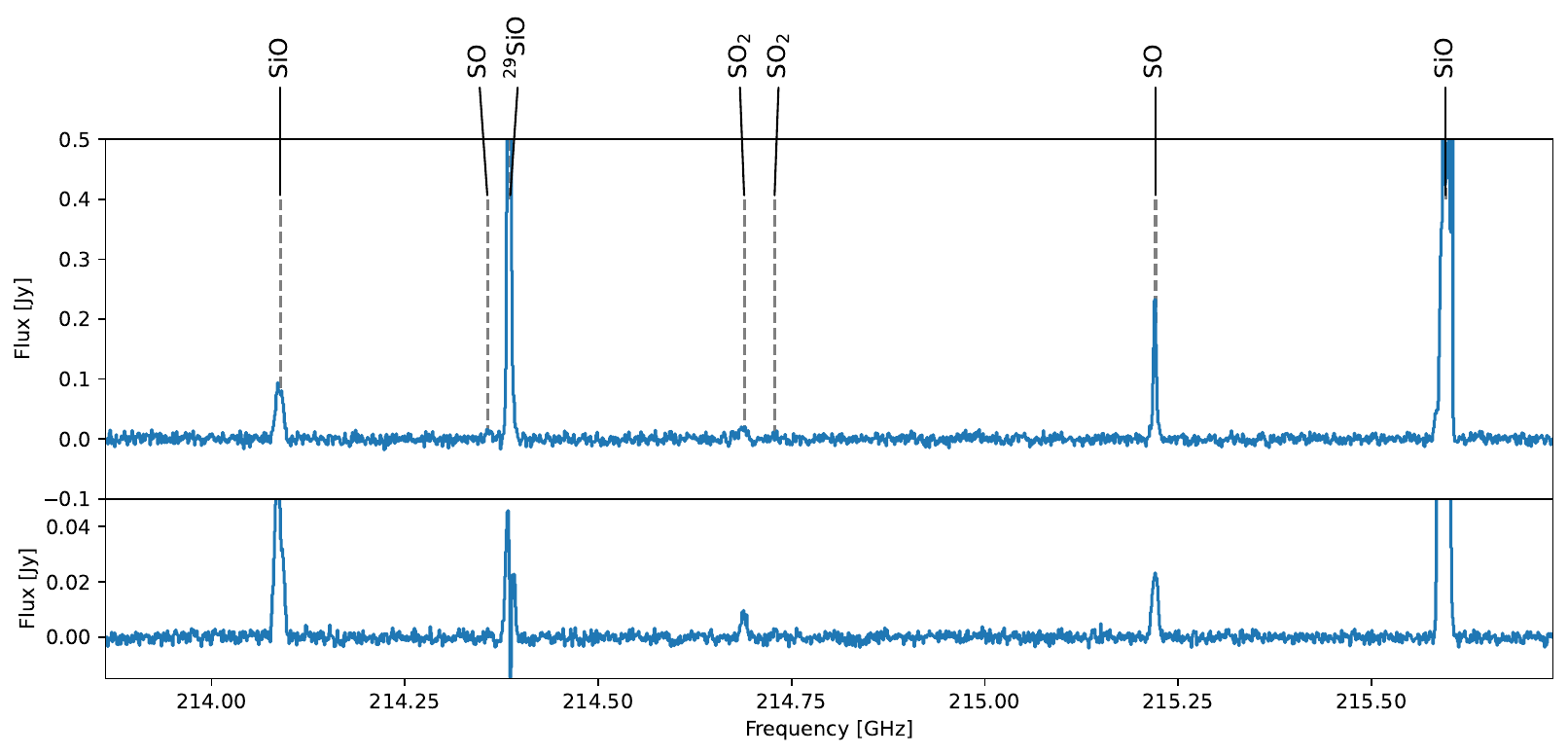}}
\\
\subfloat{\includegraphics[width=\textwidth]{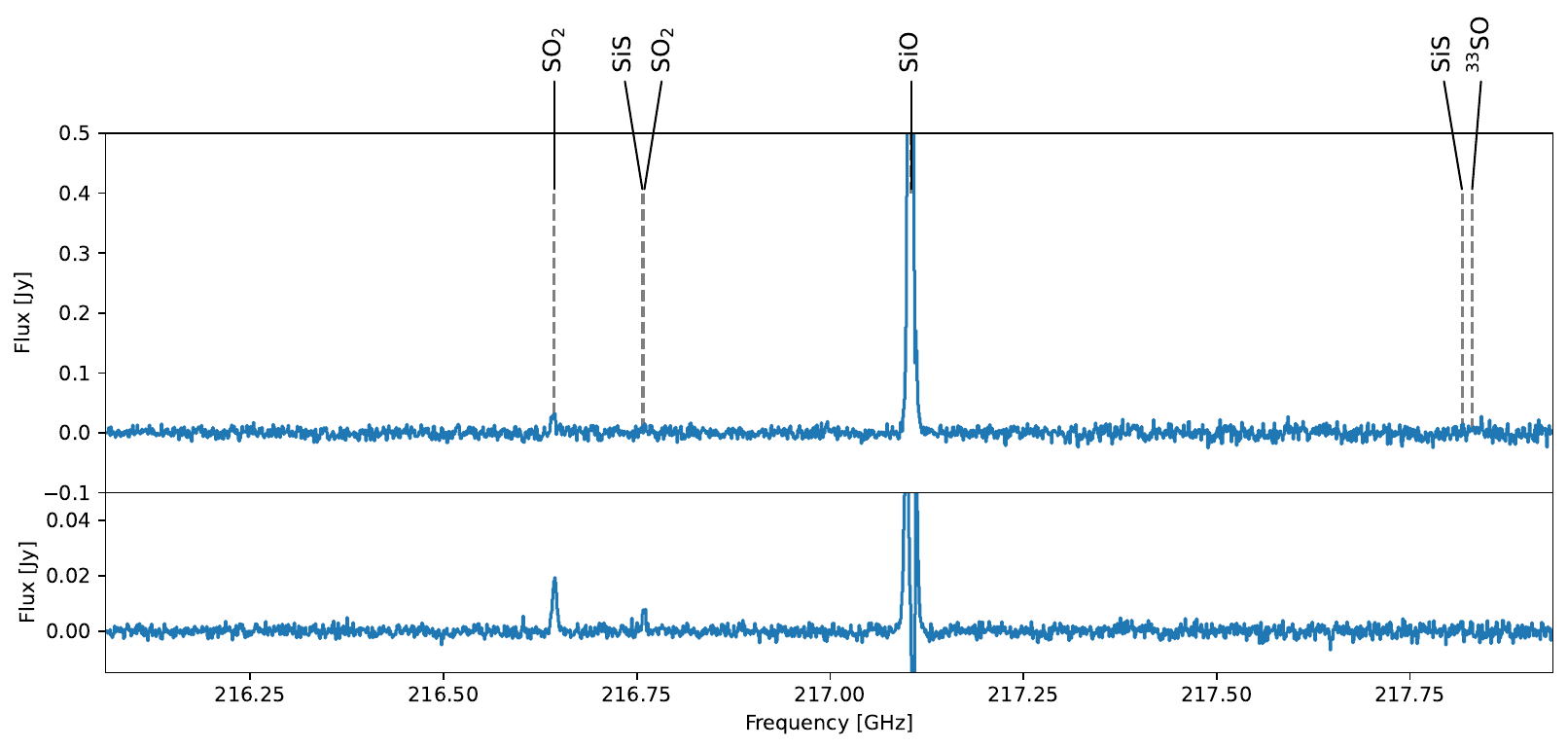}} 
\end{figure}

\begin{figure}[thbp]
\ContinuedFloat
\centering
\caption{continued}
\subfloat{\includegraphics[width=\textwidth]{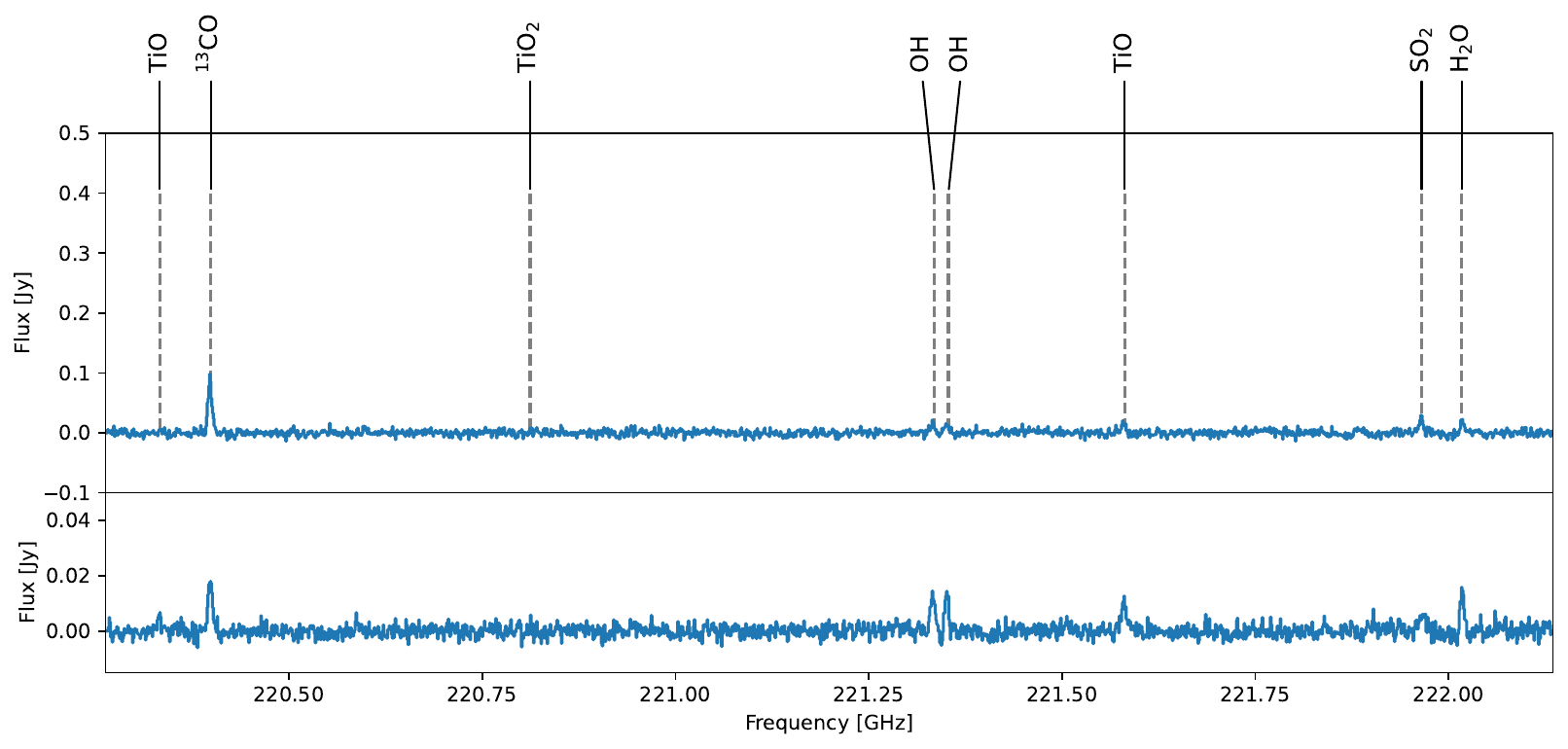}}
\\
\subfloat{\includegraphics[width=\textwidth]{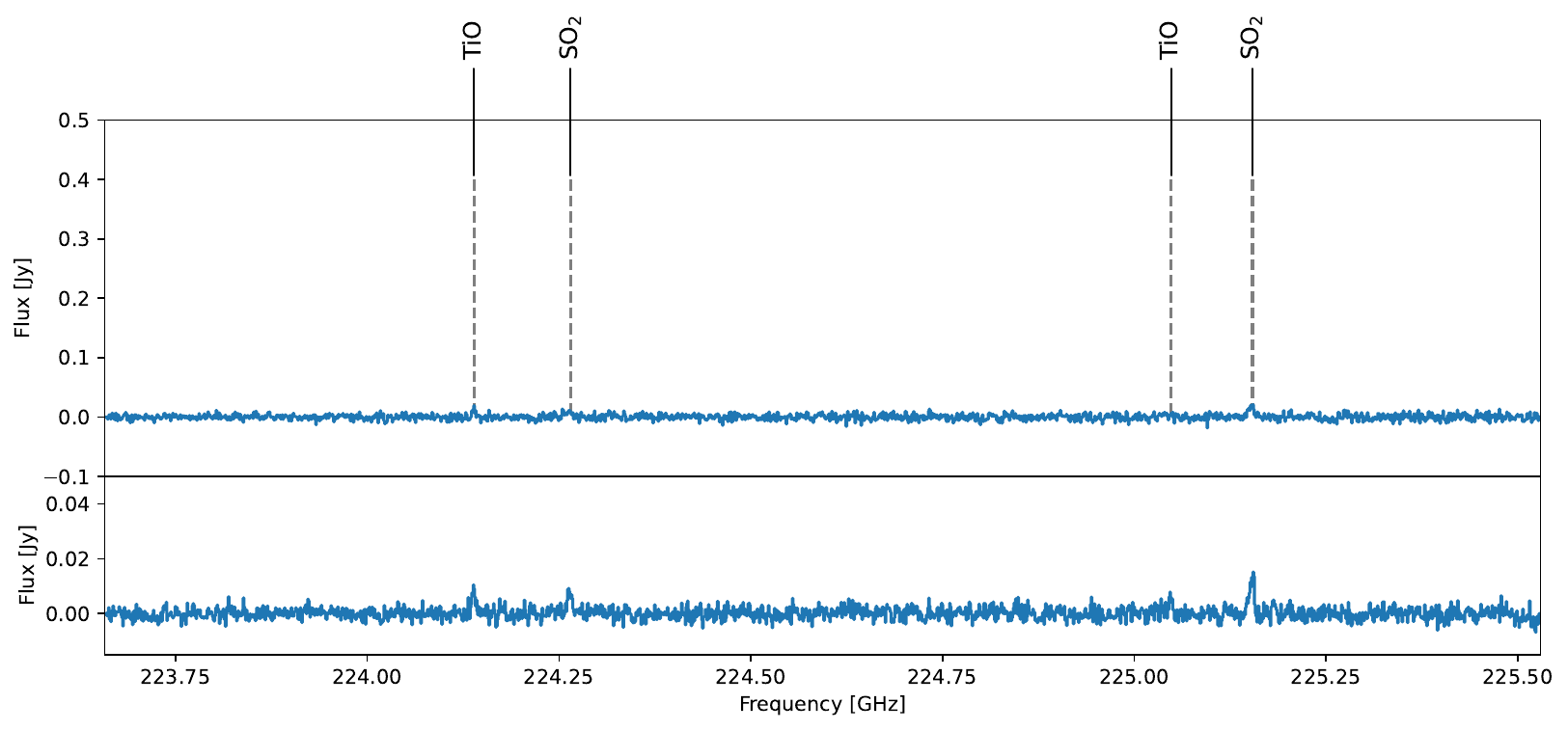}}
\end{figure}

\begin{figure}[thbp]
\ContinuedFloat
\centering
\caption{continued}
\subfloat{\includegraphics[width=\textwidth]{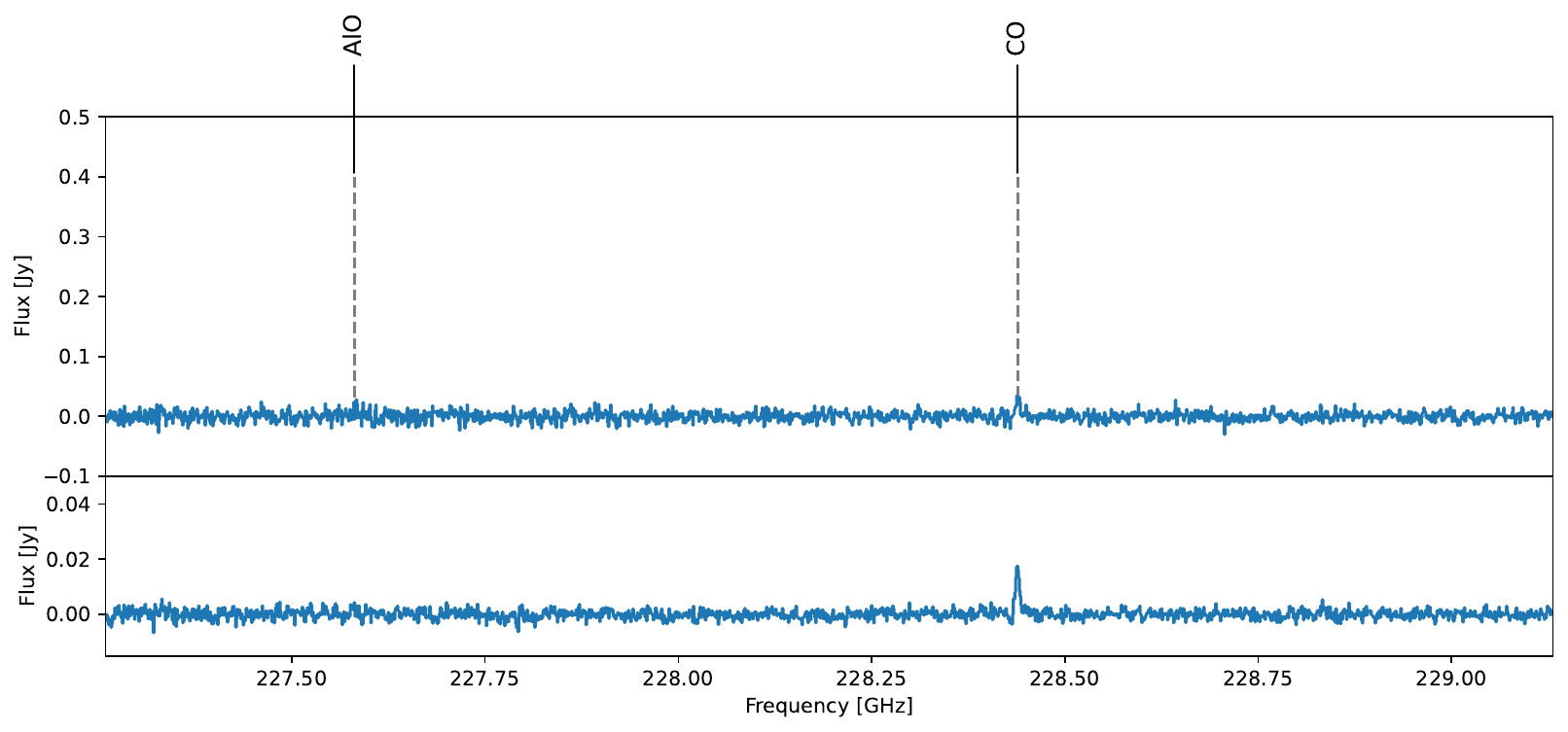}}
\\
\subfloat{\includegraphics[width=\textwidth]{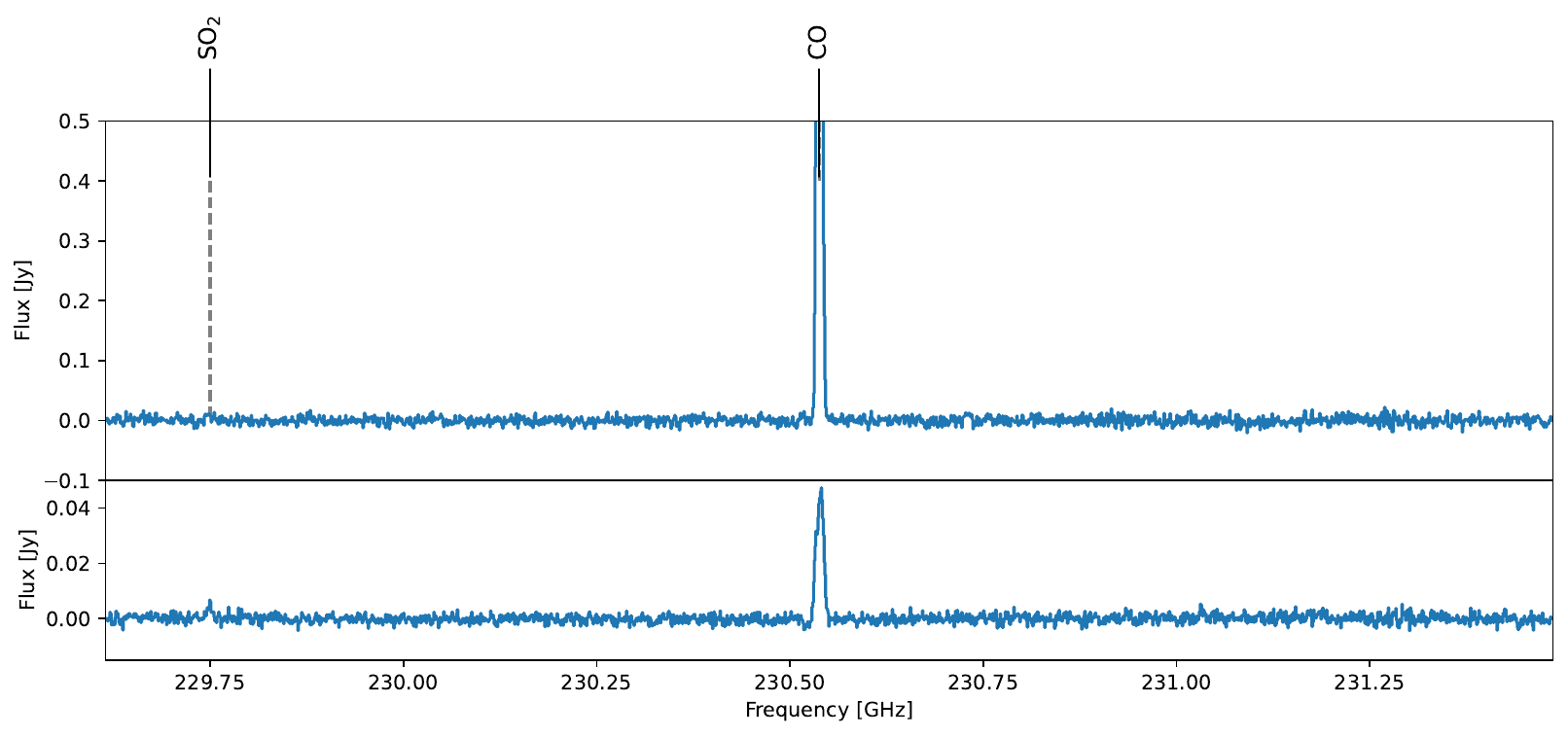}}
\end{figure}

\begin{figure}[thbp]
\ContinuedFloat
\centering
\caption{continued}
\subfloat{\includegraphics[width=\textwidth]{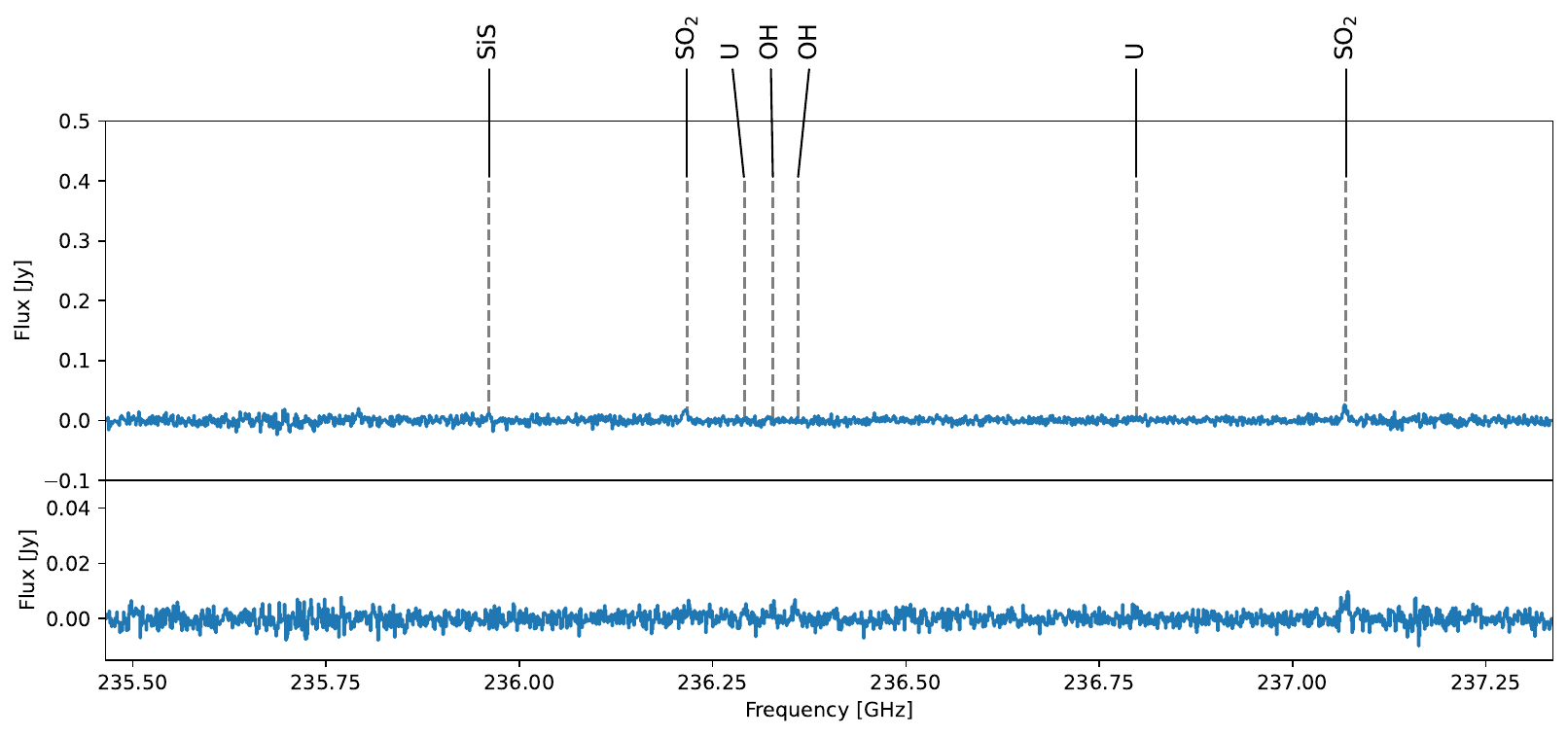}}
\\
\subfloat{\includegraphics[width=\textwidth]{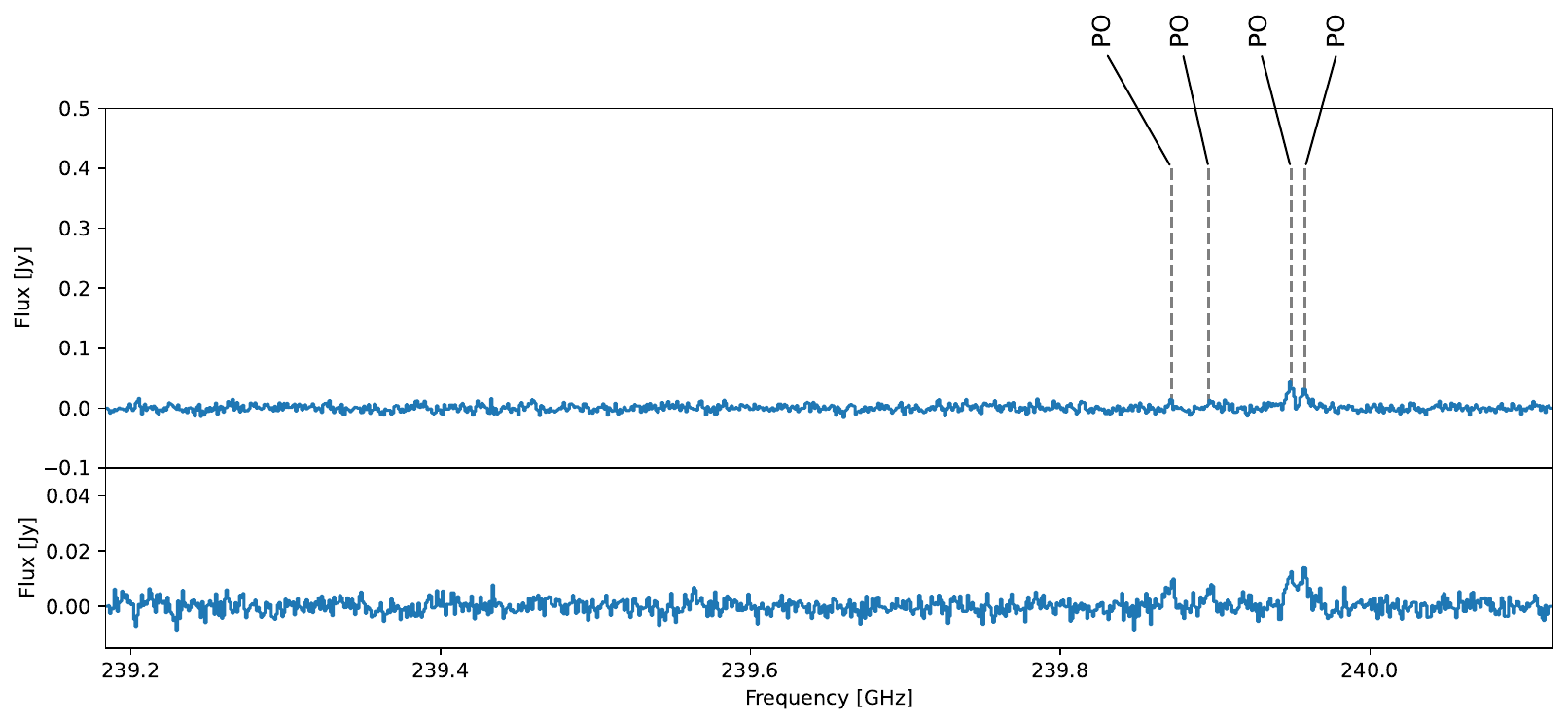}}
\end{figure}

\begin{figure}[thbp]
\ContinuedFloat
\centering
\caption{continued}
\subfloat{\includegraphics[width=\textwidth]{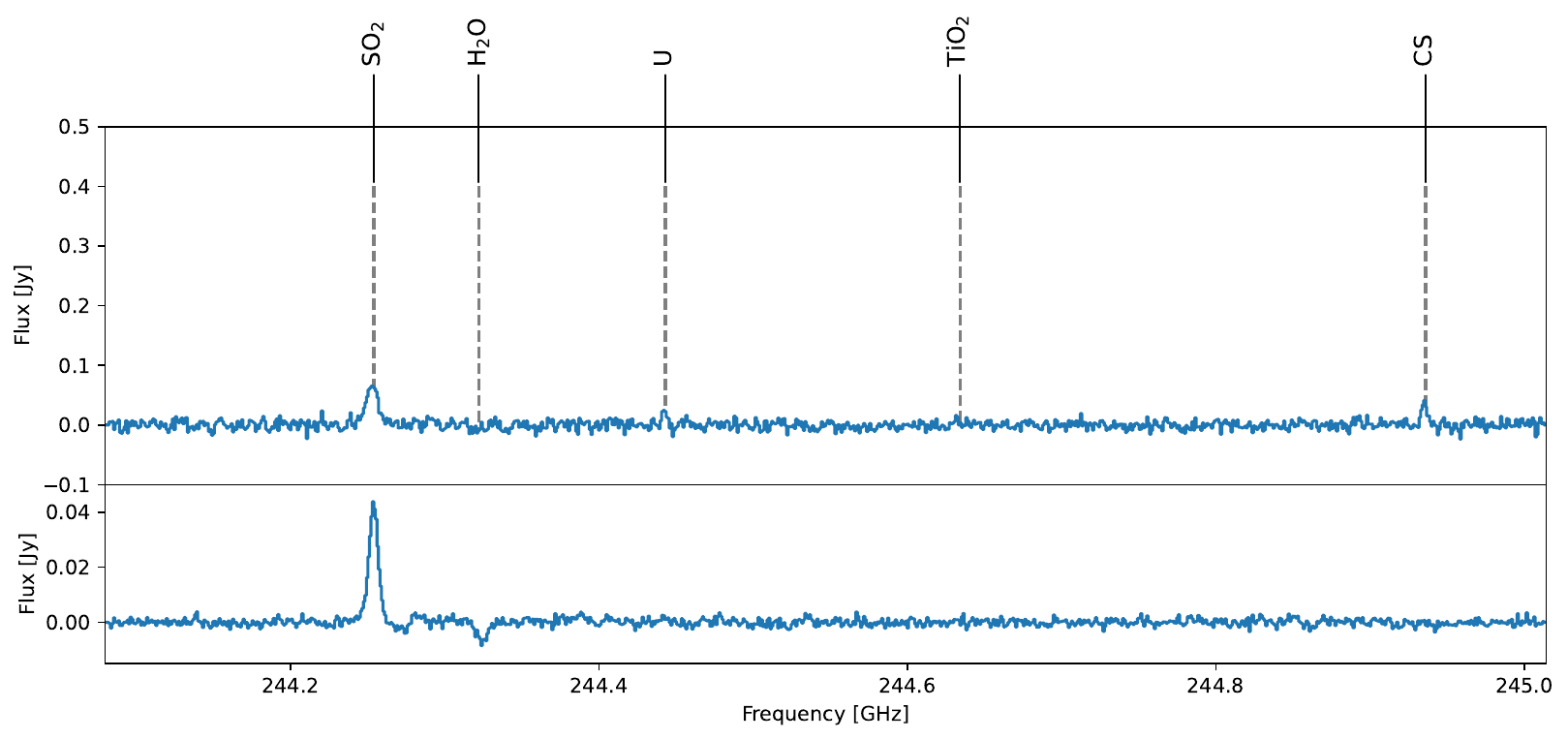}}
\\
\subfloat{\includegraphics[width=\textwidth]{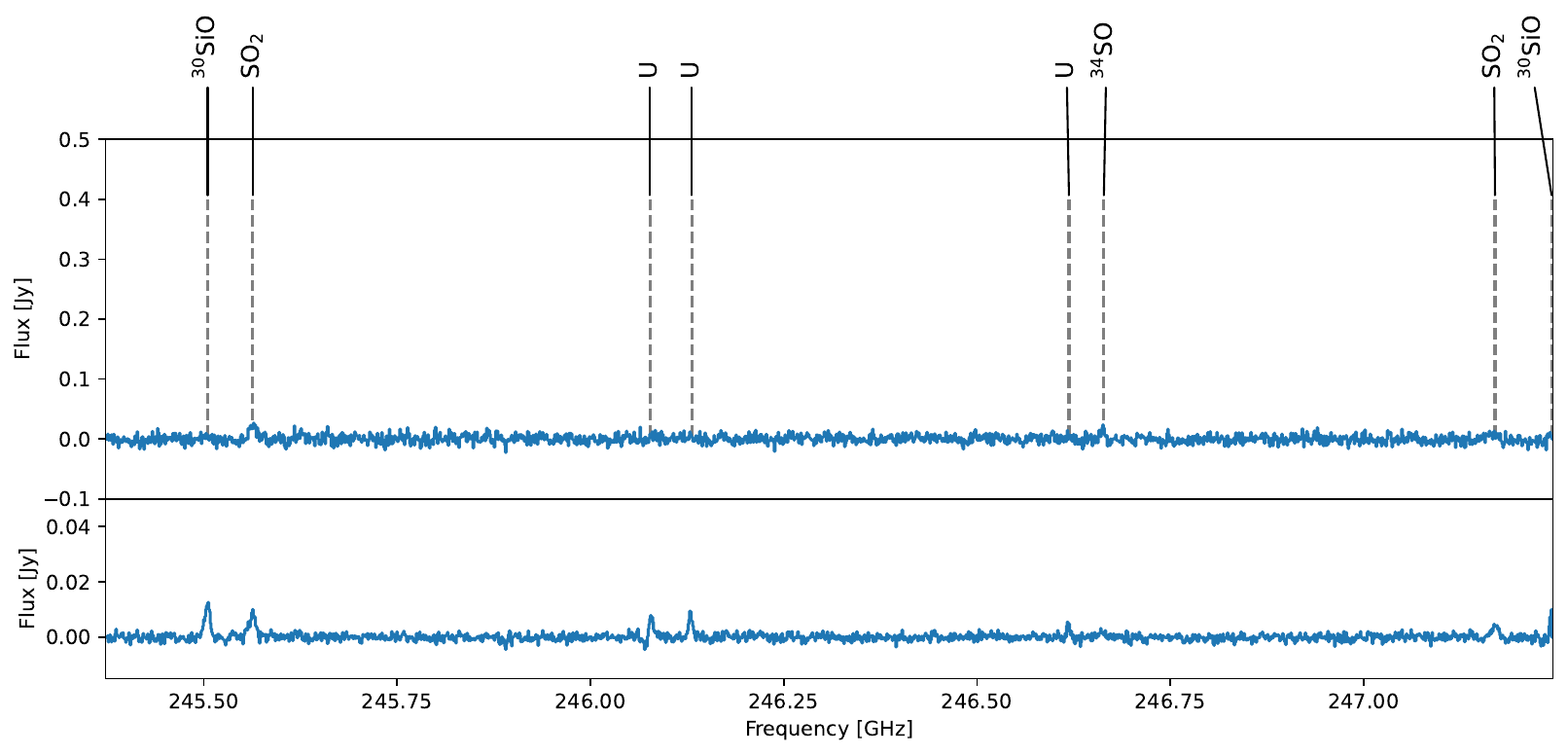}}
\end{figure}

\begin{figure}[thbp]
\ContinuedFloat
\centering
\caption{continued}
\subfloat{\includegraphics[width=\textwidth]{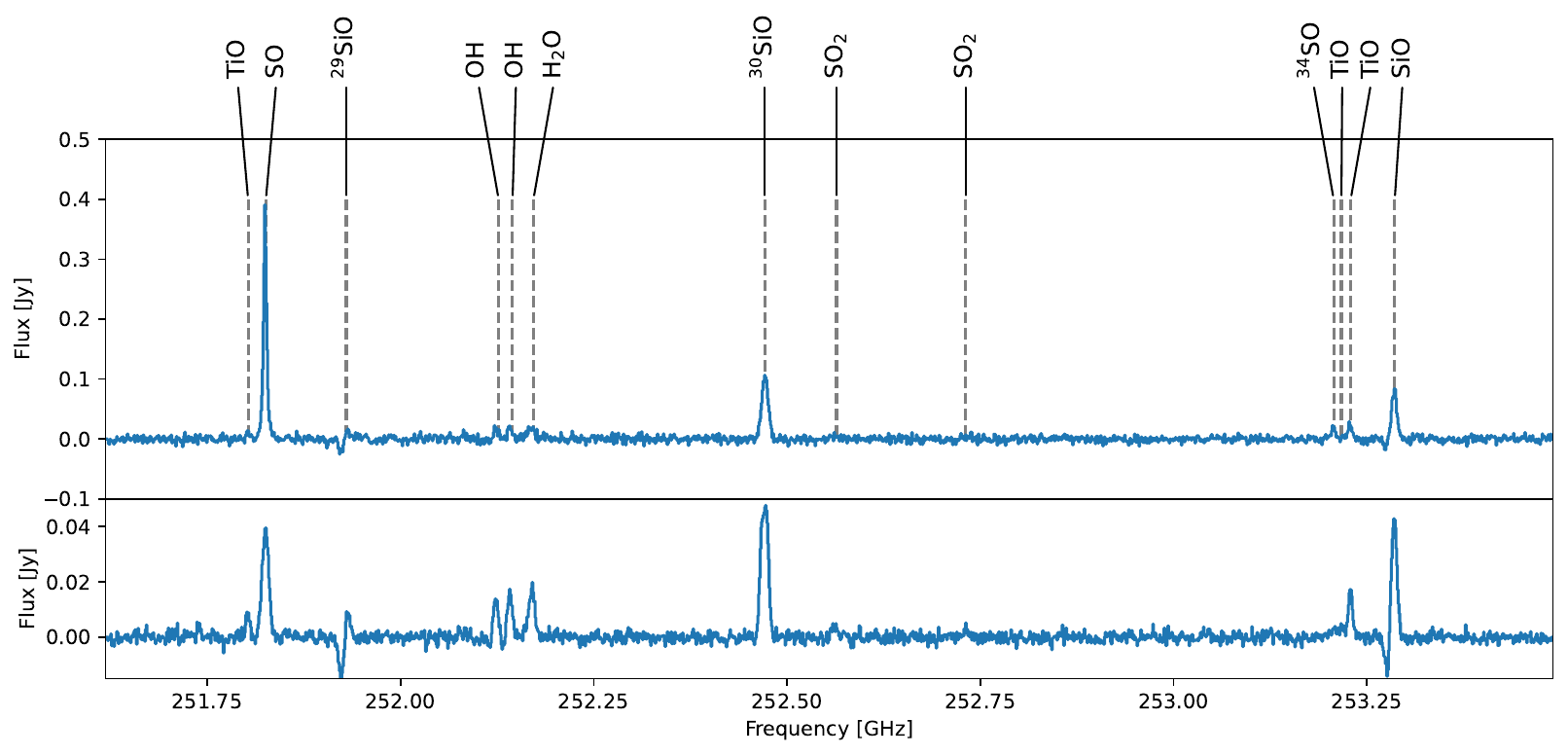}}
\\
\subfloat{\includegraphics[width=\textwidth]{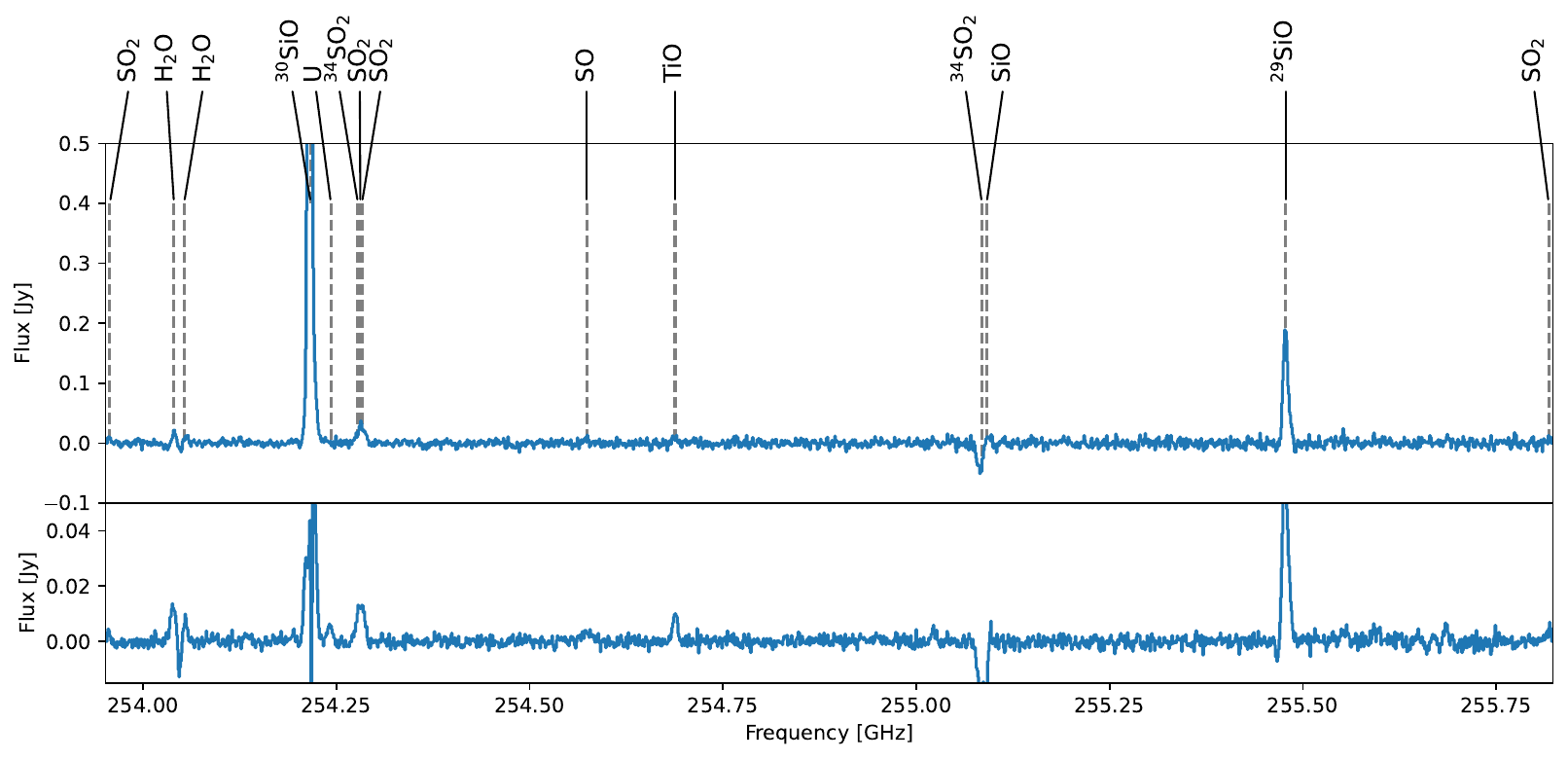}}
\end{figure}

\begin{figure}[thbp]
\ContinuedFloat
\centering
\caption{continued}
\subfloat{\includegraphics[width=\textwidth]{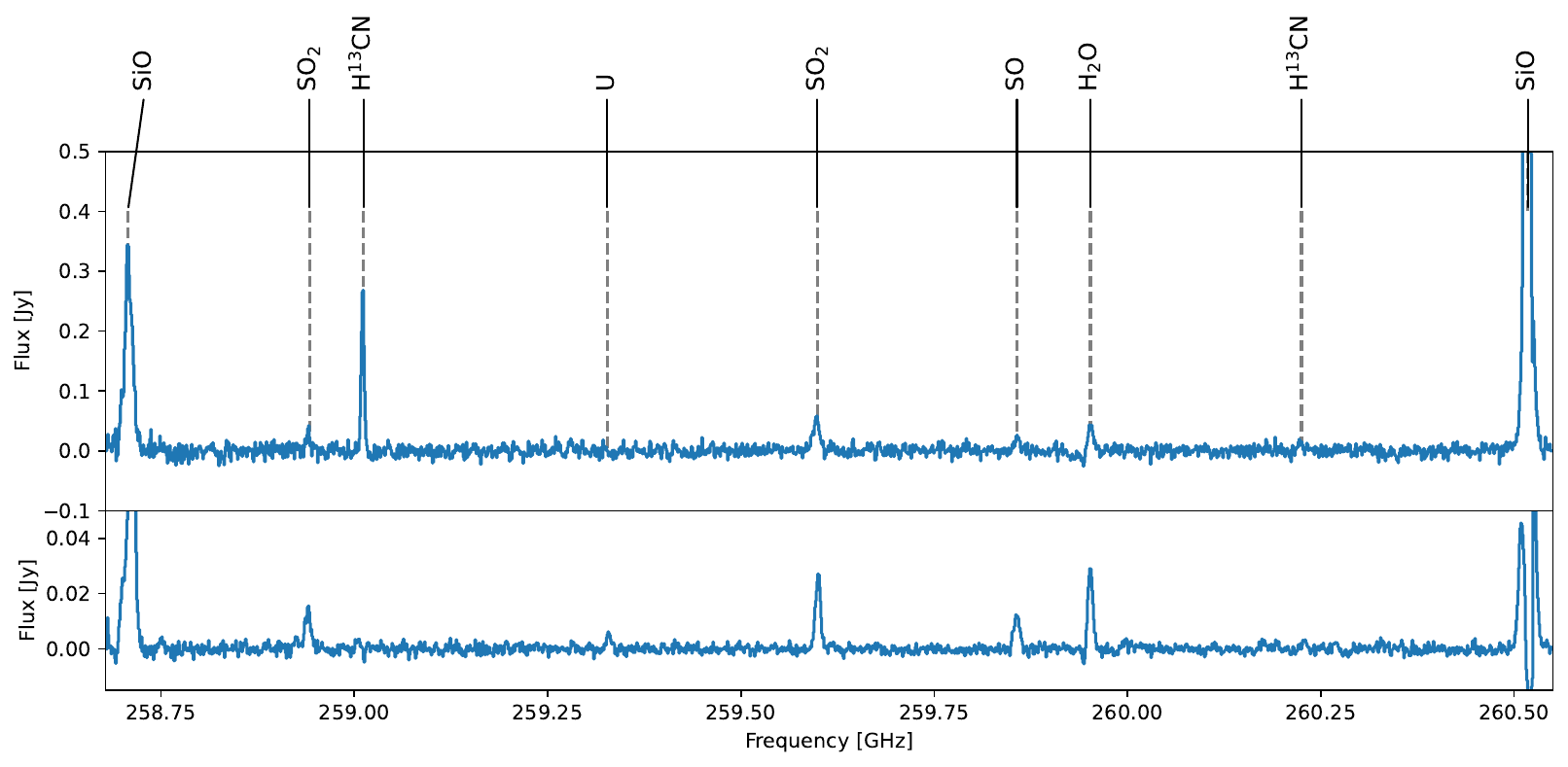}}
\\
\subfloat{\includegraphics[width=\textwidth]{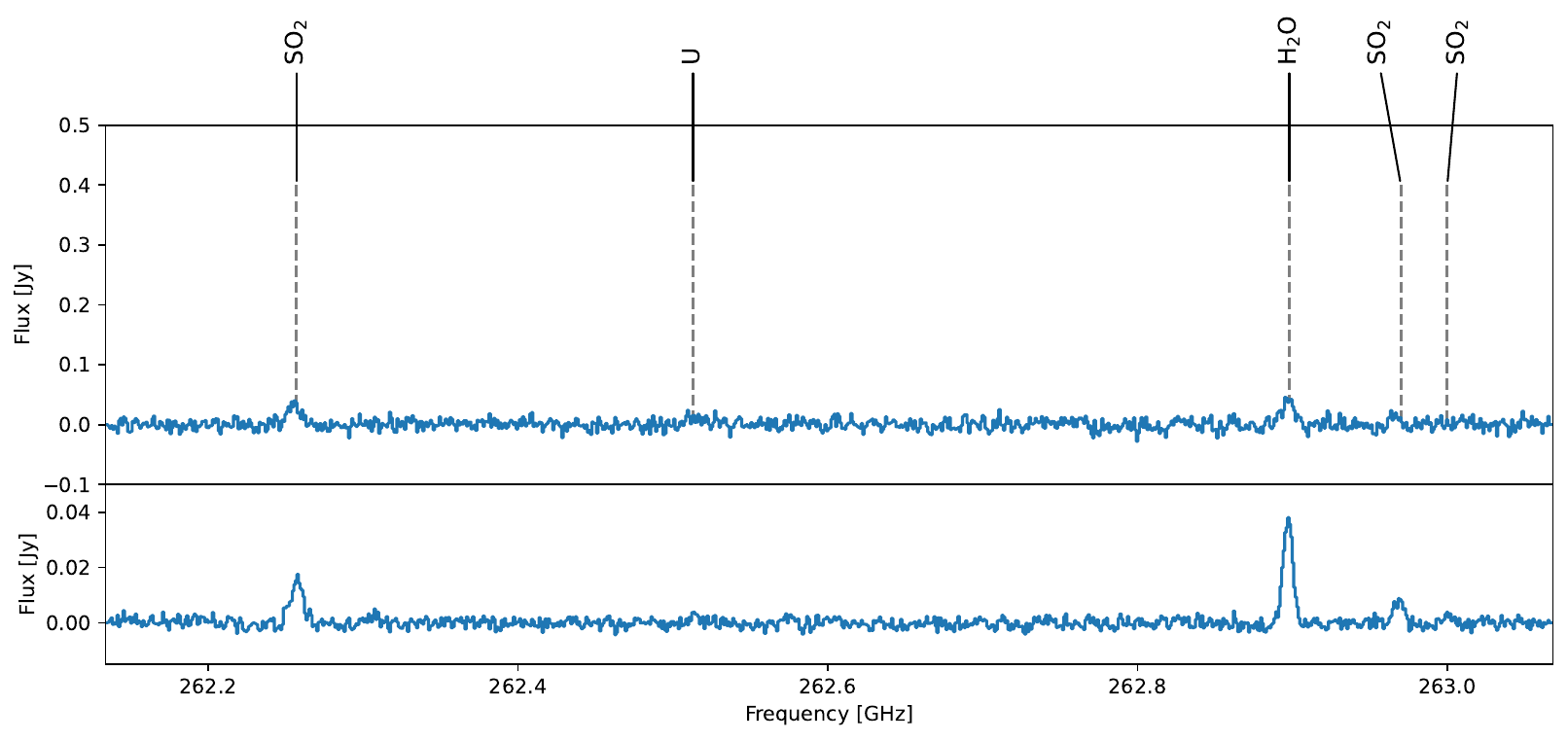}}
\end{figure}

\begin{figure}[thbp]
\ContinuedFloat
\centering
\caption{continued}
\subfloat{\includegraphics[width=\textwidth]{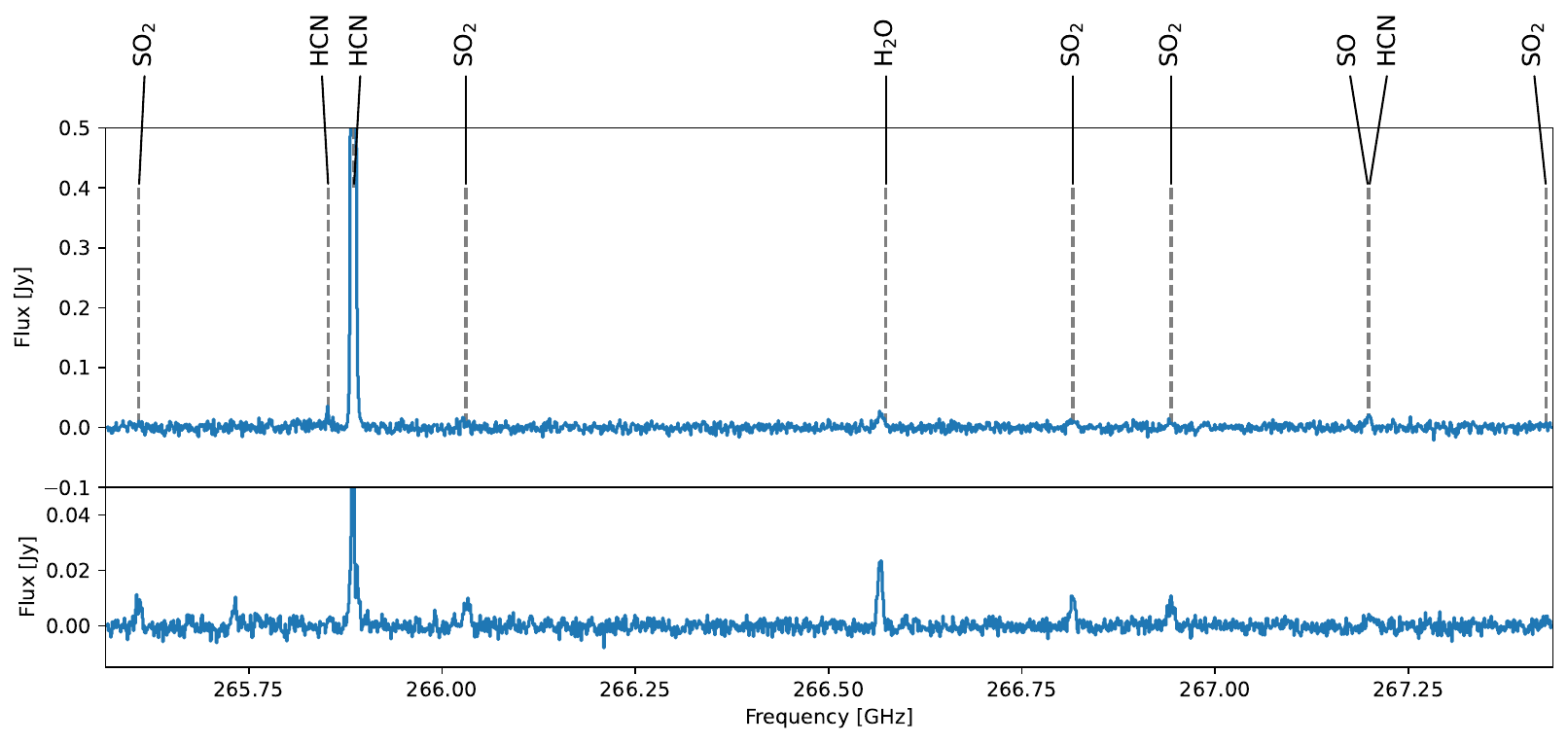}}
\\
\subfloat{\includegraphics[width=\textwidth]{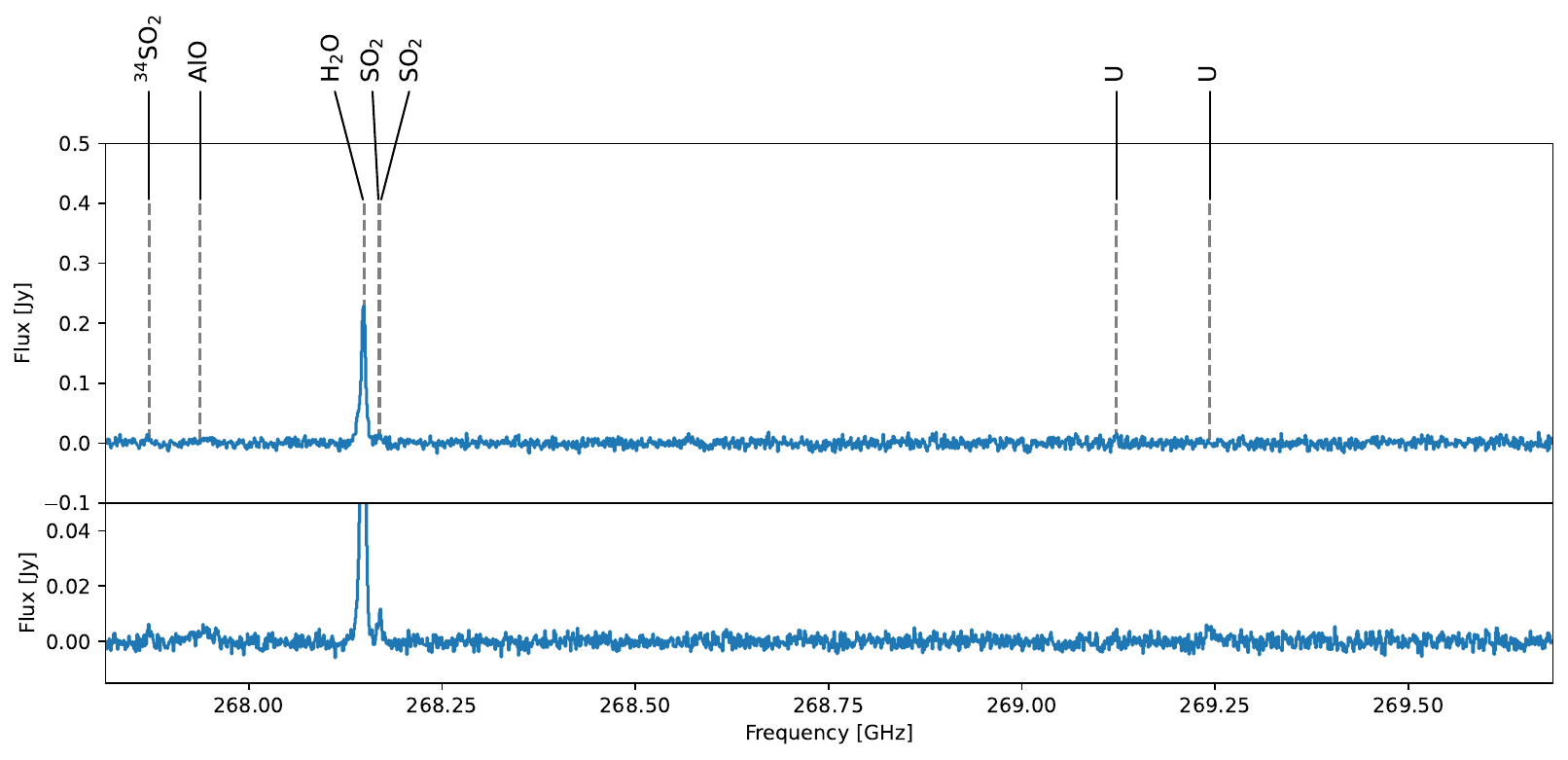}}
\end{figure}

\end{appendix}

\end{document}